\documentclass[aps,11pt,prd,groupedaddress,nofootinbib,notitlepage,eqsecnum,preprintnumbers]{revtex4-2}
\usepackage[utf8]{inputenc}
\usepackage{hyperref}
\usepackage{xcolor}
\usepackage{graphicx}
\usepackage{amsmath,amssymb}
\usepackage{bm}
\usepackage{comment}
\usepackage[shortlabels]{enumitem}
\usepackage{overpic}
\usepackage{layouts}
\usepackage{ulem}

\def\Mpc{{\rm Mpc}}

\def\keq{k_{\rm eq}}
\def\aeq{a_{\rm eq}}
\def\xNL{x_{\rm NL}}

\renewcommand{\O}[1]{\mathcal{O}\left({#1}\right)}

\begin{document}
\title{Gravitational waves from dark matter isocurvature}

\author{\textsc{Guillem Dom\`enech$^{a}$}}
    \email{{domenech}@{pd.infn.it}}
\author{\textsc{Samuel Passaglia$^{b}$}}
    \email{{samuel.passaglia}@{ipmu.jp}}
\author{\textsc{S\'{e}bastien Renaux-Petel$^{c}$}}
    \email{{renaux}@{iap.fr}}

\affiliation{\vspace{3mm}$^{a}$ \small{INFN Sezione di Padova, I-35131 Padova, Italy}}
\affiliation{$^b$ Kavli Institute for the Physics and Mathematics of the Universe (WPI), Chiba 277-8583, Japan}
\affiliation{$^c$ Institut d'Astrophysique de Paris, GReCO, UMR 7095 du CNRS et de Sorbonne Universit\'{e}, 98bis boulevard Arago, 75014 Paris, France}

\begin{abstract}
The primordial fluctuations on large scales are adiabatic, but on smaller scales this need not be the case. Here we derive the general analytical framework to compute the stochastic gravitational wave background induced by primordial cold dark matter isocurvature fluctuations on small scales. We find that large isocurvature fluctuations can yield an observable gravitational wave signal, with a spectrum distinct from the one induced by adiabatic perturbations, and we provide for the first time the exact analytic expression of the kernel necessary to compute this signal. We then forecast the constraining power of future gravitational wave detectors on dark matter isocurvature on small scales and find they will dramatically improve on existing constraints.
\end{abstract}

\maketitle
\begingroup
\setlength{\parskip}{0.0cm}
\hypersetup{linkcolor=black}
\newpage
\tableofcontents
\endgroup
\newpage

\section{Introduction \label{sec:Intro}}

	It is well established that on the largest scales the primordial fluctuations of our universe were predominantly adiabatic \cite{WMAP:2003ivt,Akrami:2018odb}. This means that energy density fluctuations in the early universe left the relative particle number densities unperturbed \cite{Langlois:2003fq}. In a geometrical view adiabatic fluctuations are curvature perturbations of the spacetime.

	Isocurvature perturbations are the orthogonal type of primordial fluctuations, corresponding to an unperturbed total energy density but inhomogeneities in the relative number densities of different particle species \cite{Kodama:1985bj,Bucher:1999re}. Adiabatic and isocurvature perturbations evolve differently in the early universe, and from the Planck observations of the Cosmic Microwave Background (CMB) anisotropies, the isocurvature modes are constrained to account for, roughly speaking, less than $1\%$ of the primordial fluctuations in the range of scales $10^{-3}\,\Mpc^{-1}\lesssim k \lesssim 10^{-1}\,\Mpc^{-1}$ \cite{Akrami:2018odb}.

	On smaller scales than those probed by the CMB, the amplitude of isocurvature fluctuations is generally subject to much weaker constraints. While in the future CMB spectral distortions offer an interesting window to test isocurvature modes in the range of scales $1\,\Mpc^{-1}\lesssim k \lesssim 10^{6}\,\Mpc^{-1}$ \cite{Chluba:2013dna,Chluba:2019kpb}, and isocurvature perturbations of the baryons are constrained to be less than $1\%$ in the range of scales $10^{-1}\,\Mpc^{-1}\lesssim k\lesssim4\times10^{8}\,\Mpc^{-1}$ because they directly affect the processes of Big Bang Nucleosynthesis \cite{Inomata:2018htm}, in this work we focus on the practically unconstrained Cold Dark Matter (CDM) isocurvature perturbations. A very large amplitude of CDM isocurvature on scales $k\gtrsim 1\, \Mpc^{-1}$, even much larger than $\O{1}$, is generally compatible with current observations up to some uncertainties on the end state of the fluctuations and possible non-trivial particle interactions \cite{Kohri:2014lza,Yang:2014lsg,Nakama:2017qac}.

	Primordial isocurvature fluctuations on small scales can be large only because of their peculiar nature. At early times, the geometry is by definition unaffected by them \cite{Kodama:1986fg}. As time evolves and the relative energy densities of matter and radiation change, isocurvature fluctuations are gradually transferred to curvature perturbations. However, when a mode enters the horizon during radiation domination the curvature perturbation starts to decay. Thus the transfer from isocurvature to curvature perturbation is suppressed for modes which enter the horizon before matter-radiation equality, with a greater suppression for the smallest scales which enter the horizon earliest \cite{Kodama:1986fg}.

	This means that even though the amplitude of isocurvature fluctuations can be much larger than unity, curvature fluctuations can remain small. Only when the local energy density of CDM overcomes the energy density of radiation does perturbation theory break down, and in this case primordial black holes (PBHs) can form \cite{Passaglia:2021jla}. This provides an alternative PBH formation channel to the typical collapse of primordially adiabatic fluctuations \cite{Zeldovich:1967lct,Hawking:1971ei,Carr:1974nx,Meszaros:1974tb,Carr:1975qj,Khlopov:1985jw,Niemeyer:1999ak} (see also Refs.~\cite{Khlopov:2008qy,Sasaki:2018dmp,Carr:2020gox,Carr:2020xqk,Green:2020jor,Escriva:2021aeh} for recent reviews). Large isocurvature fluctuations may be a consequence of phase transitions \cite{Dolgov:1992pu}, light spectator fields during inflation \cite{Chung:2017uzc,Chung:2021lfg}, or clustering of Q-balls \cite{Cotner:2019ykd}, though forming sufficiently large fluctuations for this mechanism to function may be challenging.

	In this paper, we follow up on the work of Ref.~\cite{Passaglia:2021jla} by investigating the gravitational waves (GWs) induced by large primordial CDM isocurvature fluctuations. This type of GWs is generated by spacetime oscillations induced by the evolution of the curvature perturbation when it enters the horizon \cite{Tomita,Matarrese:1992rp,Matarrese:1993zf,Ananda:2006af,Baumann:2007zm,Saito:2008jc,Saito:2009jt} (see Refs.~\cite{Yuan:2021qgz,Domenech:2021ztg} for recent reviews). However, the arguments above on the size of the curvature perturbation sourced by isocurvature show that one should expect a suppressed GW production in this scenario. Nevertheless, if the amplitude of the isocurvature fluctuations is large enough so as to compensate for the suppression factor, there can be a substantial generation of curvature fluctuations and hence gravitational waves.

	The induced GW signal studied in this work is important for two reasons. First, as the counterpart of PBHs from large initial isocurvature, it has the potential to be the distinctive signature of this new PBH production scenario. Second, regardless of whether PBHs form, it is a novel probe of the nature of the primordial fluctuations on small scales which could in general be adiabatic or isocurvature or a combination thereof. It also emphasizes that large adiabatic primordial fluctuations are not the only source of observable induced GWs (see also Refs.~\cite{Papanikolaou:2020qtd,Domenech:2020ssp,Domenech:2021wkk,Kozaczuk:2021wcl} for GWs induced by fluctuations in the density of PBHs). We find that the induced GW signal from primordial CDM isocurvature has a different shape than the one induced by adiabatic fluctuations, and therefore in the case of a detection of a GW background signal one should in general consider both contributions. 

	In the absence of a GW signal, this work provides at the moment the best prospects for constraining isocurvature fluctuations on the smallest scales. In particular, we find that future GW detectors, such as ET \cite{Punturo:2010zz,Hild:2010id,Maggiore:2019uih}, LISA \cite{Audley:2017drz,Baker:2019nia} and SKA \cite{Carilli:2004nx,Janssen:2014dka,Weltman:2018zrl}, would improve the constraints from the absence of PBHs \cite{Passaglia:2021jla} by several orders of magnitude in the range of scales $10^{7}\,\Mpc^{-1}\lesssim k \lesssim 10^{18}\,\Mpc^{-1}$. This constraints complement those on larger scales coming from CMB anisotropies \cite{Akrami:2018odb} and spectral distortions \cite{Chluba:2013dna}.	

	This paper is organized as follows. In \S\ref{sec:isocurvature} we review the evolution of cosmological perturbations in the presence of large CDM isocurvature and provide new exact analytic solutions for the evolution of primordial isocurvature perturbations during radiation domination. In \S\ref{sec:inducedGWs} we derive for the first time the spectrum of GWs induced by initial CDM isocurvature perturbations. In \S\ref{sec:constraints} we then use the sensitivity of future GW detectors to forecast future constraints on CDM isocurvature. Finally we discuss the implications of our work in \S\ref{sec:conclusions}.

\section{Curvature and isocurvature evolution\label{sec:isocurvature}}

	The evolution of primordial fluctuations in the early universe induces, in general, GWs \cite{Tomita,Matarrese:1992rp,Matarrese:1993zf,Ananda:2006af,Baumann:2007zm}. In some sense, curvature (or total energy density) fluctuations ``back-react'' onto the metric when acoustic oscillations of the total energy density fluctuations cause spacetime oscillations. At second order in cosmological perturbation theory, this appears as a scalar squared source to the equation of motion for tensor modes. These induced GWs have been shown to be a promising and unique tool to probe new physics during \cite{Garcia-Bellido:2016dkw,Gong:2017qlj,Ando:2018nge,Byrnes:2018txb,Gao:2019kto,Xu:2019bdp,Liu:2020oqe,Cai:2019amo,Ozsoy:2019lyy,Ozsoy:2020kat,Ragavendra:2020sop,Fumagalli:2020nvq,Braglia:2020eai,Braglia:2020taf,Fumagalli:2021cel,Bastero-Gil:2021fac,Fumagalli:2021mpc,Fumagalli:2021dtd,Saikawa:2018rcs} and after inflation \cite{Assadullahi:2009nf,Inomata:2019zqy,Inomata:2019ivs,Inomata:2020lmk,Papanikolaou:2020qtd,Domenech:2020ssp,Domenech:2021wkk,Dalianis:2020gup,Hajkarim:2019nbx,Bhattacharya:2019bvk,Domenech:2019quo,Domenech:2020kqm,Dalianis:2020cla,Abe:2020sqb,Witkowski:2021raz} and to map the primordial curvature spectrum on small scales \cite{Saito:2008jc,Saito:2009jt,Assadullahi:2009jc,Bugaev:2009zh,Bugaev:2009kq,Bugaev:2010bb,Inomata:2018epa}.

	GWs induced by early isocurvature fluctuations are more subtle. Isocurvature perturbations cannot directly induce GWs, as by definition they correspond to a homogeneous total energy density and an unperturbed geometry. Instead, GWs are induced by curvature perturbations sourced by the isocurvature as the CDM gains in energy relative to the radiation. However, for modes which enter the horizon before matter-radiation equality, only a fraction of the early isocurvature can be transferred into curvature fluctuations before the curvature begins to decay inside the horizon. Therefore the second-order GWs must be suppressed by the square of the transfer efficiency. 

	We now review the sourcing and evolution of curvature perturbations in the presence of early isocurvature in a matter-radiation universe. The CDM isocurvature perturbations are formally defined by \cite{Kodama:1985bj,Malik:2008im}
	\begin{equation}\label{eq:definitionS}
	S\equiv \frac{\delta\rho_m}{\rho_m}-\frac{3}{4}\frac{\delta\rho_r}{\rho_r}\,,
	\end{equation}
	where $\rho_m$ and $\delta\rho_m$ are the energy density of CDM and its fluctuations, and the subscript ``$r$'' denotes those quantities for the radiation fluid. Note that the notion of isocurvature, and so the definition \eqref{eq:definitionS}, is independent of the gauge choice. In the matter sector, there are also perturbations to the CDM and radiation fluid velocities, which we respectively denote by $V_m$ and $V_r$.  Details on the perturbation expansion of the matter sector can be found in Appendix~\ref{app:formulas}.

	Interestingly, one may restrict to linear theory even when $S$ is much larger than unity. Crudely speaking, this is because in a radiation dominated universe the quantity that matters for the validity of the perturbative expansion is $\delta\rho_m/\rho_r<1$ rather than $\delta\rho_m/\rho_m<1$. While below we will show more precisely the conditions under which linear theory holds, we direct the interested reader to Appendix \ref{app:lineartheory} for more details.

	To describe the metric perturbations, we work throughout in the Newtonian (or shear-free) gauge where the perturbed flat Friedmann-Lema\^itre-Robertson-Walker (FLRW) metric takes the form \cite{Kodama:1985bj}
	\begin{align}
	ds^2=a^2(\tau)\left[-(1+2\Psi)d\tau^2+(\delta_{ij}+2\Phi\delta_{ij}+h_{ij})dx^idx^j\right]\,,
	\end{align}
	with $a$ the scale factor, $\tau$ the conformal time, $\Psi$ the lapse perturbation, $\Phi$ the curvature perturbation, and $h_{ij}$ the tensor modes. In a matter-radiation universe, one finds an exact solution for the background given by \cite{Mukhanov:2005sc}
	\begin{align}
	\frac{a}{\aeq}=2\frac{\tau}{\tau_*}+\left(\frac{\tau}{\tau_*}\right)^2\,\quad {\rm and}\quad {\cal H}\equiv\frac{a'}{a}= \frac{2}{\tau}\frac{1+\tau/\tau_*}{2+\tau/\tau_*}\,,
	\end{align}
	where $'$ denotes a derivative with respect to the conformal time, the absolute normalization is provided by $\tau_* = 2(\Omega_{m,0} H_0^2 / \aeq)^{-1/2}$, and throughout we use the subscript ``eq'' to denote evaluation at the time of matter-radiation equality.

	In the absence of anisotropic stress
	we have that $\Psi+\Phi=0$. Pure isocurvature fluctuations imply that the total energy density is initially homogeneous\footnote{In fact, one also requires that the first derivative of the total density contrast vanishes \cite{Kodama:1986ud}.}, i.e. $\delta\rho=\delta\rho_m+\delta\rho_r=0$. Then by the Einstein equations the metric is also unperturbed and so we initially have $\Phi=0$.  The curvature and isocurvature perturbations at linear level then evolve according to \cite{Kodama:1986ud,Malik:2008im,Domenech:2020ssp}
	\begin{align}\label{eq:eomPhi}
	\Phi''+3{\cal H}(1+c_s^2)\Phi'+({\cal H}^2(1+3c_s^2)+2{\cal H}')\Phi+c_s^2k^2\Phi=\frac{a^2}{2}\rho_mc_s^2S\,,
	\end{align}
	and
	\begin{align}\label{eq:eomS}
	S''&+ 3{\cal H}c_s^2S'-\frac{3}{2a^2\rho_r}c_s^2{k^4\Phi}+\frac{3\rho_m}{4\rho_r}c_s^2k^2 S=0\,,
	\end{align}
	where 
	\begin{align}
	c_s^2\equiv\frac{4}{9}\frac{\rho_r}{\rho_m+4\rho_r/3}\,,
	\end{align}
	is the sound speed of perturbations. We suppress a $k$-marker here on the Fourier modes $\Phi$ and $S$ and only restore it when strictly necessary. The relative velocity is then obtained by
	\begin{align}\label{eq:Vreleom}
	V_{\rm rel}\equiv V_m-V_r=k^{-2}S'\,.
	\end{align}
	Note that despite the factor $k^{-2}$ in Eq.~\eqref{eq:Vreleom}, the relative velocity vanishes on superhorizon scales because the isocurvature perturbation is constant. 

	We first solve these equations assuming that linear theory holds and later confirm that the situation $S\gg1$ is allowed within linear theory. We will be interested only in modes which enter the horizon before matter-radiation equality, that is $k\gg \keq$ where $\keq={\cal H}_{\rm eq} \sim 0.01\, \Mpc^{-1}$ is the mode which enters the horizon at matter-radiation equality. These modes are the ones that may induce GWs in the sensitivity domain of future detectors. Therefore to track such modes from superhorizon scales, through horizon crossing, and deep into the subhorizon regime, we can expand Eqs.~\eqref{eq:eomPhi} and \eqref{eq:eomS} in the radiation domination limit $\tau/\tau_*\ll1$ to find
	\begin{align}\label{eq:eomPhiRD}
	\frac{d^2\Phi}{dx^2}+\frac{4}{x}\frac{d\Phi}{dx}+\frac{1}{3}\Phi+\frac{1}{4\sqrt{2}\kappa x}\left(x\frac{d\Phi}{dx}+(1-x^2)\Phi-2S\right)\simeq 0\,,
	\end{align}
	and
	\begin{align}\label{eq:eomSRD}
	\frac{d^2S}{dx^2}&+\frac{1}{x}\frac{dS}{dx}-\frac{x^2}{6}\Phi-\frac{1}{2\sqrt{2}\kappa}\left(\frac{dS}{dx}-\frac{x}{2}S-\frac{x^3}{12}\Phi\right)\simeq 0\,,
	\end{align}
	where we introduced for convenience
	\begin{align}
	x=k\tau \quad {\rm and}\quad\kappa=\frac{k}{\keq}\,.
	\end{align}
	The time coordinate $x$ splits the superhorizon ($x\ll1$) and subhorizon ($x\gg1$) regimes of the scalar perturbation, while $\kappa\gg1$ controls how deep in radiation domination the mode enters the horizon. In radiation domination we always have $x/\kappa=\keq\tau \ll1$.	One can check that Eqs.~\eqref{eq:eomPhiRD} and \eqref{eq:eomSRD} are valid up to order $(x/\kappa)^{2}$. 

	We present now an analytical solution to Eqs.~\eqref{eq:eomPhiRD} and \eqref{eq:eomSRD} for initial isocurvature conditions $\Phi(0)=0$ and $S(0) \equiv S_k(0)$. Note that we restore the subscript $k$ to emphasize that the initial conditions in Fourier space may be drawn in general from a $k$-dependent spectrum. We start with the perturbative ansatz $\Phi=\Phi_0+\kappa^{-1}\Phi_1+...$ and $S=S_0+\kappa^{-1}S_1+...$ by noting that in Eq.~\eqref{eq:eomPhiRD} there is a hierarchy between $\Phi$ and $S$ in terms of $\kappa$. This hierarchy is then preserved by Eq.~\eqref{eq:eomSRD}. The initial conditions impose $\Phi_0=0$, and then only $S_0=S_k(0)$ sources $\Phi_1$ in Eq.~\eqref{eq:eomPhiRD}. In the next step $\Phi_1$ sources $S_1$ in Eq.~\eqref{eq:eomSRD}, and with this procedure we arrive at
	\begin{align}\label{eq:analyticalsol}
	\Phi(x/\kappa \ll 1)&=\frac{3S_k(0)}{2\sqrt{2}\kappa} \frac{1}{x^3} \left(6+x^2-2\sqrt{3}x\sin(c_sx)-6\cos(c_sx)\right)+\mathcal{O}\left(x/\kappa\right)^{2}\,,\\\label{eq:analyticalsolS}
	S(x/\kappa \ll 1) &=S_k(0)+\frac{3S_k(0)}{2\sqrt{2}\kappa}\left(x+\sqrt{3}\sin(c_sx)-2\sqrt{3}{\rm Si}(c_sx)\right)+\mathcal{O}\left(x/\kappa\right)^{2}\,,
	\end{align}
	where at leading order $c_s=1/\sqrt{3}$ and ${\rm Si}(x)$ is the sine-integral function, defined in Appendix~\ref{app:usefulformulas}. From the solution for $S$, we find using Eq.~\eqref{eq:Vreleom} that the relative velocity is given by
	\begin{align}\label{eq:vrelsol}
	\keq V_{\rm rel}(x\ll x_{\rm eq})=\frac{3S_k(0)}{2\sqrt{2}\kappa^2}\left(1+\cos(c_sx)-2\frac{\sin(c_sx)}{c_sx}\right)+\mathcal{O}\left(x/\kappa\right)^3\,.
	\end{align}
	Note that the relative velocity is suppressed by $\kappa^{-1}$ with respect to $\Phi$ and $S$.

	Factoring out the initial condition $S_k(0)$, we define the transfer functions as
	\begin{align}\label{eq:transferphi}
	T_\Phi(x)\equiv\Phi(x)/S_k(0)\,, \quad T_S \equiv S(x)/S_k(0) \;, \quad T_{V_{\rm rel}}(x)\equiv V_{\rm rel}(x)/S_k(0).
	\end{align}

	\begin{figure}[t]
	\includegraphics[]{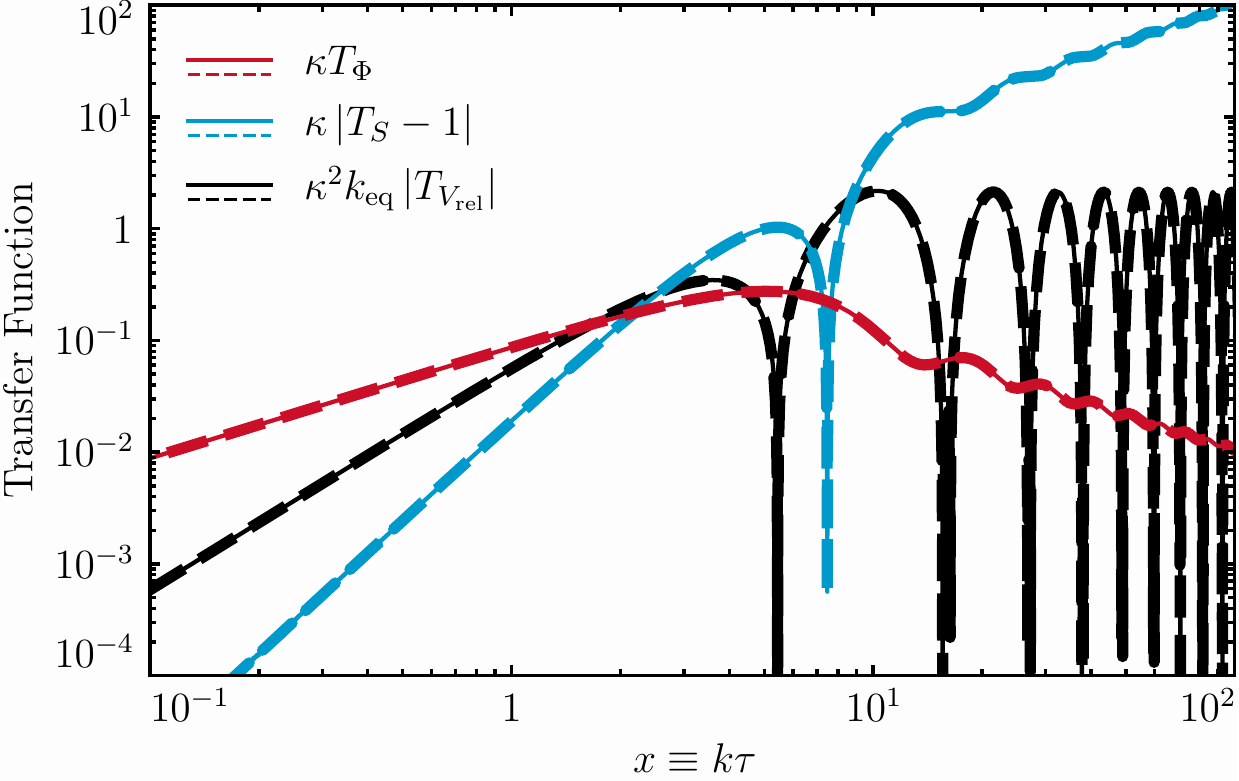}
	\caption{The numerical (solid) and analytic (dashed) transfer functions for the curvature perturbation (red), the isocurvature perturbation (blue), and the velocity perturbation (black), sourced by a primordial isocurvature perturbation on a scale $\kappa \equiv k/\keq \gg 1$ which enters the cosmological horizon ($x = 1$) in radiation domination. Note that each curve has been appropriately scaled by factors of $\kappa$ in order to show them on the same axes. For further discussion see \S\ref{sec:isocurvature}.}
	\label{fig:phi_S_V}
	\end{figure}

	We show these transfer functions in Fig.~\ref{fig:phi_S_V}, comparing the analytic solution \eqref{eq:analyticalsol}, \eqref{eq:analyticalsolS} and \eqref{eq:vrelsol} to a numerical solution of the equations of motion Eqs.~\eqref{eq:eomPhi}, \eqref{eq:eomS} and \eqref{eq:Vreleom}. We find perfect agreement between the analytics and the numerics.

	The analytic solution \eqref{eq:analyticalsol} presented here, exact during radiation domination, improves on the analytic superhorizon and subhorizon approximations provided by Ref.~\cite{Kodama:1986ud}. 
	In particular we find that in the subhorizon approximation of Ref.~\cite{Kodama:1986ud} (their equation (4.58a)), the coefficients of the sine and cosine should be, respectively, $A_\infty=-3/2$ and $B_\infty=0$. 
	Nonetheless their results are close to the exact solution found here within a factor $\sim10\%$.

	To understand the physical picture, we investigate the superhorizon ($x\ll1$) and subhorizon ($x\gg1$) regimes of our solution in more detail. On superhorizon scales, the curvature \eqref{eq:analyticalsol} and isocurvature \eqref{eq:analyticalsolS} become
	\begin{align}\label{eq:superphisuperS}
	\Phi(x\ll 1)&\simeq S_k(0) \frac{x}{8\sqrt{2}\kappa}\,,\\
	S(x\ll 1)&\simeq S_k(0)\left(1-\frac{x^3}{36\sqrt{2}\kappa}\right)\,,
	\end{align}
	and the relative velocity \eqref{eq:vrelsol} is given by
	\begin{align}
	\keq V_{\rm rel}(x\ll1)\simeq {-}S_k(0)\frac{x^2}{12\sqrt{2}\kappa^2}\,.
	\end{align}

	These limits show that on superhorizon scales the curvature perturbation and the relative velocity are in principle much smaller than unity, since $x/\kappa\ll1$. In order to induce a significant curvature perturbation at horizon crossing we therefore require a large value of $S_k(0)\gg1$. This is allowed in perturbation theory as long as the metric perturbations remain small $\Phi\ll1$, the velocity is suppressed $k xV_m\ll 1$, and 
	the local density of matter does not overcome that of radiation $\rho_mS_k(0)\ll \rho_r$. As long as this is the case the system evolves fully linearly and the universe is dominated by radiation with small radiation fluctuations.\footnote{Note that the energy density of CDM is on average unaffected by the isocurvature fluctuations.
	} We therefore expect that linear theory
	is valid as long as
	\begin{align}\label{eq:nonlinear}
	\frac{\rho_m}{\rho_r}S_k(0) = \frac{x}{\sqrt{2}\kappa}S_k(0) \ll1\,.
	\end{align}
	Note that the only $k$-dependence in Eq.~\eqref{eq:nonlinear} is through the initial condition profile $S_k(0)$.

	For any $S_k(0)$ this condition is always satisfied at early enough times. There will however be a time
	\begin{align}\label{eq:xnl}
	\tau_{\rm NL}(k)=\frac{\sqrt{2}}{\keq S_k(0)}\,,
	\end{align}
	when the local energy density of CDM overcomes that of the radiation and we enter the nonlinear regime. Our calculation is valid only for times $x<\xNL\equiv k\tau_{\rm NL}(k)$. Note that although we will use the variable $\xNL$ for convenience, the time $\tau_{\rm NL}$ when non-linearities become important only depends on $k$ through $S_k(0)$. Thus different modes reach the non-linear regime at different times only through the shape of the initial profile.

	If total density fluctuations become non-linear on superhorizon scales, i.e. $\xNL < 1$, we have that $\Phi>1$ from Eq.~\eqref{eq:superphisuperS}. We then do not know what happens to the geometry and we cannot predict the end state of the large isocurvature perturbations \cite{Passaglia:2021jla}. Though not the subject of the present paper, one possibility is that the universe may suddenly transition to a black hole dominated universe.

	We therefore require that the non-linear regime and the subsequent collapse of matter fluctuations happens inside the horizon. Therefore we require $\xNL>1$, which sets an upper limit on the amplitude of initial isocurvature for a given $k$ we can consider, 
	\begin{align}\label{eq:upperS}
	S_k(0)<\sqrt{2}\kappa \,.
	\end{align}
	This is the same condition as obtained in Ref.~\cite{Passaglia:2021jla}. We shall later see that future constraints from the absence of induced GWs are deep within this regime of validity. 

	If $\xNL>1$, then non-linearity begins on subhorizon scales. From \eqref{eq:analyticalsol} we have that $\Phi$ decays on subhorizon scales, and therefore we always have $\Phi<1$. However at $\xNL$ our condition \eqref{eq:nonlinear} is violated and signals the beginning of non-linear gravitational collapse. We are forced to stop our calculation here, but black hole formation is one possible end state of this process \cite{Passaglia:2021jla}.

	Nevertheless, as we will now see, this does not pose a problem for induced GWs since they are mainly sourced by $\Phi$ and it decays sufficiently fast for relatively large $\xNL$. Thus the overall amplitude of the GW spectrum turns out to be not directly dependent on $\xNL$ nor to the transition to the matter radiation equality if $k\gg \keq$, as long as $\xNL\gg1$.  Moreover, the induced GW spectrum can be computed solely using the linear theory solutions we have presented here, and we now turn to this.

\section{Gravitational waves induced by primordial CDM isocurvature fluctuations \label{sec:inducedGWs}}

	The formalism for analytic calculation of GWs induced by primordial adiabatic fluctuations in radiation domination was laid out in Refs.~\cite{Tomita,Matarrese:1992rp,Matarrese:1993zf,Ananda:2006af,Baumann:2007zm,Espinosa:2018eve,Kohri:2018awv} and reviewed in Ref.~\cite{Domenech:2021ztg}. For primordial isocurvature fluctuations, the main computational difference is in the behavior of the curvature perturbation, as given by the transfer functions we derived in \S\ref{sec:isocurvature}. 

	We wish to compute the spectral energy density of GWs \cite{Maggiore:1900zz},
	\begin{equation}
	\Omega_{\rm GW}(k) \equiv \frac{1}{3 H^2} \frac{d \rho_{\rm GW}}{d \ln{k}},
	\end{equation}
	today.	Deep inside the horizon, tensor perturbations behave as free GWs and their energy redshifts as radiation. Therefore if we compute $\Omega_{\rm GW}(k)$ at some time $x_c$ in radiation domination ($x_c/\kappa\ll1$) when the mode is sufficiently subhorizon ($x_c \gg 1$), then the spectral density today $\Omega_{\rm GW, 0}(k)$ can be related to it by \cite{Inomata:2016rbd} 
	\begin{align}\label{eq:spectraldensitytoday2}
	\Omega_{\rm GW,0}h^2(k)&= \Omega_{r,0}h^2 \left(\frac{g_*(T_{\rm c})}{g_{*,0}}\right)\left(\frac{g_{*s}(T_{\rm c})}{g_{*s,0}}\right)^{-4/3}\Omega_{\rm GW,c}(k)\,.
	\end{align}
	where $\Omega_{r,0}h^2\approx4.18\times 10^{-5}$ is the density fraction of radiation today \cite{Aghanim:2018eyx}, $g_{*}(T_c)$ and $g_{*s}(T_c)$ are respectively the effective number of degrees of freedom in the energy density and entropy at $T_c$, and $g_{*,0} = 3.36$ and $g_{*s,0} = 3.91$ are their values today. Eventually, for  $k \gg 5 \times  10^8\, {\rm Mpc}^{-1}$, corresponding to modes of interest in this paper, one has $g_{*}(T_{\rm c}) = g_{*s}(T_{\rm c}) =106.75$ \cite{Fumagalli:2021mpc}. In the general case, we use the fitting formulas for $g_{*}(T_{\rm c})$ and $g_{*s}(T_{\rm c})$ provided by Ref.~\cite{Saikawa:2018rcs}. The subscript ``c'' denotes evaluation at $x_c$ here and throughout.

	Before proceeding to the computation of the induced GW spectrum, we highlight a distinctive feature of this work with respect to typical calculations of induced GWs. As we discussed earlier, to have significant GW production we will require large initial isocurvature, that is $S_k(0)\gg1$. From Eq.~\eqref{eq:definitionS}, we see that imposing that the energy density of CDM is non-negative yields $S>-1$ and therefore that $S$ should follow a non-Gaussian, very highly skewed, probability distribution. Most of the universe must be filled with $S\sim -1$ (devoid of CDM) regions in order to allow for regions of substantial $S\gg1$.\footnote{Note that this situation is not unique to our set up. For instance, assuming $S_k(0)$ peaks at a given scale, a similar situation would be a universe where CDM is concentrated in very compact halos or in black holes, or when there are bubbles of CDM. For example, see the discussion in Refs.~\cite{Dolgov:1992pu} and \cite{Cotner:2019ykd} for creating black holes respectively from large baryon isocurvature and from clustering of Q-balls. In all these cases, a linear treatment of the perturbations in the fluid description is valid until we enter the non-linear regime where we stop our calculations.}

	We argue in Appendix \ref{app:NG} that such a highly {skewed} non-Gaussian probability distribution enhances the production of GWs. Roughly speaking, the universe is very inhomogeneous in certain places and the gradients of the gravitational potential there, which are the source of induced GWs, are {likely} larger than in the Gaussian case (for which positive regions can be compensated with negative regions). However, the probability distribution function for $S$ must be extremely non-Gaussian which hinders a rigorous description of the four point correlation function. Nevertheless, we propose a reasonable estimate for the amplitude of such a four point correlation based on a motivated example.

	Thus for simplicity we first proceed by assuming Gaussian isocurvature fluctuations despite the fact that strictly speaking it is inconsistent with the condition $S>-1$. The amplitude of the derived induced GW spectrum can then be thought of as a conservative estimate, and at the end of this section in \S\ref{subsec:NG} we use our example from Appendix~\ref{app:NG} to estimate the enhancement due to the non-Gaussianity.

	Assuming now that the isocurvature fluctuations are Gaussian, the spectral density of induced GWs at $x_c$ is given by
	\begin{align}\label{eq:Phgaussian}
	\Omega_{\rm GW,c}(k)=\frac{2}{3}\int_0^\infty dv\int_{|1-v|}^{1+v}du\left(\frac{4v^2-(1-u^2+v^2)^2}{4uv}\right)^2\overline{I^2(x_c,k,u,v)}{{\cal P}_{S}(ku)}{{\cal P}_{S}(kv)}\,,
	\end{align}
	where ${\cal P}_{S}(k)$ is the initial dimensionless spectrum of isocurvature fluctuations.\footnote{The dimensionless power spectrum is defined by
	\begin{align}
	\langle S_\mathbf{k}(0)S_{\mathbf{k}'}(0)\rangle=\frac{2\pi^2}{k^3}{\cal P}_S(k)\times(2\pi)^3\delta^{(3)}\left(\mathbf{k}+\mathbf{k}'\right)\,.
	\end{align}
	} $\overline{I^2}$ is the oscillation average of the square of the kernel
	\begin{equation}\label{eq:kernel1}
	I(x_c,k,u,v)\equiv x_{c} \,\int_{x_i}^{x_c}d\tilde x  \,G(x_{c},\tilde x)f(\tilde x,k,u,v)\,,
	\end{equation}
	which is an integral of the tensor modes Green's function in radiation domination
	\begin{align}
	G(x,\tilde x)= \frac{a(\tilde x)}{a( x )}\left(\sin x \cos \tilde x-\cos x \sin \tilde x \right)\,,
	\end{align}
	against the scalar source term
	\begin{align}\label{eq:f}
	f(x,k, u,v)=&T_\Phi(v x)T_\Phi(u x)+ \frac{3}{2}c_s^2 a^2\rho_mT_{V_{\rm rel}}(v x )T_{V_{\rm rel}}(u x)\nonumber\\&+\frac{3}{2}c_s^2\left(1+\frac{\rho_m}{\rho_r}\right)\left(T_\Phi(v x)+\frac{T'_\Phi(v x)}{\cal H}\right)\left(T_\Phi(u x)+\frac{T'_\Phi(u x)}{\cal H}\right)\,,
	\end{align}
	which involves the transfer functions of the curvature perturbation and relative velocity, $T_\Phi$ and $T_{V_{\rm rel}}$, given in Eqs.~\eqref{eq:analyticalsol}, \eqref{eq:vrelsol} and \eqref{eq:transferphi}. In Eq.~\eqref{eq:kernel1}, $x_i=k\tau_i$ is an initial time which we take as $x_i\to0$ since we begin the calculation when all modes of interest are superhorizon. Since Eq.~\eqref{eq:vrelsol} shows $V_{\rm rel}$ is suppressed by $\kappa^{-1}\ll1$ with respect to $\Phi$, we can safely neglect its contribution to $f$ from now on. These formulas have been derived using the second order equations of motion for tensor modes focusing only on the scalar source, which can be found in Appendix \ref{app:formulas}. 

	In the radiation dominated regime,
	 we can decompose the kernel into a sine and cosine piece
	\begin{align}\label{eq:transferI}
	I(x,k,u,v)=I_c(x,k,u,v)\sin x-I_s(x,k,u,v)\cos x\,,
	\end{align}
	where
	\begin{align}\label{eq:Ics}
	I_{c/s}(x,k,u,v)\equiv\int_{0}^{x}d\tilde x  \, \tilde x\,\left\{
	\begin{aligned}
	&\cos \tilde x\\
	&\sin \tilde x
	\end{aligned}
	\right\} f(\tilde x,k,u,v)\,.
	\end{align}
	Then $x_c$ should be a  time late enough that we can compute the oscillation averaged kernel as
	\begin{equation}\label{eq:kernel2}
	\overline{I^2(x_c, k, u, v)} \simeq \frac{1}{2}\left(I_{c,\infty}^2(k,u,v)+I_{s,\infty}^2(k,u,v)\right)\,,
	\end{equation}
	where $I_{c/s, \infty}(k, u, v) \equiv \lim_{x \rightarrow \infty} I_{c/s}(x,k,u,v)$ while remaining in radiation domination.

	So far these equations are very similar to the ones derived from adiabatic initial conditions. However as we shall now see the explicit form of the kernel \eqref{eq:transferI} is very different.

	\subsection{The general kernel} 
	\label{subsec:kernel}

	We wish to compute the general integration kernel $I(x, k, u, v)$ \eqref{eq:transferI}. To do so, we first compute the source term $f(x, k, u,v)$ by plugging in our analytic solution for $\Phi$, Eq.~\eqref{eq:analyticalsol}, into Eq.~\eqref{eq:f} at leading order in $\kappa^{-1}$. 

	\begin{figure}[t]
	\includegraphics[]{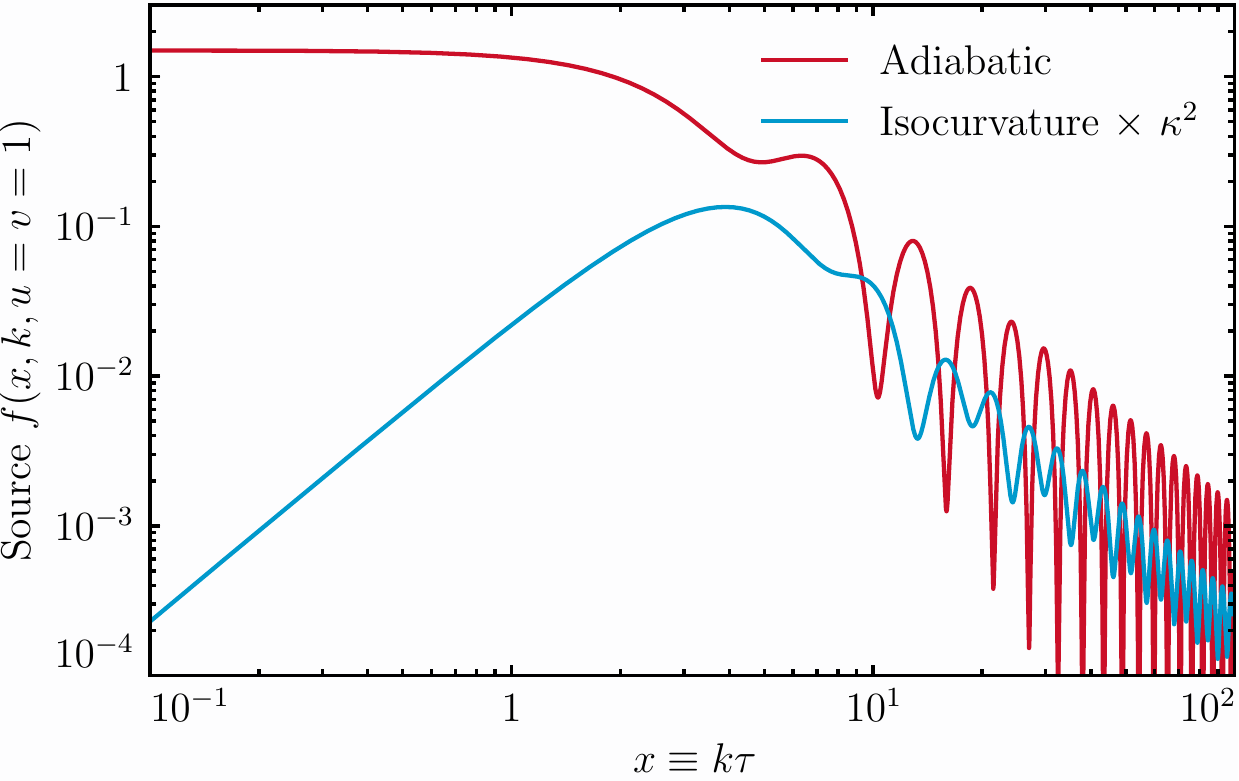}
	\caption{The GW production source $f(x, k, u, v)$ from adiabatic (red, Eq.~\eqref{eq:fsimple}) and isocurvature (blue, Eq.~\eqref{eq:f}) initial conditions for $u=v=1$ and $\kappa \equiv k/\keq \gg 1$. The isocurvature curve is scaled by $\kappa^2$. Note that the oscillations have opposite phase, as is also the case in the studies of CMB anisotropies \cite{Hu:1994jd,Hu:1995em}. For further discussion see \S\ref{subsec:kernel}.}
	\label{fig:f}
	\end{figure}

	We show the resulting source $f$ in Fig.~\ref{fig:f} while relegating the expression itself to Appendix \ref{app:explicit}. We compare it to the source term from purely adiabatic initial conditions which can also be found in that appendix. To facilitate comparison we scale out the $\kappa^{-2}$ suppression of the isocurvature case.

	While in the adiabatic case the source term is constant on superhorizon scales, in the isocurvature case the source is proportional to $x^2$ at early times. In both cases the source begins to oscillate inside the horizon, but in the case of adiabatic initial conditions the source can be written as the square of sine and cosine functions, with oscillations that hit zero and have a frequency twice the frequency of $\Phi$, i.e. $2 c_s$. In contrast, the source from isocurvature initial conditions decays as $x^{-2}$ but has no zero crossings. Moreover, the oscillatory modulation in $f$ \eqref{eq:f} is out of phase with the adiabatic one and has two frequencies, $c_s$ and $2c_s$ as is clear from looking at Fig.~\ref{fig:f}. This implies that while some oscillations have twice the frequency of $\Phi$ and there might be resonances in the GW production, for primordial isocurvature initial conditions there can be no full destructive interference of the kind seen in the adiabatic case. 

	With this source we have all ingredients necessary to compute the $I_{c, s}(k, x, u, v)$ cosine and sine components of the kernel $I$ by appropriately integrating over $f$ using Eq.~\ref{eq:Ics} until $I_{c/s}$ approach constants at late times. At that point we can compute the oscillation averaged kernel with Eq.~\eqref{eq:kernel2}. Evaluating the limits of the kernels, we find
	\begin{align}
	\label{eq:Icinf}
	\kappa^{2}I_{c,\infty}(u,v)&=\frac{9}{32u^4v^4}\Bigg\{-3u^2v^2+\left(-3+u^2\right)\left(-3+u^2+2v^2\right)\ln\left|1-\frac{u^2}{{3}}\right|\nonumber\\&
	+\left(-3+v^2\right)\left(-3+v^2+2u^2\right)\ln\left|1-\frac{v^2}{{3}}\right|\nonumber\\&
	-\frac{1}{2}\left(-3+v^2+u^2\right)^2\ln\left[\left|1-\frac{(u+v)^2}{{3}}\right|\left|1-\frac{(u-v)^2}{{3}}\right|\right]\Bigg\}\,,
	\end{align}
	and
	\begin{align}
	\label{eq:Isinf}
	\kappa^{2}I_{s,\infty}(u,v)&=\frac{9\pi}{32u^4v^4}\Bigg\{9-6v^2-6u^2+2u^2v^2\nonumber\\&+\left(3-u^2\right)\left(-3+u^2+2v^2\right)\Theta\left(1-\frac{u}{\sqrt{3}}\right)\nonumber\\&
	+\left(3-v^2\right)\left(-3+v^2+2u^2\right)\Theta\left(1-\frac{v}{\sqrt{3}}\right)\nonumber\\&
	+\frac{1}{2}\left(-3+v^2+u^2\right)^2\left[\Theta\left(1-\frac{u+v}{\sqrt{3}}\right)+\Theta\left(1+\frac{u-v}{\sqrt{3}}\right)\right]\Bigg\}\,.
	\end{align}

	A particularity of the kernel for isocurvature-induced GWs with respect to the adiabatic case is that it presents a $\kappa^{-2}$ dependence due to the fact that long-wavelength scalar modes enter the horizon at later times and are therefore less suppressed. We can gain intuition for the effect of this $\kappa^{-2}$ dependence by introducing an effective induced-curvature power spectrum
	\begin{align}\label{eq:Pphieffectve}
	{\cal P}_\Phi^{\rm eff}(k)\equiv\kappa^{-2}{\cal P}_{\cal S}(k)\,,
	\end{align}
	which we could in principle use in the $\Omega_{\rm GW,c}$ equation \eqref{eq:Phgaussian} with the kernels \eqref{eq:Icinf} and \eqref{eq:Isinf} without the $\kappa^{-2}$ and with one less factor of $1/(uv)$ in the prefactor. The resulting kernels, although explicitly very different from the adiabatic case, should then have qualitatively similar features as in the adiabatic case such as no explicit $k$-dependence, a resonant peak at $u+v=\sqrt{3}$ and a high momenta tail for $u\sim v\gg1$ that goes as $v^{-2}\ln v$. Consider for example that ${\cal P}_S$ is a scale invariant power spectrum, with some infra-red cut-off not to violate constraints from CMB. Then, because ${\cal P}_\Phi^{\rm eff}(k)\propto k^{-2}$ the resulting induced GW spectrum goes as $\Omega_{\rm GW}\propto k^{-4}$. If instead ${\cal P}_S$ is peaked, then ${\cal P}_\Phi^{\rm eff}$ is also peaked and the enhancement of long-wavelength modes gradually disappears for narrower and narrower ${\cal P}_S$. As we shall shortly see, for an infinitely narrow peak only a single scalar mode is the source of induced GWs and the resulting GW spectrum resembles that of the adiabatic case. While this effective curvature power spectrum ${\cal P}_\Phi^{\rm eff}$ is useful to gain intuition in some situations, in this paper we still use ${\cal P}_S(k)$ to do all the calculations. 

	We have glossed over here that we cannot self-consistently integrate the source $f(x,k,u,v)$ past the point when linear theory breaks down in the transfer functions. This occurs, for a scalar mode of wavenumber $q$, at \eqref{eq:xnl}
	\begin{align}\label{eq:taunl}
	\tau_{\rm NL}(q)=\frac{\sqrt{2}}{\keq\sigma(q)}\,,
	\end{align}
	where
	\begin{align}
	\sigma^2(q)=\int \frac{dk}{k}W^2_q(k){\cal P}_S(k)\,,
	\end{align}
	is the variance smoothed over a scale $q$ and $W_q$ is some window function. The smoothing is necessary to go from the power spectrum to the density fluctuations just like in PBH calculations, which in our case determines when fluctuations at a give scale enter on average the non-linear regime. For simplicity, we consider $W_q(k)$ a top-hat.
	This means that in the $I_{c, s}$ integrals \eqref{eq:Ics}, we cannot take $x_c$ to be larger than
	\begin{align}\label{eq:xcut}
	x_{\rm cut}=k\tau_{\rm cut}\equiv k \ {\rm min}\left(\tau_{\rm NL}(kv),\tau_{\rm NL}(ku)\right)\,,
	\end{align}
	as $\tau_{\rm cut}$ corresponds to the time when we have to cut off the source term. Connecting to the notation of \S\ref{sec:isocurvature}, $\tau_{\rm cut}$ is related to $\tau_{\rm NL}$ \eqref{eq:xnl} for the mode $k u$ or $k v$, whichever is earlier. Note that in Eq.~\eqref{eq:xcut}, the wavenumber $k$ is the tensor mode wavenumber.

	If ${\cal P}_S$ is very large, this cut-off time may be early and the $x_c\rightarrow\infty$ limits taken in Eqs.~\eqref{eq:Isinf} and \eqref{eq:Icinf} maybe not be valid.  This does not significantly limit our calculation: if the finite $x_{\rm cut}$ is not large enough, we can approximate that the production of induced tensor modes abruptly ends before all relevant modes are deep inside the horizon. From that moment on, tensor modes propagate freely and still eventually become GWs once deep enough inside the horizon. Thus, the averaging procedure is still valid, although we should use the general analytical form of the kernel \eqref{eq:Ics} for finite $x_{\rm cut}$. The full analytic expression for the kernel $I(k,x_{\rm cut}, u, v)$ is lengthy and not particularly enlightening and we therefore omit it, but we discuss the effect in detail in \S\ref{subsec:cutoff}. Until then, however, we assume that the cutoff is sufficiently late that Eqs.~\eqref{eq:Icinf} and \eqref{eq:Isinf} are valid.

	We now proceed to compute the induced gravitational wave spectrum for certain primordial isocurvature spectra in \S\ref{subsec:dirac} and \S\ref{subsec:lognormal}, before returning to some of the technicalities in \S\ref{subsec:cutoff} and  \S\ref{subsec:NG}.

	\subsection{Dirac delta peak}
	\label{subsec:dirac}

	We consider first a Dirac delta function in $\ln k$ for the isocurvature power spectrum,
	\begin{align}
	\label{eq:Pkdelta}
	{\cal P}_{S}(k)={\cal A}_{S}\delta(\ln(k/k_p))\,.
	\end{align}
	This distribution is characterized by a spectral amplitude ${\cal A}_S$ and a peak position $k_p$.

	The delta function picks out $u = v = k_p/k$ in the $\Omega_{\rm GW}$ integral \eqref{eq:Phgaussian}, with the resulting GW spectrum
	\begin{align}\label{eq:omegac}
	\Omega_{\rm GW,c}&= \frac{1}{3} \left(\frac{k}{k_p}\right)^{-2} \left(1-\frac{k^2}{4k_p^2}\right)^2  \ (I^2_{c,\infty}(u,v) + I^2_{s,\infty}(u,v)) {\cal A}_S^2  \Theta\left(k-2 k_p\right),
	\end{align}
	which we have also confirmed by numerical integration of numerical solutions of the equations of motion.

	\begin{figure}[t]
	\includegraphics[]{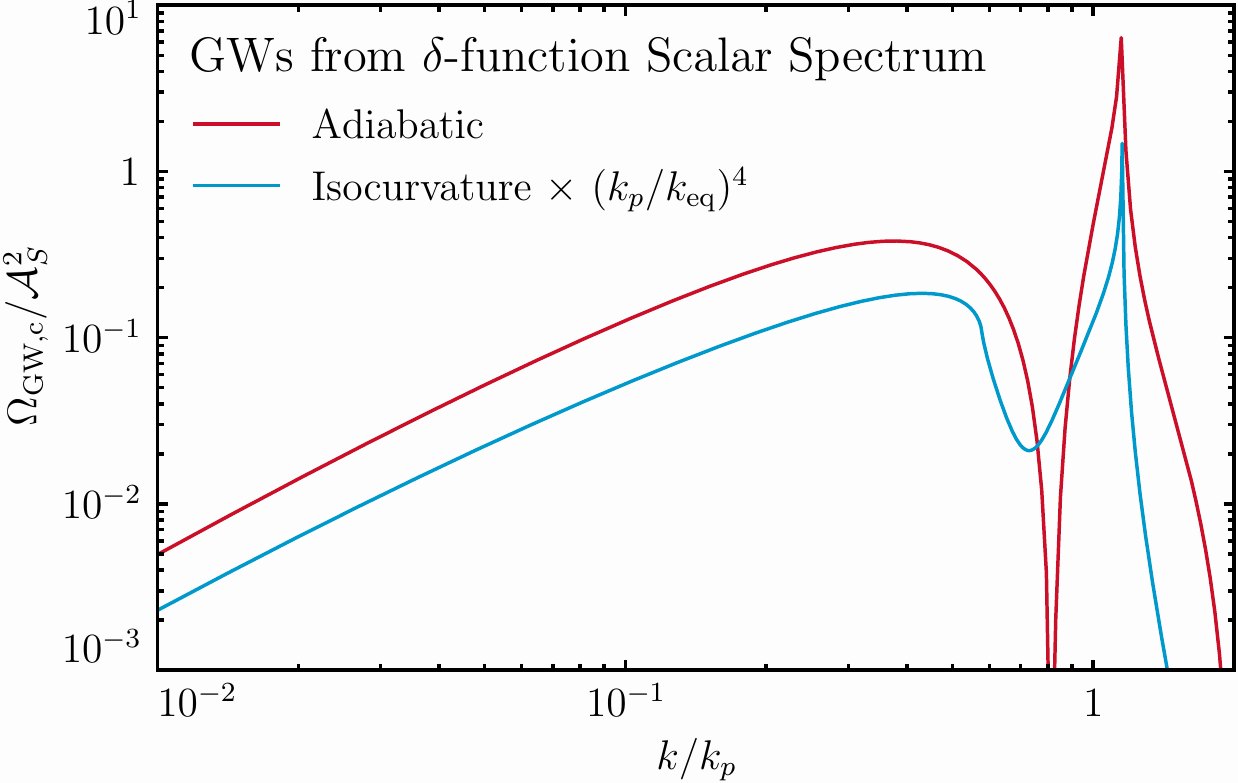}
	\caption{The gravitational wave spectrum induced by a $\delta$-function peak \eqref{eq:Pkdelta} of amplitude ${\cal A}_S$ at scale $k_p$ in the primordial adiabatic (red, Eq.~\eqref{eq:omegacad}) and CDM isocurvature (blue, Eq.~\eqref{eq:omegac}) power spectra evaluated at some time deep in the subhorizon regime in radiation domination. For identical input spectra the isocurvature induced GWs are suppressed by $(k_p/\keq)^4$, and thus producing an observable signal from CDM isocurvature requires a much larger input spectrum. Beyond the amplitude difference, the adiabatic and isocurvature spectra show some common features, such as the IR scaling \eqref{eq:ir_scaling_delta} and the divergent peaks at $k/k_p = 2/\sqrt{3}$, as well as some differences such as the structure around the peak and the presence or absence of a zero in the spectrum. For further discussion see \S\ref{subsec:dirac}.}
	\label{fig:power_spectrum_deltafunction}
	\end{figure}

	We show this result in Fig.~\ref{fig:power_spectrum_deltafunction} and compare it to the spectrum produced by the same feature in the adiabatic power spectrum, Eq.~\eqref{eq:omegacad}. For the same input ${\cal A}_S$, the isocurvature mode induces GWs with an amplitude suppressed by $(k_p/\keq)^{-4}$, as we previously described, and therefore producing an observable GW abundance requires much larger input isocurvature fluctuations than adiabatic fluctuations.

	Beyond the overall amplitude, however, the induced GWs from isocurvature and adiabatic fluctuations have different spectral shapes. While both spectra have resonant peaks at $k/k_p = 2 c_s  = 2/ \sqrt{3}$, due to the same resonance phenomenon when a tensor mode has a wavenumber twice the frequency of the scalar modes just like a harmonic oscillator with a periodic force, the shape around the divergent peak is different. Furthermore, the adiabatic-induced spectrum shows a zero at $k/k_p = \sqrt{2/3}$ whereas the isocurvature-induced one shows a non-zero minimum at $k/k_p\approx 0.73$. The shift in the position of the minimum of the GW spectrum in the isocurvature case is due to the different phase of the oscillations of the induced $\Phi$. The zero in the adiabatic-induced GW spectrum is due to precise cancellations in the GW integrals in radiation domination, and for a general equation of state parameter $w=c_s^2=p/\rho$ the minimum occurs at $k=\sqrt{2}c_sk_p$ and is in general non-zero \cite{Domenech:2021ztg}. The resonant peak-dip relation, that is $k_{\rm peak}/k_{\rm dip}\approx \{1.58,\sqrt{2}\}$ for, respectively, the isocurvature and adiabatic cases, might be one possible way to distinguish both scenarios.

	The low frequency tail goes as
	\begin{align}
	\label{eq:ir_scaling_delta}
	\Omega_{\rm GW,c}(k\ll k_p) \simeq \frac{27}{64}{\cal A}_{S}^2\kappa_p^{-4}\frac{k^2}{k_p^2}\left(\pi^2+\left(\frac{3}{4}+2\ln\frac{\sqrt{6}k}{k_p}\right)^2\right)\,
	\end{align}
	where $\kappa_p=k_p/\keq$ and which, although it has the same scaling as the adiabatic result, i.e. for $k\ll k_p$ it goes as $k^2\ln^2 k$, for the same initial isocurvature and adiabatic primordial power spectra it has a lower amplitude
	\begin{align}
	\frac{\Omega^{\rm iso}_{\rm GW,c}(k\ll k_p)}{\Omega^{\rm ad}_{\rm GW,c}(k\ll k_p)}=\frac{9}{16\kappa_p^4}\,.
	\end{align}

	The divergence of the induced GW spectrum at $k/k_p = 2/\sqrt{3}$ makes a $\delta$-function peak in the primordial spectrum rather unphysical. We therefore consider instead, as a toy model, a log-normal peak in the scalar spectrum which in the narrow peak limit approaches a $\delta$-function while regulating the divergence.

	\subsection{Log-normal peak}
	\label{subsec:lognormal}

	We consider a log-normal peak in the primordial isocurvature power spectrum with functional form \cite{Pi:2020otn}
	\begin{align}\label{eq:Pklog}
	{\cal P}_{S}(k)=\frac{{\cal A}_S}{\sqrt{2\pi}\Delta}\exp\left(-\frac{\ln^2(k/k_p)}{2\Delta^2}\right)\,,
	\end{align}
	characterized by an amplitude $A_S$, a position $k_p$, and a width $\Delta$. In the $\Delta\rightarrow0$ limit, we expect to return to the $\delta$-function results of \S\ref{subsec:dirac}. The normalization factor in \eqref{eq:Pklog} has been chosen such that $\int_{0}^\infty d\ln k \,{\cal P}_{S}(k)={{\cal A}_S}$.

	For such an input spectrum a complete analytic solution for $\Omega_{\rm GW, c}$ evades us. Nonetheless $\Omega_{\rm GW, c}$ can be obtained straightforwardly by numerical integration of Eq.~\eqref{eq:Phgaussian} with the analytic expression for the source $I^2$ from Eqs.~\eqref{eq:kernel2}, \eqref{eq:Icinf}, and \eqref{eq:Isinf}. 

	\begin{figure}[t]
	\includegraphics[]{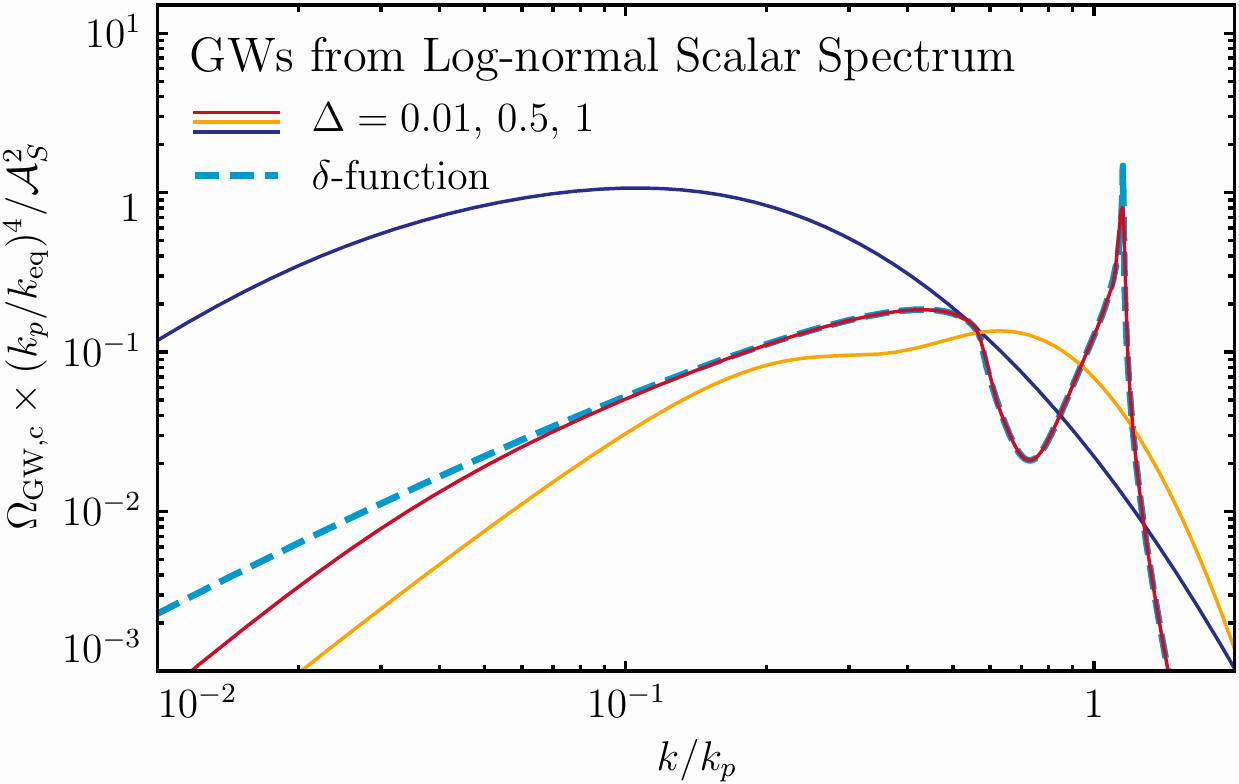}
	\caption{The gravitational wave spectrum induced by a log-normal peak \eqref{eq:Pklog} in the primordial CDM isocurvature power spectrum, of amplitude ${\cal A}_S$ and width $\Delta$ at scale $k_p$. The dashed blue line shows the GW spectrum from a Dirac delta, Eq.~\eqref{eq:omegac}. Then, in red, orange and purple we show the GW spectrum from log-normal peaks of width, respectively, $\Delta=0.01$, $0.5$ and $1$. In the narrowest case, the induced spectrum approaches the $\delta$-function result while regulating the divergence of the induced spectrum. Wide peaks wash out the structure completely, as in the case of initial adiabatic fluctuations. However, the peak of the GW spectrum shifts to larger scales with respect to the peak in the isocurvature spectrum, according to Eq.~\eqref{eq:Pphieffectve}. For further discussion see \S\ref{subsec:lognormal}.
	}
	\label{fig:power_spectrum_lognormal}
	\end{figure}

	We show in Fig.~\ref{fig:power_spectrum_lognormal} the resulting induced GW spectrum for log-normal widths $\Delta=0.01$, $0.5$, and $1$. While a wide scalar width $\Delta = 1$ leads to an effectively structureless GW spectrum, as the width $\Delta$ is taken smaller and smaller the GW spectrum approaches the fully analytic results for the $\delta$-function scalar power spectrum of \S\ref{subsec:dirac}. As the peak in the scalar spectrum broadens, the peak in the GW shifts to lower wavenumbers. This is because although the isocurvature power spectrum peaks at $k=k_p$, the effective induced curvature spectrum \eqref{eq:Pphieffectve} peaks at $k_{\rm max}=e^{-2\Delta^2}\,k_p$. For $\Delta=1$ we have that $k_{\rm max}\sim 0.13 k_p$.

	For any small but finite $\Delta$ there are two important differences relative to the $\delta$-function case. First, the finite nature of the peak inevitably suppresses the induced spectrum in the IR below $k/k_p \sim \Delta$, with the asymptotic behavior $\propto (k/k_p)^3\ln^2(k/k_p)$. This has consequences for the range of $k_p$ which can be probed with GW observations at a given frequency.

	More important for our purposes, however, is that the finite amplitude of the scalar source regulates the divergent peak at $k/k_p = 2/\sqrt{3}$ and yields a physically reasonable induced GW spectrum which we can use to extract physical constraints on the primordial spectrum.

	The peak amplitude can be estimated analytically by smoothing out the $\delta$-function's resonant peak as explained in detail in Refs.~\cite{Pi:2020otn,Fumagalli:2021cel}. This procedure can be applied to either the adiabatic or isocurvature case as long as the GW spectrum can be written as in Eq.~\eqref{eq:Phgaussian}. If the GW spectrum sourced by a Dirac delta is known, the peak amplitude of the GW spectrum induced by a narrow scalar spectrum with dimensionless width $\Delta$ is well estimated by \cite{Pi:2020otn,Fumagalli:2021cel}
	\begin{align}
	\Omega_{\textrm{GW}}^{\textrm{peak}}(\Delta) \simeq \frac{1}{2 \Delta} \int_{2/\sqrt{3}-\Delta}^{2/\sqrt{3}+\Delta} \textrm{d}\epsilon\, \Omega_{\textrm{GW}}^{(\delta)}(\epsilon)\,,
	\label{eq:narrow-averaging}
	\end{align}
	where we used $k=(2/\sqrt{3}+\epsilon) k_p$ and expanded for small $\epsilon$ and $\Omega_{\textrm{GW}}^{(\delta)}$ is the GW spectrum for the $\delta$-function peak at $k_{p}$. For the induced GWs produced by primordial isocurvature fluctuations, this smearing procedure yields
	\begin{align}
	\Omega^{\rm peak}_{\rm GW,c}\approx \frac{1}{16}{\cal A}_{S}^2\left(\frac{\keq}{k_p}\right)^4\left(1 + \frac{\pi^2}{2}+ \left(\frac{1}{2}+3\ln\frac{4}{3} + \frac{1}{2}\ln\left(3\Delta^2\right)\right)^2\right)\,,
	\end{align}
	which for, e.g., $\Delta\sim 0.01$ leads to 
	\begin{align}\label{eq:estimateamplitude}
	\Omega^{\rm peak}_{\rm GW,c}(\Delta=0.01)\approx 0.8\times{{\cal A}_{S}^2\left(\frac{\keq}{k_p}\right)^4}\,.
	\end{align}
	We have checked that this estimate is in good agreement with our numerical results.

	\subsection{Dependence on the nonlinear cutoff}
	\label{subsec:cutoff}

	We now return to the issue of the finite non-linear cutoff $x_{\rm cut}$ by examining its effect on the GW spectra we computed in \S\ref{subsec:dirac} and \S\ref{subsec:lognormal}. For a Dirac delta scalar spectrum the cut-off \eqref{eq:xcut} is given by
	\begin{align}\label{eq:deltacut}
	x^\delta_{\rm cut}=\frac{1}{v_p}\frac{\sqrt{2}k_p}{\keq{\cal A}_S^{1/2}}=\frac{\xNL^{\rm peak}}{v_p}\,,
	\end{align}
	where $\xNL^{\rm peak}$ denotes $\xNL$ \eqref{eq:xnl} evaluated for a scalar mode of wavenumber $k_p$ and amplitude ${\cal A}_S^{1/2}$.\footnote{Note that although $x^\delta_{\rm cut}$ \eqref{eq:deltacut} becomes strictly zero for $v_p=k_p/k\to\infty$ it is not a problem for our formalism. First, it means that far superhorizon tensor modes did not have time to grow when we cut-off the source. Second, we are always considering $k\gg \keq$ and, therefore, we never reach the zero limit in \eqref{eq:deltacut}.} For the log-normal scalar spectrum, on the other hand, we have
	\begin{align}
	x^\Delta_{\rm cut}=x^\delta_{\rm cut}\times\sqrt{2}\,{\rm Min}\left\{
	\left({\rm Erfc}\left[\frac{\ln (v_p/v)}{\sqrt{2}\Delta}\right]\right)^{-1/2},v\to u\right\}\,,
	\end{align}
	where ${\rm Erfc}[x]$ is the complementary error function.	

	\begin{figure}[t]
	\includegraphics[]{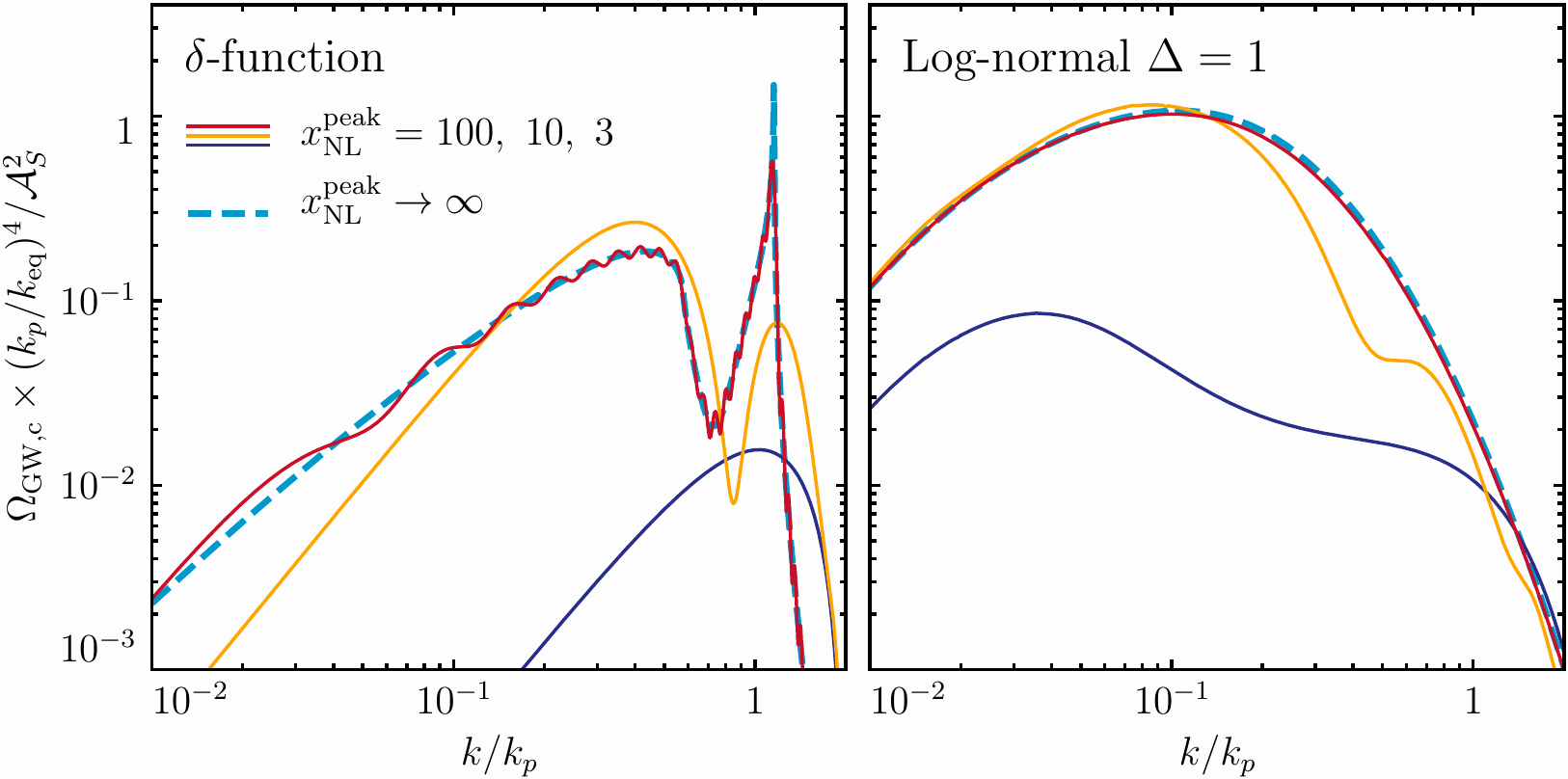}
	\caption{The onset of non-linearity at $\xNL^{\rm peak}$, controlled by the amplitude of the scalar isocurvature fluctuations through Eq.~\eqref{eq:xcut}, cuts off the production of gravitational waves and modifies the asymptotic $\xNL^{\rm peak} \rightarrow \infty$ (dashed blue) results we computed in Figs.~\ref{fig:power_spectrum_deltafunction} and \ref{fig:power_spectrum_lognormal} for the Dirac $\delta$-function isocurvature spectrum (left panel) and broad Log-normal spectrum (right panel). For a very early non-linear onset such as $\xNL^{\rm peak} = 3$ (purple), the predominant effect is to suppress the induced spectrum. This suppression is reduced for $\xNL^{\rm peak}=10$ (yellow) and further for $\xNL^{\rm peak} = 100$ (red), with the dominant effect becoming oscillations induced by the sudden shut-off of the oscillating source. The broad spectrum is less sensitive to the non-linear cutoff. In practice $x_{\rm NL}^{\rm peak}$ will usually be large enough that we can use the asymptotic spectrum. For further discussion see \S\ref{subsec:cutoff}.}
	\label{fig:power_spectrum_xNL}
	\end{figure}

	We show in Fig.~\ref{fig:power_spectrum_xNL} how finite values of $x_{\rm cut}$ induced by finite values of $\xNL^{\rm peak}$ change the induced GWs produced by the $\delta$-function and broad log-normal scalar spectrum. For a small value $\xNL^{\rm peak} = 3$, the primary effect is to suppress the induced GW spectrum. As $\xNL^{\rm peak}$ increases this suppression becomes less important, and already for $\xNL^{\rm peak} = 10$ the overall amplitude of the GW spectrum is similar to the asymptotic limit $\xNL^{\rm peak} \rightarrow \infty$. The most visible effect is then the presence of oscillations in the GW spectrum due to the sudden cut-off of the source, reflecting the oscillations of $\Phi$ which in our approximation suddenly stop sourcing the GWs once non-linearity begins. We note that this discussion is predicated on our conservative choice to turn off all GW production after $x_{\rm cut}$. In practice we may expect some GWs to be produced in the non-linear regime.

	Such early cutoffs erase the resonant peak in the GW spectrum in the $\delta$-function case because there has been no time for the resonance to develop. Furthermore, the deep low-frequency tail of the spectrum at $v_p\gg \xNL^{\rm peak}$ now becomes $k^2$ in the $\delta$-function case as the logarithmic correction seen in Eq.~\eqref{eq:ir_scaling_delta} is a result of integrating over many oscillations.\footnote{In more mathematical detail, what occurs is that the general kernel contains the cosine integral functions ${\rm Ci}[x_{\rm cut}]$, ${\rm Ci}[(1-(u-v)/\sqrt{3})x_{\rm cut}]$ and ${\rm Ci}[(1-(u+v)/\sqrt{3})x_{\rm cut}]$ in addition to the usual logarithmic terms. When $x_{\rm cut}$ is not large enough, the logarithmic terms that appear for vanishing argument, i.e. ${\rm Ci}(x_{\rm cut}\sim 0)\sim \ln x$, cancel the typical logarithmic corrections. A similar discussion would apply to the adiabatic case if one were to suddenly shut-off the source term to the induced GWs.} For $\xNL^{\rm peak} = 100$, sufficient oscillations are averaged over that the resonant peak returns and the GW spectra approach their asymptotic limit.

	This discussion holds qualitatively for both the $\delta$-function and broad scalar peaks, but the induced spectrum is less sensitive to the non-linear cutoff in the broad spectrum case than the narrow-peak case. This is because in the broad peak case there are several scalar modes sourcing induced GWs and while some might enter the regime earlier than others, there is overall a similar production of induced GWs. The resulting amplitude of the GW spectrum is similar regardless for all $\xNL^{\rm peak}\gtrsim 10$ and presents less oscillatory behavior.

	If we consider a sharply peaked primordial isocurvature spectrum at $k_p$, we may gain an idea of the value of $\xNL^{\rm peak}$ needed to enter the observational window of future GW detectors. Using the estimate for the resonant peak amplitude in Eq.~\eqref{eq:estimateamplitude}, we find that
	\begin{align}\label{eq:asgausspeaked}
	{\cal A}_S\approx2.7\kappa_p^2 \times 10^{-5}\left(\frac{\Omega_{\rm GW,0}h^2}{10^{-14}}\right)^{1/2}\,,
	\end{align}
	for a given value of $\Omega_{\rm GW,0}h^2$. Once we know the amplitude of the primordial isocurvature spectrum in terms of $\Omega_{\rm GW,0}h^2$, we find that 
	\begin{align}\label{eq:xnlgausspeaked}
	\xNL^{\rm peak}\approx270\left(\frac{\Omega_{\rm GW,0}h^2}{10^{-14}}\right)^{-1/4}\,.
	\end{align}
	Note that our formalism is only valid for $\xNL> 1$ which implies we must have $\Omega_{\rm GW,0}h^2<5\times {10^{-5}}$. While we will study the constraints on $\Omega_{\rm GW,0}$ in detail in \S\ref{sec:constraints}, we can already use Big Bang Nucleosynthesis constraints\footnote{Strictly speaking the BBN constraints are integral constraints. However, for a peaked spectrum it already provides a good order of magnitude estimate. See \S\ref{sec:constraints} for further discussion.} $\Omega_{\rm GW,0}h^2\lesssim{10^{-6}}$ \cite{Caprini:2018mtu}, to find a lower bound of $\xNL\gtrsim 3$. Thus, our formalism is completely valid in the allowed observable range. 

	From now on, we will assume that we can take $\xNL^{\rm peak}$ and therefore $x_{\rm cut}$ to $\infty$ when we forecast constraints on ${\cal A}_S$ in \S\ref{sec:constraints}. While in detail the finite value of $x_{\rm cut}$ may affect our constraints, qualitatively our results will be robust to it.

	\subsection{Accounting for non-Gaussianity}
	\label{subsec:NG}

	The calculation thus far has assumed that the CDM isocurvature fluctuations are Gaussian distributed. In fact, for large $S$ they must be {highly skewed} since $S$ is bounded from below by $-1$. In Appendix~\ref{app:NG}, we argue that such a non-Gaussian spectrum might be described by a $\chi^2$-like distribution with power-spectrum 
	\begin{align}
	\label{eq:chi2_compatible}
	{\cal P}_{S}(k)=3{\cal A}_S\sqrt{\frac{6}{\pi}}\frac{k^3}{k_p^3}\,\exp\left(-\frac{3}{2}\frac{k^2}{k_p^2}\right),
	\end{align}
	and that due to the enhancement of the $4$-point function in the non-Gaussian case the induced GW signal might be enhanced relative to the Gaussian calculation by an extra factor of ${\cal A}_S$, namely
	\begin{align}
	\label{eq:NG_enhancement}
	\Omega_{\rm GW,c}^{\chi^2}&\approx {\cal A}_S\Omega^g_{\rm GW,c}.
	\end{align}
	where $\Omega^g_{\rm GW,c}$ denotes the result of the Gaussian calculation we have performed so far. 

	Since the primordial spectrum \eqref{eq:chi2_compatible} is fairly broad, we do not expect that the spectral shape of the non-Gaussian $\Omega_{\rm GW,c}^{\chi^2}$ will differ much from the Gaussian calculation $\Omega^g_{\rm GW,c}$. For instance, the effect of large local non-Gaussianity on the GW spectrum induced by adiabatic fluctuations has been studied in Refs.~\cite{Cai:2018dig,Unal:2018yaa,Atal:2021jyo,Adshead:2021hnm}. They found (see, e.g., Fig. 8 of Ref.~\cite{Adshead:2021hnm}) that the overall spectral shape is barely modified by the local non-Gaussian contribution. This provides indications that our amplitude enhancement \eqref{eq:NG_enhancement} might be sufficient to describe the effect of non-Gaussianity for broad peaks in the primordial spectrum. However, let us clarify that our estimate is based on one working example and that further investigation is needed to draw any general conclusion. This is out of the scope of the paper and we leave it for future work.

	If our estimate \eqref{eq:NG_enhancement} gives a good order of magnitude estimate, we can re-derive the estimate \eqref{eq:asgausspeaked} in the non-Gaussian case to find that the scalar spectrum should have amplitude
	\begin{align}\label{eq:asNGpeaked}
	{\cal A}_S\approx9\kappa_p^{4/3} \times 10^{-4}\left(\frac{\Omega_{\rm GW,0}h^2}{10^{-14}}\right)^{1/3}
	\end{align}
	to produce a given GW abundance $\Omega_{\rm GW,0}h^2$. Since we know the amplitude of the primordial isocurvature spectrum in terms of $\Omega_{\rm GW,0}h^2$, we find that the non-linear cutoff $\xNL^{\rm peak}$ is 
	\begin{align}\label{eq:xnlNGpeaked}
	\xNL^{\rm peak}\approx47\kappa_p^{1/3}\left(\frac{\Omega_{\rm GW,0}h^2}{10^{-14}}\right)^{-1/6}\,.
	\end{align}
	This time BBN constraints impose $\xNL^{\rm peak}>2\kappa_p^{1/3}$, and recall that we always have $\kappa_p\gg1$. Therefore the limit $\xNL^{\rm peak}\to\infty$ is an excellent approximation in the non-Gaussian case. In other words, the {highly skewed} non-Gaussianity significantly enhances the induced GW abundance, rendering it observable even for much smaller values ${\cal A}_S$ which in turn lead to a much later non-linearity time $\xNL^{\rm peak}$.

	We now proceed to present constraints on the primordial isocurvature spectrum from induced gravitational waves using both the conservative Gaussian calculation, as well as by attempting to capture the non-Gaussian enhancement using Eq.~\eqref{eq:NG_enhancement}.

\section{Future constraints on dark matter isocurvature from GWs\label{sec:constraints}}

	We now forecast how well upcoming GW detectors can constrain the primordial CDM isocurvature power spectrum. We use the peak integrated method developed by Ref.~\cite{Schmitz:2020syl}, which takes advantage of having a peaked spectral shape of GWs to derive constraints on the parameters of the scalar power spectrum.
	The peak-integrated sensitivity is defined as
	\begin{align}
	\Omega_{\rm PIS}(k_p)\equiv\left[n_{\rm det}t_{\rm obs}\int_{k_{\rm min}}^{k_{\rm max}}\frac{d k}{2 \pi} \left(\frac{{\cal S}(k,k_p)}{\Omega_{\rm noise}(k)}\right)^2\right]^{-1/2}\,,
	\end{align}
	where $n_{\rm det}=1$ if the measurement is by auto-correlation and $n_{\rm det} = 2$ if by cross-correlation, $t_{\rm obs}$ is the observing time of the experiment, 
	$\Omega_{\rm noise}(k)$ is the noise spectrum of the detector as a function of $k = 2 \pi f$, which extends from some $k_{\rm min}$ to $k_{\rm max}$, and ${\cal S}(k,k_p)$ is defined as
	\begin{equation}
	{\cal S}(k,k_p) \equiv \frac{\Omega_{\rm GW,0}(k, {\cal A}_S, k_p)}{\Omega_{\rm peak}({\cal A}_S)},
	\end{equation}
	where $\Omega_{\rm GW,0}(k, {\cal A}_S, k_p)$ is the induced GW spectrum at comoving wavenumber $k$ for a scalar spectrum with amplitude ${\cal A}_S$ and peak $k_p$, suppressing here the dependence on any parameters which are held fixed such as the scalar width $\Delta$. $\Omega_{\rm peak}({\cal A}_S) $ is an arbitrarily normalized function which encodes the scaling of the GW amplitude with the parameter which we wish to constrain, ${\cal A}_S$, such that ${\cal S}(k,k_p)$ does not depend on it. Note that $\Omega_{\rm peak}$ is not necessarily the maximum amplitude of $\Omega_{\rm GW,0}$. For the Gaussian calculation we will take $\Omega_{\rm peak}({\cal A}_S) = {\cal A}_S^2$, while for the non-Gaussian estimate we will take $\Omega_{\rm peak}({\cal A}_S) = {\cal A}_S^3$, reflecting Eq.~\eqref{eq:NG_enhancement}.

	We then claim a detection of a scalar peak of amplitude ${\cal A}_S$ at $k_p$ if the signal to noise ratio
	\begin{align}
	{\varrho}=\frac{\Omega_{\rm peak}({\cal A}_S)}{\Omega_{\rm PIS}(k_{p})}
	\end{align}
	exceeds unity.

	\begin{figure}[t]
	\includegraphics[]{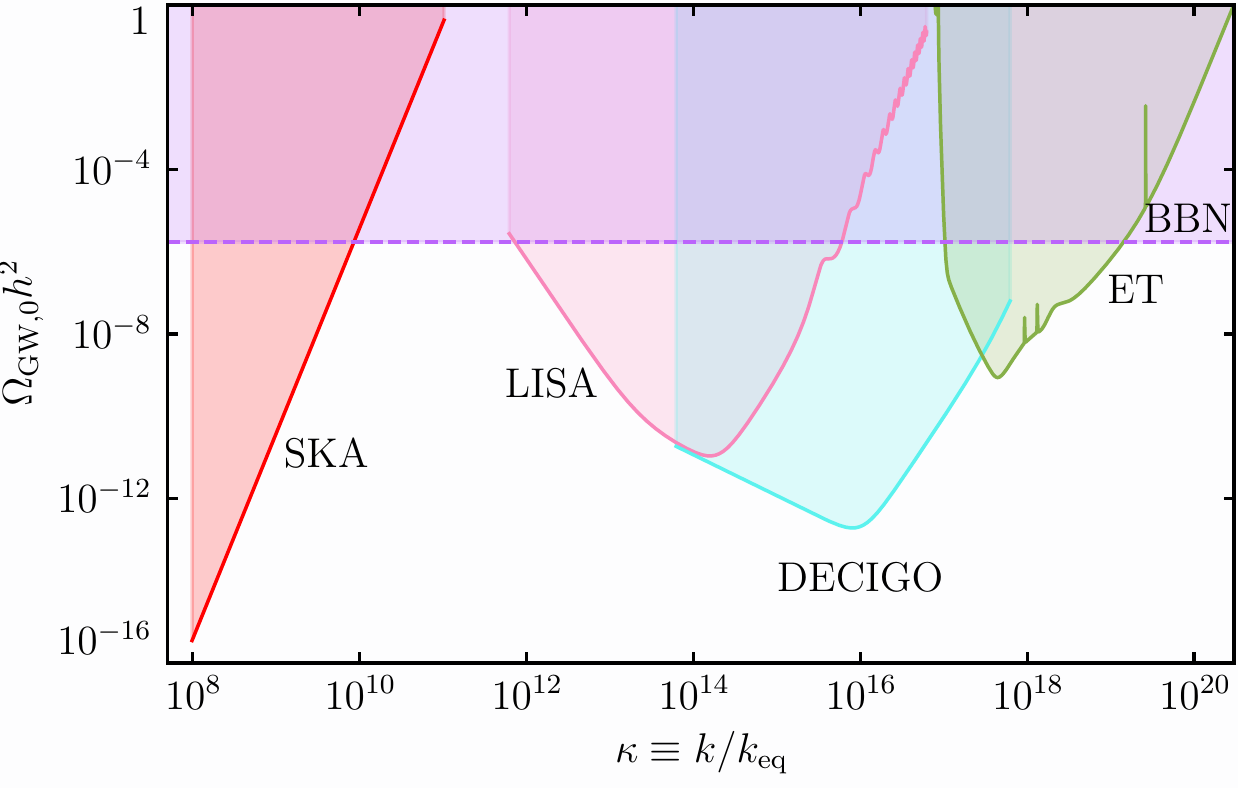}
	\caption{Projected constraints on the spectral energy density in GWs from next generation experiments SKA, LISA, DECIGO, and ET, which we use in this work to produce forecasted constraints on primordial CDM isocurvature. We also show the existing constraint from Big Bang Nucleosynthesis. For further discussion see \S\ref{sec:constraints}.
	}
	\label{fig:omegaGW0_constraints}
	\end{figure}

	We summarize in Fig.~\ref{fig:omegaGW0_constraints} the gravitational wave spectral sensitivities we use to obtain constraints in this work, all of which are obtained from Ref.~\cite{Schmitz:2020syl}. For clarity we have plotted them as a function of the comoving scale $\kappa=k/\keq$ probed, and we see they all lie in the $\kappa\gg1$ regime where our analytic calculation is valid. These noise curves are based on
	\begin{itemize}
	\item the forecasted sensitivity of a pulsar timing array constructed from the Square Kilometer Array (SKA) inspired by assumptions in Ref.~\cite{Janssen:2014dka},
	\item the forecasted sensitivity of the Laser Interferometer Space Antenna (LISA) from Ref.~\cite{Robson:2018ifk},
	\item the forecasted sensitivity of the Deci-Hertz Interferometer Gravitational-Wave Observatory (DECIGO) from Ref.~\cite{Kuroyanagi:2014qza},
	\item the forecasted sensitivity of the Einstein Telescope (ET), labeled ET-D, from Ref.~\cite{ET:sensitivities}.
	\end{itemize}
	
	We also show the integrated constraint from the contribution of GWs to the total energy density of the universe during Big Bang nucleosynthesis (BBN), $\int^{\infty}_{k_{\rm min}} d\ln k \,\Omega_{\rm GW,0}h^2(k)\lesssim {10^{-6}}$ \cite{Caprini:2018mtu}. This constraint applies for GWs which are present before BBN and thus for GWs on scales smaller than $k_{\rm min}\approx 10^3\,{\rm Mpc}^{-1}$, the scale that enters the horizon near the end of BBN. Although it is an integral constraint, the evaluation of the integrand provides an upper bound to the GW spectrum which we show in Fig.~\ref{fig:omegaGW0_constraints} and which we used in \S\ref{subsec:cutoff} to set a lower bound on the non-linearity scale $x_{\rm NL}$.  Comparable gravitational wave bounds extending to larger scales can be obtained from the CMB \cite{Smith:2006nka,Pagano:2015hma}. The constraints shown here are only a sampling of the wide variety of  present and upcoming constraints on the gravitational wave abundance \cite{Arzoumanian:2020vkk,KAGRA:2013rdx,Blas:2021mqw}.

	\begin{figure}[t]
	\includegraphics[]{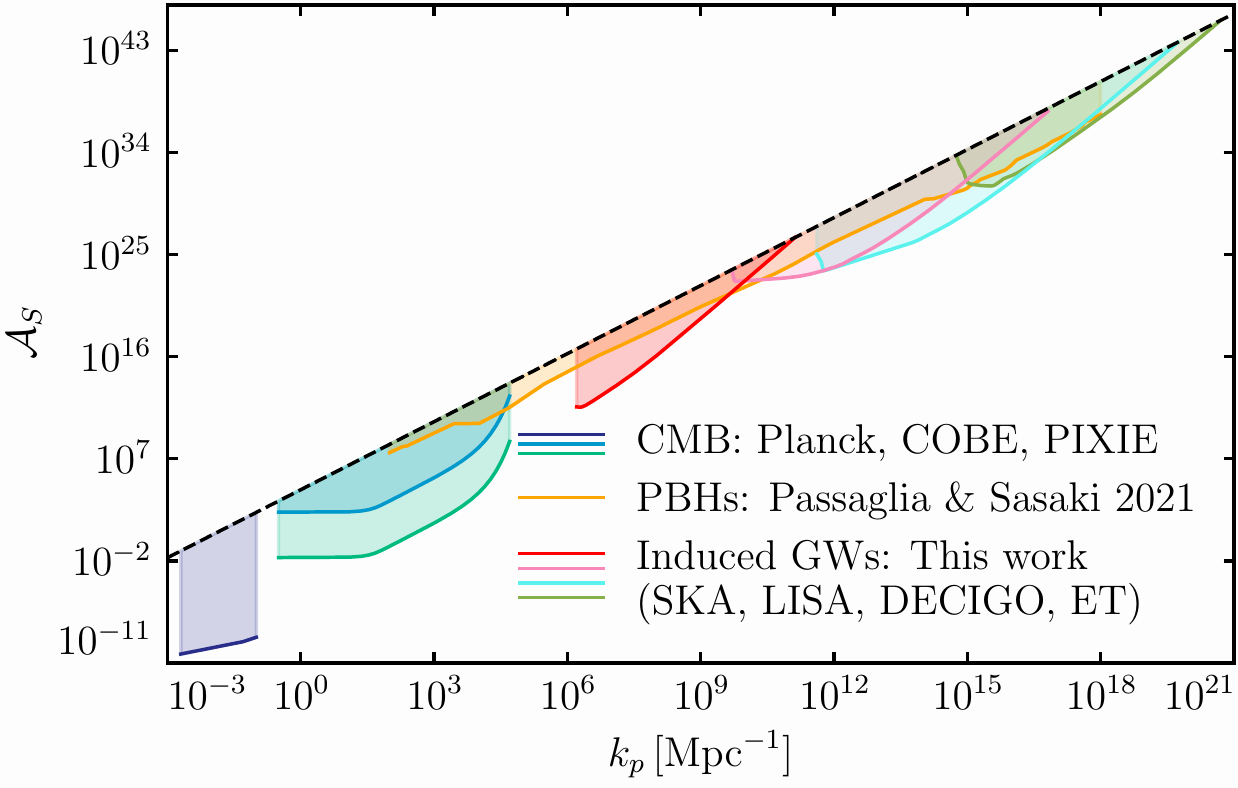}
	\caption{Constraints on primordial isocurvature from the CMB (Appendix \ref{app:other}), from PBHs (Ref.~\cite{Passaglia:2021jla}), and now from induced gravitational as forecasted in this work. We assume a sharply peaked primordial spectrum ($\Delta = 0.01$) and compute the GW signal in the Gaussian approximation. Above the dashed black line, which scales as $\left(k_p/\keq\right)^2$, the isocurvature perturbations become non-linear before horizon crossing and we cannot describe them in our formalism. For further discussion see \S\ref{sec:constraints}.}
	\label{fig:constraints}
	\end{figure}

	In Fig.~\ref{fig:constraints}, we use the projected sensitivities of the upcoming detectors to forecast constraints on the primordial CDM isocurvature power spectrum, using a sharp log-normal peak with $\Delta=0.01$ in the conservative Gaussian approximation as a fiducial model. These constraints extend from $k\sim 10^{6}\, \Mpc^{-1}$ to $k\sim 10^{20}\, \Mpc^{-1}$, reflecting the range of the spectral constraints shown in Fig.~\ref{fig:omegaGW0_constraints}. The constraints on ${\cal A}_S$ degrade as $(k_p/\keq)^2$, as we described in \S\ref{sec:inducedGWs}, due to the decreasing importance of CDM at horizon crossing for increasing $k$. We show with a dashed black line the constraint imposed by $\xNL^{\rm peak} \geq 1$ which is necessary for the validity of our calculation. Recall that for $\xNL^{\rm peak} < 1$ the fluctuations become non-linear before horizon crossing and cannot be described in our formalism.

	We overplot more traditional constraints on the CDM isocurvature, namely those  from CMB temperature anisotropies (Plank) and spectral distortions (COBE, PIXIE forecast), details of which are provided in Appendix~\ref{app:other}. These are significantly stronger than our induced GW constraints, but they extend only up to $k \sim 10^{5} \, \Mpc^{-1}$. In contrast, constraints from induced GWs have the potential to extend to much smaller scales.

	We also show in Fig.~\ref{fig:constraints} constraints on the isocurvature spectrum from the absence of PBHs, as computed in Ref.~\cite{Passaglia:2021jla}. That work found that for PBHs produced by CDM isocurvature perturbations to not overclose the universe, the amplitude of the primordial isocurvature power spectrum should be less than\footnote{We have corrected here an order unity factor neglected in the initial version of Ref.~\cite{Passaglia:2021jla}.}
	\begin{equation}
	\label{eq:sigmaR_powerlaw_k}
	{\cal A}_S(k_p) \lesssim \left({\frac{2^{1/2}}{3^{1/4}}} \frac{k_p}{\keq} \left(\frac{1}{b}\right)^{\frac{1}{2q}} \right)^{\frac{4q}{2q+1}},
	\end{equation}
	where $q$ and $b$ are constants which encode the PBH formation probability, with $q=13/2$, $b= 0.02$ reasonable choices. Because $q$ is large, the same weakening of constraints as $(k_p/\keq)^2$ which we found for the induced GWs is present in the PBH constraints. The slight deviation from this scaling for finite $q$ is due to the favorable redshifting of PBHs formed during radiation domination. While this formula encodes the rough scaling of the constraints, in Fig.~\ref{fig:constraints} we show the full constraints from Ref.~\cite{Passaglia:2021jla} computed using a variety of existing experimental constraints on PBH abundances.

	As Fig.~\ref{fig:constraints} shows, induced GW constraints have the potential to improve on the constraints from PBHs by orders of magnitude. Moreover, the constraints from GWs can extend to smaller scales than the PBH constraints because small PBHs can evaporate while GWs persist. Similar conclusions have been reached in forecasts of constraints on the amplitude of the adiabatic spectrum on small scales (see, e.g., Ref.~\cite{Gow:2020bzo}).

	The constraints in Fig.~\ref{fig:constraints} were obtained for a sharply peaked log-normal spectrum of the form \eqref{eq:Pklog}. More generally, we can immediately see from this figure that for a primordial isocurvature power spectrum ${\cal P}_S$ which leads to an observable GW signal in some future GW detector, the low-$k$ tail of ${\cal P}_S$ should decay faster than $k^2$. Otherwise, it would be in conflict with CMB constraints. In other words, the effective curvature power spectrum \eqref{eq:Pphieffectve} should always decay on large scales.

	\begin{figure}[t]
	\includegraphics[]{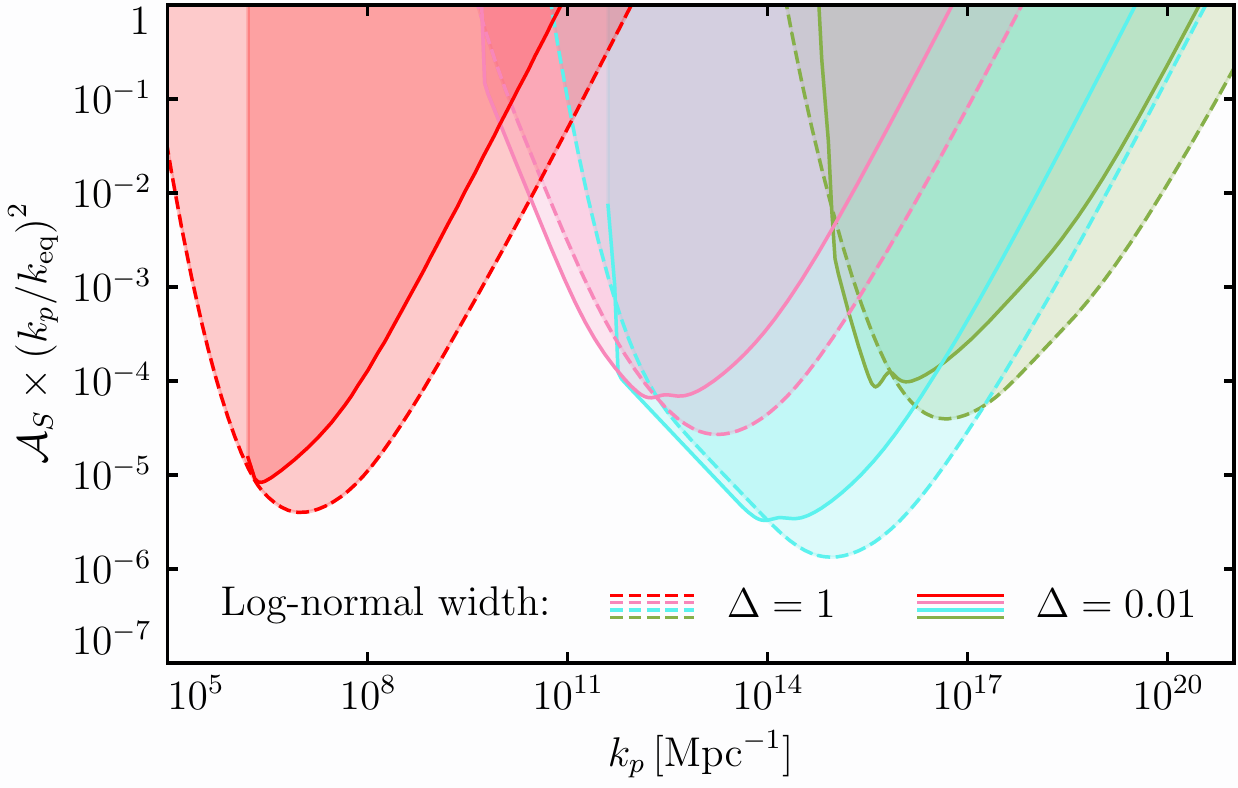}
	\caption{The constraints on the primordial isocurvature power spectrum depend on the shape of the spectrum. For a broad $\Delta = 1$ log-normal peak in the scalar power spectrum (dashed lines), the constraints strengthen and shift to the right relative to a narrow $\Delta = 0.01$ peak (solid lines), reflecting the larger effective curvature power spectrum \eqref{eq:Pphieffectve} in the broad case due to the $k^{-2}$ factor in the mapping between isocurvature and effective curvature. For further discussion see \S\ref{sec:constraints}.}
	\label{fig:constraints_variety}
	\end{figure}

	In Fig.~\ref{fig:constraints_variety}, we show explicitly how the constraints on the CDM isocurvature depend on the shape of the primordial spectrum. We compare the narrow-peak $\Delta=0.01$ scalar spectrum we showed in Fig.~\ref{fig:constraints} to a wider peak $\Delta=1$. These correspond to the induced GW spectra we showed in Fig.~\ref{fig:power_spectrum_lognormal}. Because the general scaling of the constraints is independent of the shape of the primordial isocurvature spectrum, we here factor out the overall $(k_p/\keq)^2$ dependence for legibility. Nonetheless the detailed structure of the constraints does depend on the spectral shape of the primordial isocurvature. The broader $\Delta = 1$ scalar spectra are more strongly constrained than the narrow $\Delta = 0.01$ results we showed in Fig.~\ref{fig:constraints}, both in terms of the amplitude of the constraint at a given scale  and in the range of scales constrained. This reflects the wider induced spectra we found in Fig.~\ref{fig:power_spectrum_lognormal} as well as the shift and enhancement of the peak of the induced GWs due to the enhanced effective curvature power spectrum \eqref{eq:Pphieffectve}. For a broad peak, there exists more power on large scales which is then favored by the $k^{-2}$ factor in the effective curvature power spectrum.

	\begin{figure}[t]
	\includegraphics[]{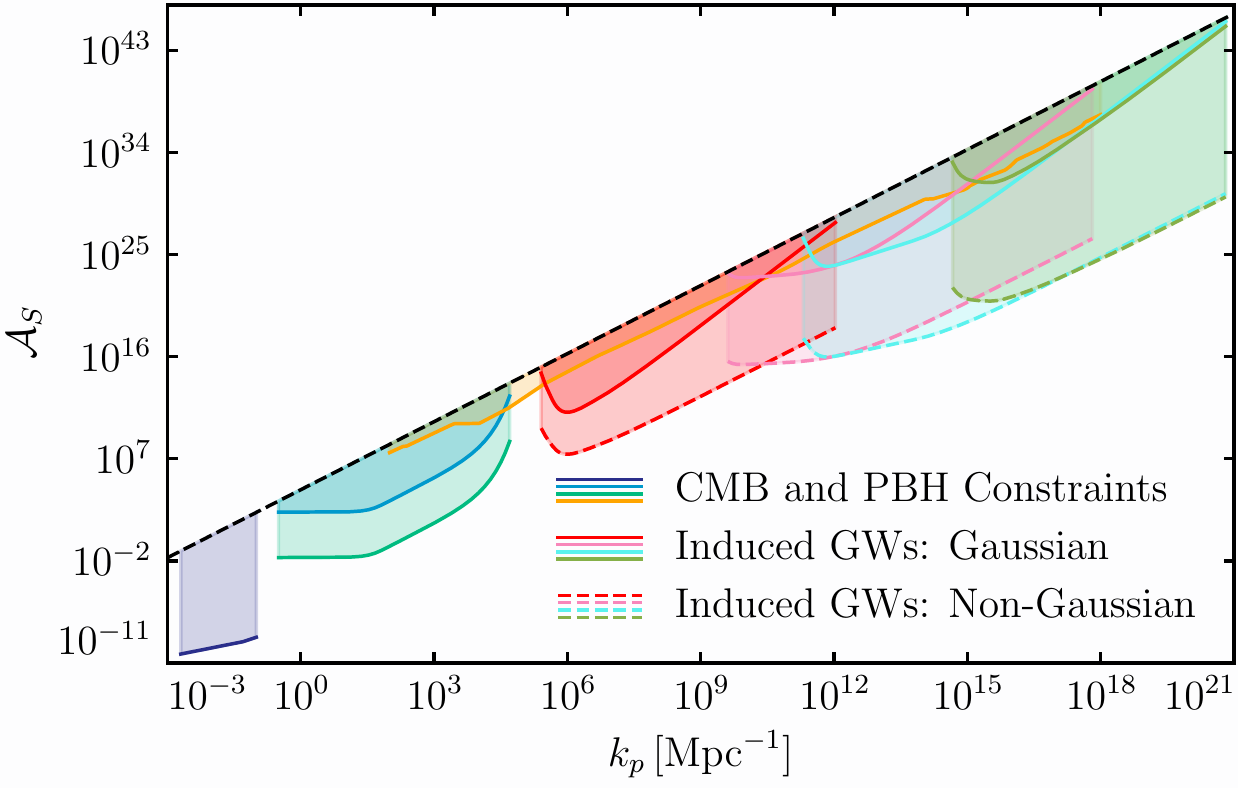}
	\caption{The non-Gaussianity of the isocurvature perturbations should increase the GW signal relative to the Gaussian approximation we have employed so far and therefore strengthen the constraint from induced GWs. We estimate the effect by using the $\chi^2$-{like} isocurvature power spectrum Eq.~\eqref{eq:chi2_compatible} and then enhancing the induced GW spectrum by an additional factor ${\cal A}_S$, Eq.~\eqref{eq:NG_enhancement}. While the constraints degraded as  $(k_p/\keq)^2$ in the Gaussian approximation, they degrade only as $(k_p/\keq)^{4/3}$ in the non-Gaussian estimate. The actual result may lie somewhere between the two estimates. For further discussion see \S\ref{sec:constraints}.}
	\label{fig:constraints_bump}
	\end{figure}

	So far we have assumed that the isocurvature fluctuations follow a Gaussian distribution. As we discussed in \S\ref{subsec:NG}, this cannot be the case and our Gaussian calculation is likely conservative. We attempt in Fig.~\ref{fig:constraints_bump} to estimate the effect of the {highly skewed} non-Gaussianity on the constraints by using the non-Gaussian enhancement of the signal proposed in \S\ref{subsec:NG} and in Appendix~\ref{app:NG}. We use the $\chi^2$-like isocurvature power spectrum Eq.~\eqref{eq:chi2_compatible}, and then account for the non-Gaussianity by increasing the amplitude of the induced GW spectrum by an additional factor ${\cal A}_S$ with respect to the Gaussian case, Eq.~\eqref{eq:NG_enhancement}. Since ${\cal A}_S$ must be very large to produce observable GWs, the forecasted constraints are now therefore much stronger.  While in the Gaussian case the constraints on ${\cal A}_S$ weakened as $(k_p/\keq)^2$, in the non-Gaussian case the constraints weaken only as $(k_p/\keq)^{4/3}$. We leave a more detailed investigation of the effects of such highly non-Gaussian probability distribution function on the induced GW spectrum for future work. 

	Finally, one might wonder about the effect of possible mixed adiabatic-isocurvature initial conditions. Since the GW spectrum for a very sharp peak in the primordial isocurvature spectrum is similar to the one from primordial curvature fluctuations, we can anticipate that for mixed initial conditions the peak of the induced GW spectrum would be given by a sum of terms of the type
    \begin{align}
	\Omega^{\rm peak}_{\rm GW,c}\approx \gamma_n \left({\cal A}_S\kappa_p^{-1}\right)^{n}{\cal {\cal A}_{\cal R}}^{4-n}\,,
	\end{align}
	where ${\cal A}_{\cal R}$ is the amplitude of the primordial curvature perturbation, $n\in [0,4]$, $\gamma_{0,2,4}\sim1$, and $\gamma_{1,3} \leq 1$ take into account possible correlations between $S$ and ${\cal R}$, with $\gamma_{1,3}=0$ in absence of correlations. It is interesting to note that additional factors of ${\cal R}$ do not necessarily help in boosting the amplitude of induced GWs, as from primordial black hole constraints one already has ${\cal A}_{\cal R}<0.1$. Even in the case when ${\cal A}_{\cal S}\kappa_p^{-1}\ll 1$, correlations with ${\cal R}$ do not help much as the dominant contribution would be the ${\cal R}^4$ term. We leave for future work a detailed study of mixed initial conditions. 

\section{Discussion and conclusions\label{sec:conclusions}}

	Gravitational waves generated in the early universe by the scalar perturbations have been recognized as an exciting probe of the adiabatic mode of the primordial fluctuations. In this work, we have shown that observable gravitational wave signals can also be generated from the cold dark matter isocurvature mode if its amplitude is sufficiently large. These induced GWs might be the counterpart to the PBHs generated by large isocurvature studied in Ref.~\cite{Passaglia:2021jla}.

	We derived for the first time the general analytical kernel necessary to compute isocurvature-induced GWs in radiation domination, presented in Eqs.~\eqref{eq:Icinf} and \eqref{eq:Isinf}, enabling the use of gravitational wave astronomy to constrain the nature of the primordial perturbations on small scales in the early universe.

	We then studied the GWs induced by sharp and broad peaks in the primordial isocurvature power spectrum and found three main differences with respect to the case of adiabatic fluctuations. 

	First, since the transfer from isocurvature to curvature effectively stops at horizon crossing, the isocurvature-induced GW spectrum is suppressed by a factor $\left(\keq/k_p\right)^4$, where $k_p$ is the scale of the peak in the primordial spectrum and $\keq\approx 0.01\,{\rm Mpc}^{-1}$. This suppression can be compensated for with a large amplitude of the primordial isocurvature spectrum while respecting the confines of linear theory, and the induced GW spectrum \eqref{eq:Phgaussian} can be rewritten to absorb this suppression by introducing an effective primordial curvature power spectrum \eqref{eq:Pphieffectve} ${\cal P}_\Phi^{\rm eff}(k)=\left(\keq/k\right)^2{\cal P}_S(k)$ times an effective kernel. The effective kernel, given by Eqs.~\eqref{eq:Icinf} and \eqref{eq:Isinf} times a factor $(uv)$, presents qualitative similarities with the case of adiabatic-induced GWs: it has the same resonance phenomena and infrared behavior.

	The effective primordial curvature power spectrum \eqref{eq:Pphieffectve} gives us intuition for the second main difference associated with primordial isocurvature perturbations: the peak in the induced GW spectrum becomes shifted to large scales relative to the peak in the isocurvature power spectrum by the additional $k^{-2}$ factor. This is especially important for broad primordial spectra, as we saw in Fig.~\ref{fig:power_spectrum_lognormal}. 

	The third difference appears for very sharp peaks. As is clear from Fig.~\ref{fig:power_spectrum_deltafunction}, the isocurvature-induced GW spectrum does not present an exact zero and a dip appears slightly below $k=\sqrt{2}c_sk_p$, which is the position of the dip for adiabatic fluctuations. This effect is due to the difference in phase in the isocurvature-induced curvature fluctuations. Although the resonance does lead to a peak in the GW spectrum at $k=2c_sk_p$, the overall shape is different than in the adiabatic case. The resonant peak-dip relation for sharp peaks may be helpful in distinguishing the GW spectrum induced by isocurvature and adiabatic initial conditions.

	Our work shows that if a stochastic GW signal is detected in future experiments, one must consider the possibility that these GWs were induced by either primordial adiabatic or isocurvature fluctuations. GWs induced by different initial conditions are in principle distinguishable,
	and additional information from the PBH mass function may also help to tell both scenarios apart \cite{Passaglia:2021jla}.

	In the absence of a future induced GW signal, our work provides at the moment the best prospects to constrain dark matter isocurvature fluctuations on small scales. We show our forecast in Fig.~\ref{fig:constraints} for future GW detectors such as LISA, ET, DECIGO and SKA. We expect an improvement of several orders of magnitude with respect to the constraints from the absence of PBHs \cite{Passaglia:2021jla}.

	There are several aspects that deserve further exploration. For instance, a crucial difference with respect to the adiabatic case is that to produce significant gravitational waves the isocurvature fluctuations must have an amplitude vastly exceeding unity. This implies that: $(i)$ the probability density distribution of isocurvature fluctuations must be highly non-Gaussian (see Appendix \ref{app:NG}), since the local energy density must also be positive, and $(ii)$ such large dark matter density fluctuations enter the non-linear regime well before matter-radiation equality.

	We found that point $(i)$ may have a large effect on the amplitude of the induced GW spectrum. Though for analytic viability we have assumed in our calculations of induced GWs \eqref{eq:Phgaussian} that primordial isocurvature fluctuations are drawn from a Gaussian distribution, we expect that the non-Gaussianity enhances the amplitude of induced GWs by enhancing the 4-point correlation function. Our result can then be thought of as a conservative estimate for the GW production from large isocurvature.
	By analogy with a $\chi^2$-like distribution with a very large variance, where the 4th moment of the distribution is proportional to the ${\rm (variance)}^3$, we estimated that $\Omega_{\rm GWs}^{\rm NG}\sim {\cal A}_S \Omega_{\rm GWs}^{\rm g}$, where ${\cal A}_S$ is the amplitude of the primordial isocurvature spectrum. This non-Gaussian enhancement leads to the strengthening of the constraints as shown in Fig.~\ref{fig:constraints_bump}. A related issue is that we lack a concrete model which could lead to such large isocurvature and, therefore, our estimates for the effects of non-Gaussianity are a first, educated, guess. In any case, we expect that the result of a more accurate calculation should lie somewhere within the Gaussian and non-Gaussian results.

	About point $(ii)$, we argued that despite such large amplitude, linear theory is valid until density fluctuations of dark matter locally overcome the radiation. When dark matter density fluctuations enter the non-linear regime, we cut-off the production of induced GWs. This leads to a conservative amplitude of the resulting GWs as we expect GWs to be generated even in the non-linear regime. Nevertheless, because most of the gravitational waves are produced around horizon crossing, we find that the overall amplitude of the resulting GW spectrum is rather insensitive to the cut-off unless the onset of the non-linear regime occurs very close to that epoch. To derive the full induced GW spectrum inside the non-linear regime, numerical simulations may be required.

	In addition to the above, we have primarily considered pure primordial isocurvature fluctuations. However, our formulation also applies directly to any mixed initial condition, and in fact our results might be important in the presence of correlations between large adiabatic and isocurvature fluctuations. 

	Before we conclude our work, let us comment on another important and related aspect to explore, which is the possible formation of small-scale CDM halos. Since the small-scale dark matter distribution is likely to be highly inhomogeneous, small-scale CDM halos may form(either during radiation domination\footnote{During radiation domination the density fluctuations of matter grow logarithmically which makes structure formation less likely. When we reach matter domination density fluctuations collapse into virialized objects. However, some of the formed halos will be disrupted by interaction with, e.g., stars \cite{Kavanagh:2020gcy}.} or the subsequent matter domination). This is in contrast with PBH formation from adiabatic fluctuations in a radiation dominated universe, where any fluctuations that do not end up forming a black hole are washed away by the free streaming of radiation. The formation of microhalos from adiabatic and isocurvature fluctuations has been studied in the context of axions, with the so-called axion miniclusters \cite{Hogan:1988mp,Fairbairn:2017sil,Eggemeier:2019khm,Kavanagh:2020gcy}, and recently by long range scalar forces \cite{Savastano:2019zpr,Flores:2020drq} and CDM self-interactions \cite{Erickcek:2020wzd,Blinov:2021axd}. Since, in general, the CDM halos are rather disperse, some constraints applicable to PBHs do not apply or apply weakly to microhalos \cite{Savastano:2019zpr,Dai:2019lud}. Interesting recently proposed future detection methods are irregularities in the microlensing caustics of highly magnified stars \cite{Dai:2019lud} and fluctuations in pulsar timing arrays \cite{Dror:2019twh}. In the future, these methods may restrict the amount of microhalos to be less than $1\%$ of CDM in the range of roughly $10^{-12} - 100 M_\odot$. We note, however, that current and future constraints depend on the density profile of the halos, their halo mass function and tidal disruption by other compact objects such as stars \cite{Kavanagh:2020gcy}. The study of the formation of such microhalos in our present model and their evolution until today is an interesting topic but out of the scope of this paper. We leave it for future work.\\

	\textit{Note Added:} After submission, Ref.~\cite{Yoo:2021fxs} showed with numerical simulations that large initial isocurvature fluctuations of a massless scalar field can collapse to form PBHs. This is a strong indication of the validity of the results of Ref.~\cite{Passaglia:2021jla}, where PBHs form from the collapse of large CDM isocurvature fluctuations.

\section*{Acknowledgments} 
	We would like to thank Misao Sasaki for many useful discussions and for providing an example of a power spectrum with positive 2-point function. G.D. would also like to thank V.~Atal for helpful discussions on the peak integrated sensitivity curves, J.~Chluba, D.~Grin and S.~Patil for useful correspondence on CMB spectral distortions and J.~Shelton for useful correspondence on dark matter microhalos. S.RP would also like to thank J.~Lesgourgues, L.~Pinol and L.T.~Witkowski for useful discussions.	G.D. as a Fellini fellow was supported by the European Union's Horizon 2020 research and innovation programme under the Marie Sk{\l}odowska-Curie grant agreement No 754496. S.P. was supported by the World Premier International Research Center Initiative (WPI), MEXT, Japan. S.RP is supported by the European Research Council under the European Union's Horizon 2020 research and innovation programme (grant agreement No 758792, project GEODESI).
	
\appendix

\section{Fundamentals of linear perturbation theory \label{app:formulas}}

In this appendix we explicitly state the convention used for the perturbative expansion and write down the main formulas needed in the text.

The metric in the Newtonian gauge is given by
\begin{align}
ds^2=a^2(\tau)\left[-(1+2\Psi)d\tau^2+(\delta_{ij}+2\Phi\delta_{ij}+h_{ij})dx^idx^j\right]\,,
\end{align}
and the matter and radiation sectors are described by their energy momentum tensors, respectively given by
\begin{align}
T_{m\mu\nu}&=\rho_m u_{m\mu}u_{m\nu}\,,\\
T_{r\mu\nu}&=(\rho_r+p_r)u_{r\mu}u_{r\nu}+p_r g_{\mu\nu}\,,
\end{align}
where $g_{\mu\nu}$ is the metric, $\rho$ and $p$ respectively are the energy density and pressure and $u_{\mu}$ the fluid $4$-velocity. The subscripts ``$m$'' and ``$r$'' respectively refer to CDM and radiation.

\subsection{Background}
In the background we have that $u_{m0}=u_{r0}=-a$ and $u_{mi}=u_{ri}=0$. Then, the Friedmann and energy conservation equations in terms of conformal time are given by
\begin{align}
3{\cal H}^2=a^2(\rho_m+\rho_r)\equiv a^2\rho\,\quad,&
\quad{\cal H}^2+2{\cal H}'=-\frac{1}{3}a^2\rho_r\,,\\
\rho_m'+3{\cal H}\rho_m=0\,\quad&{\rm and}\quad
\rho_r'+4{\cal H}\rho_r=0\,.
\end{align}
Interestingly, an exact solution to the scale factor is given by
\begin{align}
\frac{a(\tau)}{\aeq}=2\left(\frac{\tau}{\tau_*}\right)+\left(\frac{\tau}{\tau_*}\right)^2\,,
\end{align}
where $(\sqrt{2}-1)\tau_*=\tau_{\rm eq}$ and the subscript ``eq'' refers to the time of matter-radiation equality. Then we have that
\begin{align}
\rho_m=\rho_{\rm eq}\left(\frac{a}{\aeq}\right)^{-3}\quad{\rm and}\quad \rho_r=\rho_{\rm eq}\left(\frac{a}{\aeq}\right)^{-4}\,,
\end{align}
where by the Friedmann equations $3{\cal H}_{\rm eq}^2=2\rho_{\rm eq}$ and also 
\begin{align}
{\cal H}_{\rm eq}=\keq=\frac{2\sqrt{2}}{\tau_*}\,.
\end{align}
$\keq$ refers to the comoving wavenumber that entered the horizon right at matter-radiation equality.

\subsection{First order}

At the linear perturbation level, we have $u_{m0}=u_{r0}=-a\left(1+\Psi\right),~u_{mi}=a\partial_iV_m$ and $u_{ri}=a\partial_iV_r$. In the absence of anisotropies the $ij$ component of Einstein equations leads to
\begin{align}
\Phi+\Psi=0\,.
\end{align}
Since we are interested in isocurvature, we introduce the following variables:
\begin{align}
S\equiv \frac{\delta\rho_m}{\rho_m}-\frac{3}{4}\frac{\delta\rho_r}{\rho_r}\quad{\rm and}\quad V_{\rm rel}\equiv V_m-V_r\,,
\end{align}
where $V_{\rm rel}$ are the relative velocities of the fluids. Using these new variables, we find that the curvature perturbation obeys
\begin{align}
\Phi''+3{\cal H}(1+c_s^2)\Phi'+({\cal H}^2(1+3c_s^2)+2{\cal H}')\Phi-c_s^2\Delta\Phi=\frac{a^2}{2}\rho_mc_s^2S\,,
\end{align}
where we defined as usual
\begin{align}
c_s^2\equiv\frac{4}{9}\frac{\rho_r}{\rho_m+4\rho_r/3}\,.
\end{align}
The isocurvature mode follows
\begin{align}\label{eq:isoeom}
S'=-\Delta V_{\rm rel}\,,
\end{align}
and the relative velocity evolves according to
\begin{align}\label{eq:eomsvrel}
V'_{\rm rel}&+3c_s^2{\cal H}V_{\rm rel}+\frac{3}{2a^2\rho_r}c_s^2{\Delta\Phi}+\frac{3\rho_m}{4\rho_r}c_s^2S=0\,.
\end{align}

\subsection{Second order}

At second order in perturbation theory, scalar modes act as a source to the linear tensor modes. The tensor mode equation of motion the reads
\begin{align}
h_{ij}''+2{\cal H}h'_{ij}+\Delta h_{ij}={\cal P}_{ij}\,^{ab}S_{ab}\,,
\end{align}
where ${\cal P}_{ij}\,^{ab}$ is the transverse-traceless projector which can be found in, e.g., Refs.~\cite{Domenech:2020ssp,Domenech:2021ztg}, and
\begin{align}
S_{ij}&=4\partial_i\Phi\partial_j\Phi+6c_s^2\frac{\rho}{\rho_r}\partial_i\left(\frac{\Phi'}{{\cal H}}
+\Phi\right)\partial_j\left(\frac{\Phi'}{{\cal H}}+\Phi\right)+6a^2c_s^2\rho_m\partial_iV_{\rm rel}\partial_jV_{\rm rel}\,.
\end{align}
From the equation above it is clear that induced GWs are mainly sourced by the curvature perturbation $\Phi$. As we show in the text, the contribution from the relative velocities is always suppressed except for modes $k\sim \keq$. For a study of the effect of relative velocities in the baryon-CDM fluids see Ref.~\cite{Gurian:2021rfv}.

\section{Validity of linear theory for large isocurvature \label{app:lineartheory}}

In this appendix we give more details on why one can treat large isocurvature $S\gg1$ within linear cosmological perturbation theory. $S\gg1$ essentially implies that the local matter density is much larger than the background value, $\delta\rho_m \gg\rho_m$. This might seem to indicate that perturbation theory no longer holds. 

However, a similar situation commonly occurs in the study of structure formation deep inside the cosmological horizon. There CDM density fluctuations grow above the mean background density to form halos, but the gravitational potential is much smaller than unity. One then employs the Newtonian limit, where the gravitational potential $\Phi$ is treated as a metric perturbation, and only the matter is considered at the non-linear level. If the matter velocity perturbation remain small enough, however, even perturbations in the matter sector can be treated within linear perturbation theory.

Let us apply this reasoning to the case of large isocurvature perturbations on superhorizon scales using the Einstein and the conservation equations
\begin{align}
M_{\rm pl}^2 G_{\mu\nu}&=T_{\mu\nu}\,,\nonumber\\
\nabla^\mu T_{\mu\nu}&=0\,.
\end{align}

By definition, isocurvature fluctuations do not affect the metric far on superhorizon scales. Linear perturbations in the metric are therefore well justified and we may expand $G_{\mu\nu}$ linearly in $\Phi$. Note that the expansion of $G_{\mu\nu}$ on cosmological scales entails factors proportional to $H^2$. Therefore, we crudely estimate that in the situation we are discussing we have $G_{\mu\nu}\sim \mathcal{O}(H^2)$. 

Turning our attention to the matter sector, for CDM $T_{\mu\nu}$ is linear in $\rho_m+\delta\rho_m$ and quadratic in the velocities $V_m$. This means that as long as velocities are small,
we may treat the matter sector linearly in $\rho_m+\delta\rho_m$ and $V_m$. This does not pose a problem to the metric sector as long as $\rho_m+\delta\rho_m\ll H^2M_{\rm pl}^2$. If radiation dominates the universe so that $H^2M_{\rm pl}^2\sim \rho_r$, then this leads to the requirement that $\rho_m S\ll \rho_r$, which is the one used in the main text to define the time $\xNL$ when our calculation breaks down. 

In the matter conservation equations, we may look for illustrative purposes to the continuity equation for the energy density of CDM in the Newtonian limit, namely
\begin{align}
\delta\rho_m'+2{\cal H}\delta\rho_m+\partial_i\left((\rho_m+\delta\rho_m) \partial^i V_m\right)=0\,,
\end{align}
where we neglected time derivatives of $\Phi$. We see that we can treat the system in the linear regime as long as 
\begin{align}
\delta\rho_m',{\cal H}\delta\rho_m\gg k^2 \delta\rho_m V_m\,.
\end{align}
On one hand, for scales close to the horizon we have that $\delta\rho_m'\sim{\cal H}\delta\rho_m$ and thus we need
\begin{align}
\frac{k^2V_m}{{\cal H}}\ll 1\,,
\end{align}
even if we have $\delta\rho_m\gg \rho_m$. This requirement is satisfied throughout our work, with the solution for the velocity Eq.~\eqref{eq:vrelsol} indicating that
\begin{align}
\frac{k^2V_{\rm rel}}{{\cal H}}\propto \frac{x}{\kappa}S(0)\,,
\end{align}
which is $\mathcal{O}(1)$ at the non-linear regime when $x=\xNL$. On the other hand, for scales much smaller than the horizon we have that $\delta\rho_m'\sim k\delta\rho_m$, and the requirement becomes $kV_m\sim S(0)/\kappa\ll1$. We plot the behavior of $V_m$ in Fig.~\ref{fig:phi_S_V}.

\section{Non-Gaussian nature of large isocurvature fluctuations \label{app:NG}}
In this appendix we elaborate on a consistent statistical nature of fluctuations. We note that by definition one must have $S>-1$, since initially
\begin{align}
S\approx \frac{\delta\rho_m}{\rho_m}=\frac{\rho_m(x)-\rho_m}{\rho_m}\,.
\end{align}
This means that in order to have $S\gg1$, $S$ cannot be distributed according to Gaussian statistics. Most of the fluctuations must lie at $S\sim -1$ (or $\rho_m(x)\sim 0$) to compensate for the very large rare fluctuations elsewhere. So the universe is basically empty of CDM except for special regions where CDM highly accumulates. The CDM distribution is therefore highly inhomogeneous on small scales, and although the exact distribution will depend on the underlying mechanism to generate large isocurvature, we can expect that the realistic non-Gaussian distribution leads in general to an enhanced induced GW spectrum with respect to the Gaussian case we have computed in the main text. In this appendix, we develop this argument by discussing two examples in real space and then the implications for Fourier modes.

\subsection{Scalar field without mean value}
\label{subapp:nomean}

In the early days of CMB studies, isocurvature perturbations presented a promising scenario for structure formation (see, e.g., Ref.~\cite{Hu:1995xs} and discussion therein). This was definitively abandoned after WMAP measured the detailed structure of the anisotropy spectrum and constrained the amplitude of isocurvature fluctuations to be less than $1\%$ of the adiabatic ones on CMB scales, but nevertheless this period led to an interesting discussion on the generation of isocurvature.  A class of models discussed in Refs.~\cite{Bucher:1996gg,Linde:1996gt,Peebles:1997av} (see also Ref.~\cite{Komatsu:2001ysk}) is relevant here.

The main idea is that, for example, CDM isocurvature fluctuations are sourced by a massive scalar field with zero mean. This leads to a CDM energy density 
\begin{align}\label{eq:rhomvarphi2}
\rho_m(x)= {\cal A}\,\varphi^2(x)\,,
\end{align}
where $\varphi(x)$ is generated during inflation and follows a Gaussian distribution. Thus, $\rho_m$ follows a Gaussian squared distribution and is of course always positive definite. With our definition of isocurvature, we have that
\begin{align}\label{eq:Svaprhi2}
S=\frac{\varphi(x)^2-\langle\varphi^2\rangle}{\langle\varphi^2\rangle}\,,
\end{align}
and by definition $S>-1$. Note that equations \eqref{eq:rhomvarphi2} and \eqref{eq:Svaprhi2} look like the functional form for a perturbation with local non-Gaussianity, the induced GW signal of which has been studied in, e.g., Refs.~\cite{Garcia-Bellido:2017aan,Cai:2018dig,Unal:2018yaa,Atal:2021jyo,Adshead:2021hnm}. Note, however, that the current situation is different, as in Refs.~\cite{Cai:2018dig,Unal:2018yaa,Adshead:2021hnm} the size of the non-Gaussian piece is determined by the parameter $f_{\rm NL}$, which is absent in \eqref{eq:Svaprhi2}. Nevertheless, we consider the results of \cite{Cai:2018dig,Unal:2018yaa,Atal:2021jyo,Adshead:2021hnm} as an indication that the shape of the GW spectrum due to a non-Gaussian distribution is crudely speaking similar to the Gaussian case if the primordial scalar spectrum is broad. We use this when estimating the constraints in \S\ref{sec:constraints} when accounting for non-Gaussianity.

The problem with the present example is that $S$ is not large enough to induce a significant GW signal. Indeed, in real space one finds that since $\langle\rho_m(x)\rangle=\rho_{m}$ then
\begin{align}
\langle\rho_m^2\rangle\sim \langle\rho_m\rangle^2\,.
\end{align}
This means that typical fluctuations have a dispersion $\langle S^2\rangle\sim \mathcal{O}(1)$ which is not enough for our purposes. Thus, we turn to another example.

\subsection{\texorpdfstring{$\chi^2$}{chi}-distribution}
\label{subapp:chi2}

Consider that $\rho_m(x)$ is drawn from a $\chi_n^2$ distribution, so that $\rho_m(x)>0$. The probability density of the $\chi_n^2$ is given 
\begin{align}\label{eq:chi2}
P_{\chi^2}[\rho_m]=\frac{1}{2\Gamma[n/2]\sigma} \left(\frac{\rho_m}{\sigma}\right)^{n/2-1}e^{-\rho_m/(2\sigma)}\,,
\end{align}
where $\sigma>0$ and $n$ is often taken to be a positive integer which indicates the order of distribution.
We shall instead consider the generalization $n\in \mathbb{R}^+$ which from a
probability distribution point of view is well defined. With Eq.~\eqref{eq:chi2} we can study the moments of the distribution in real space. The first moment is
\begin{align}
\rho_{m,0} \equiv \langle \rho_m\rangle = \int_0^\infty d\rho_m \,\rho_m P_{\chi^2}[\rho_m]= n\sigma\,,
\end{align}
so from now on we replace $\sigma$ with $\rho_{m,0}/n$. We also find that
\begin{align}
\langle \rho_m^2\rangle = \frac{2+n}{n}\rho_{m,0}^2 \,,
\end{align}
\begin{align}
\langle \rho_m^3\rangle = \frac{(2+n)(4+n)}{n^2}\rho_{m,0}^3  \,,
\end{align}
and
\begin{align}
\langle \rho_m^4\rangle = \frac{(2+n)(4+n)(6+n)}{n^3}\rho_{m,0}^4\,.
\end{align}
from which we conclude that the isocurvature perturbation has moments
\begin{align}\label{eq:S2S4}
\langle S^2\rangle= \frac{2}{n}\,
\quad {\rm and}\quad 
\langle S^4\rangle= \frac{12(4+n)}{n^3}=6\langle S^2\rangle^3+3\langle S^2\rangle^2\,.
\end{align}
We see that for $n\ll1$ we have a large dispersion, i.e. $\langle S^2\rangle\gg1$. We will refer to this distribution as $\chi^2$-like or $\chi^2_n$ since with non-integer $n$ it does not correspond to a sum of squares of Gaussian variables. Let us compare our results with the Gaussian case and let us call $S_g$ when $S$ is drawn from a Gaussian distribution. In this case we have that
\begin{align}\label{eq:S4Gaussian}
\langle S_g^4\rangle= 3\langle S_g^2\rangle^2\,.
\end{align}

We therefore conclude that in the $\chi_n^2$ example with $n\ll1$ we have from Eq.~\eqref{eq:S2S4} that the fourth moment of the distribution is much larger than the Gaussian case. We also extract that the amplitude of $\langle S^4\rangle$ is proportional to $\langle S^2\rangle^3$. This will be important in the following discussion in Fourier space.

\subsection{Amplitude of the four point correlator}

In the calculations of induced GWs, one encounters the four point correlator of the isocurvature perturbation. Roughly, the induced GW spectrum is
\begin{align}
\langle h_k h_{k'}\rangle \sim \int\frac{d^3q}{(2\pi)^3}\frac{d^3q'}{(2\pi)^3}\langle S_{\mathbf{q}}S_{\mathbf{k}-\mathbf{q}}S_{\mathbf{q}'}S_{\mathbf{k}'-\mathbf{q}'}\rangle\,,
\end{align}
where $h_k$ is the induced tensor modes and we neglected the Green and transfer functions for simplicity. \footnote{Note that although $S$ may be highly non-Gaussian, induced GWs still depend primarily on the 4-point function since they are generated during the regime where the perturbative expansion is valid. Next order terms would lead to a 5-point function of $\Phi$ which would contain a further $\kappa^{-1}$ suppression relative to the 4-point function of $\Phi$. This is in contrast to PBH formation, which being a fully non-linear process may depend on higher order statistics if fluctuations are highly non-Gaussian.}
In general, one can decompose the four point correlator into a connected and a disconnected part \cite{Yuan:2021qgz}
\begin{align}\label{eq:phi4}
\langle S_{\mathbf{q}}S_{\mathbf{k}-\mathbf{q}}S_{\mathbf{q}'}S_{\mathbf{k}'-\mathbf{q}'}\rangle=&\langle S_{\mathbf{q}}S_{\mathbf{k}-\mathbf{q}}S_{\mathbf{q}'}S_{\mathbf{k}'-\mathbf{q}'}\rangle_c +\langle S_{\mathbf{q}}S_{\mathbf{k}-\mathbf{q}}\rangle\langle S_{\mathbf{q}'}S_{\mathbf{k}'-\mathbf{q}'}\rangle\nonumber\\&
+\langle S_{\mathbf{q}}S_{\mathbf{q}'}\rangle\langle S_{\mathbf{k}-\mathbf{q}}S_{\mathbf{k}'-\mathbf{q}'}\rangle+\langle S_{\mathbf{q}}S_{\mathbf{k}'-\mathbf{q}'}\rangle\langle S_{\mathbf{q}'}S_{\mathbf{k}-\mathbf{q}}\rangle\,,
\end{align}
where the connected part can be expressed in terms of the trispectrum ${\cal T}_S$,
\begin{align}
\langle S_{\mathbf{q}}S_{\mathbf{k}-\mathbf{q}}S_{\mathbf{q}'}S_{\mathbf{k}'-\mathbf{q}'}\rangle_c = (2\pi)^3 \delta^3(\mathbf{k}+\mathbf{k'}){\cal T}_S\left(\mathbf{q},\mathbf{q}',\mathbf{k},\mathbf{k}'\right)\,.
\end{align}
Due to symmetries, i.e., homogeneity and isotropy of the background, a generic connected four point function depends on six variables, that one can take to be the norms of the four external momenta and the norms of any two sums of them.
To compute it, however, we usually need a probability distribution for the density which is related to a Gaussian in order to be able to easily go to Fourier space. We have seen that the Gaussian squared case is not suitable for our purposes, and the working example of the $\chi^2_n$ with $n\ll1$ is not writable in terms of a Gaussian. 
This is out of the scope of the present paper and deserves a separate study.

Instead, since we expect from Appendix \ref{subapp:nomean} and Appendix \ref{subapp:chi2} that the Gaussian calculation Eq.~\eqref{eq:Phgaussian} is an underestimate, and since we are mainly interested in obtaining an order of magnitude estimate of the non-Gaussian enhancement, we proceed with an analogy.

Looking at equations \eqref{eq:S2S4}, \eqref{eq:S4Gaussian} and \eqref{eq:phi4} we see that in the $\chi^2$ case the connected four point correlator should dominate over the disconnected one. We also know that $\langle S^4\rangle\sim 6\langle S^2\rangle^3$ and we use this result in real space to estimate the amplitude in Fourier space. With a two-point correlator of the form
\begin{align}
\langle S_{\mathbf{q}}S_{\mathbf{q}'}\rangle=\frac{2\pi^2}{k^3}{\cal P}_S(q)(2\pi)^3\delta^3(\mathbf{q}+\mathbf{q}')\,,
\end{align}
we estimate that the four-point correlator is
\begin{align}\label{eq:S4estimate}
\langle S_{\mathbf{q}}S_{\mathbf{k}-\mathbf{q}}S_{\mathbf{q}'}S_{\mathbf{k}'-\mathbf{q}'}\rangle&\sim \frac{2\pi^2}{q^3}\frac{2\pi^2}{|\mathbf{k}-\mathbf{q}|^3}{\cal P}_S(q){\cal P}_S(\mathbf{k}-\mathbf{q})\left({\cal P}_S(q)+{\cal P}_S(\mathbf{k}-\mathbf{q})\right)\nonumber\\&\times (2\pi)^6\delta^3\left(\mathbf{q}+\mathbf{q}'\right)\delta^3\left(\mathbf{k}+\mathbf{k}'-\mathbf{q}-\mathbf{q}'\right)+\left\{\mathbf{q}\leftrightarrow \mathbf{k}-\mathbf{q}\right\}\,,
\end{align}
where we used all possible combinations compatible with the dimensions and with Eq.~\eqref{eq:S2S4} that could result in two delta functions and that is symmetric with respect to $\mathbf{q}\leftrightarrow\mathbf{k}-\mathbf{q}$. Of course this is a very crude estimate and in general the connected four point correlator is not proportional to two delta functions, but in the absence of a concrete model and to avoid unnecessary complications in the integrals we expect that \eqref{eq:S4estimate} provides a rough order of magnitude estimate while keeping the simplicity of the Gaussian results.

The last step is to find a shape of ${\cal P}_S(q)$ that is compatible with the $\chi^2_n$ distribution. This means that we need a power spectrum compatible with the positivity requirement $\rho_m>0$, and in particular with  $\langle\rho_m(x)\rho_m(y)\rangle>0$. For this purpose, we consider that the CDM energy density dimensionless power spectrum is given by
\begin{align}\label{eq:Pm}
{\cal P}_m(k)=3{\cal A}\rho_{m,0}^2\sqrt{\frac{6}{{\pi}}}(k R_0)^3e^{-\frac{3}{2}k^2R_0^2}\,,
\end{align}
where $R_0$ is a given scale. Then, the 2-point correlation function reads
\begin{align}
\langle \rho_m(x)\rho_m(y)\rangle = \int \frac{dk}{k} {\cal P}_m(k) \frac{\sin(kr)}{kr}={\cal A}\rho_{m,0}^2e^{-\frac{r^2}{6R_0^2}}\,,
\end{align}
where $r=|x-y|$.
It then follows that $\langle \rho_m(x)\rho_m(y)\rangle>0$ and in particular we have that
\begin{align}
\langle \rho_m^2(x)\rangle = {\cal A}\rho_{m,0}^2\,.
\end{align}
With this example we obtain that
\begin{align}
\langle S^2(x)\rangle=\frac{\langle (\rho_m(x)-\rho_{m,0})^2\rangle}{\rho_{m,0}^2}={\cal A}-1\,,
\end{align}
where we used that $\langle\rho_m(x)\rangle=\rho_{m,0}$. For ${\cal A}\gg1$ we have a large positive dispersion. Associating this spatial average with the ensemble average of Eq.~\eqref{eq:S2S4} yields that ${\cal A}=1+{2}/{n}$. We shall take as an example for ${\cal P}_S(k)$ the following shape
\begin{align}\label{eq:psapp}
{\cal P}_S(k)=3{\cal A}_S\sqrt{\frac{6}{{\pi}}}\frac{k^3}{k_p^3}e^{-\frac{3}{2}\frac{k^2}{k_p^2}}\,,
\end{align}
which peaks at $k=k_p$. The resulting induced GW spectrum may then be given by
\begin{align}\label{eq:nongaussian}
		\Omega^{\chi_n^2}_{\rm GW,c}\sim \frac{2}{3}\int_0^\infty dv\int_{|1-v|}^{1+v}du&\left(\frac{4v^2-(1-u^2+v^2)^2}{4uv}\right)^2\overline{I^2(x_c,k,u,v)}\nonumber\\&\times{{\cal P}_{S}(ku)}{{\cal P}_{S}(kv)}\left({{\cal P}_{S}(kv)}+{{\cal P}_{S}(ku)}\right)\,.
	\end{align}
However, since we do not know the explicit shape of the four point correlator we simply rescale the amplitude of the induced GW spectral density in the main text by
\begin{align}\label{eq:amplitudeapp}
\Omega^{\chi_n^2}_{\rm GW,c}\approx {\cal A}_S\Omega^{g}_{\rm GW,c}
\end{align}
where the subscript $g$ refers to the Gaussian calculation. Both \eqref{eq:nongaussian} and \eqref{eq:amplitudeapp} are consistent within an $O(1)$ factor for ${\cal P}_S(k)$ given by \eqref{eq:psapp}.

\section{Explicit expressions omitted in main text \label{app:explicit}}

In this appendix we provide some explicit expressions which are too long to be included in the main text. First, the source term \eqref{eq:f} at leading order in $\kappa^{-1}$ after using \eqref{eq:analyticalsol} and \eqref{eq:vrelsol} is given by
\begin{align}\label{eq:fapp}
\kappa^2f(x,u,v)\approx&\frac{9}{8 u^2 v^2 x^2}+\frac{27 \left(u^2+v^2\right)}{4 u^4 v^4 x^4}+\frac{243}{2 u^4 v^4 x^6}\nonumber\\&
   %%%%%%%%%%%
   +\frac{27}{2}\left(\frac{\left(2 u^2-v^2\right)}{2u^4 v^4
   x^4}-\frac{9}{u^4 v^4 x^6}\right) \cos \left({uc_s
   x}\right)-\frac{27}{2}\left(\frac{\left(u^2-2 v^2\right)}{2 u^4 v^4
   x^4}+\frac{9}{u^4 v^4 x^6}\right) \cos \left({v
   c_sx}\right)\nonumber\\&
   %%%%%%%%%%%
   -\frac{9 \sqrt{3}}{2}\left(\frac{1}{2 u^2
   v^3 x^3}+\frac{9}{u^4 v^3 x^5}\right) \sin \left({vc_s x}\right)-\frac{9 \sqrt{3}}{2}\left(\frac{1}{2 u^3 v^2 x^3}+\frac{9
   }{u^3 v^4 x^5}\right) \sin \left(uc_s
   x\right)\nonumber\\&
   %%%%%%%%%%%
	+\frac{9 \sqrt{3} (u-v)}{4}\left(\frac{1}{u^3 v^3 x^3}+\frac{9}{u^4 v^4 x^5}\right) \sin \left({
   (u-v)}c_sx\right)\nonumber\\&
   %%%%%%%%%%%
   -\frac{9 \sqrt{3}(u+v)}{4}\left(\frac{1}{ u^3 v^3 x^3}-\frac{9}{u^4 v^4 x^5}\right) \sin \left({
   (u+v)}c_sx\right)\nonumber\\&
   %%%%%%%%%%%
   %%%%%%%%%%%
   +\frac{9}{4}\left(\frac{1}{2 u^2 v^2 x^2}-\frac{3 \left(u^2-3 u v+v^2\right)}{ u^4 v^4
   x^4}+\frac{27}{ u^4 v^4 x^6}\right) \cos \left({
   (u-v)}c_sx\right)\nonumber\\&
   %%%%%%%%%%%
   %%%%%%%%%%%
   +\frac{9}{4}\left(\frac{1}{2 u^2 v^2 x^2}-\frac{3 \left(u^2+3 u v+v^2\right)}{ u^4 v^4
   x^4}+\frac{27}{ u^4 v^4 x^6}\right) \cos \left({
   (u+v)}c_sx\right)\,,
\end{align}
which is to be contrasted with the source term for initial adiabatic fluctuations \cite{Domenech:2021ztg}
\begin{align}\label{eq:fsimple}
f_{\rm ad}(x,u,v)=&\frac{3}{2}\left(j_{0}(c_svx)j_0(c_sux)+2j_2(c_svx)j_2(c_sux)\right)\,,
\end{align}
where $j_0(x)$ and $j_2(x)$ are the spherical Bessel functions of the first kind of order $0$ and $2$. In this formula we neglected the effect of the relative velocity.

Second, to find the general kernels $I_{c, s}$ provided in Eqs.~\eqref{eq:Icinf} and \eqref{eq:Isinf}, the source function \eqref{eq:fapp} has to be further multiplied by $x$ and either $\cos x$ or $\sin x$ as shown in Eq.~\eqref{eq:Ics}. The necessary integrals are given in Appendix~\ref{app:usefulformulas}. 

With the general kernels, we were able to compute the induced gravitational wave spectrum $\Omega_{\rm GW,c}$ for a $\delta$-function peak in the primordial isocurvature spectrum fully analytically, Eq.~\eqref{eq:omegac}. The corresponding result for a $\delta$-function peak in the adiabatic spectrum is \cite{Kohri:2018awv,Domenech:2021ztg}
\begin{align}\label{eq:omegacad}
\Omega_{\rm GW,c}^{\rm ad}&=3{\cal A}_{\cal R}^2\left(\frac{k}{k_p}\right)^2\left(1-\frac{k^2}{4k_p^2}\right)^2\left(1-\frac{3k^2}{2k_p^2}\right)^2\Theta(k-2k_p)\nonumber\\&
\times\Bigg\{\frac{\pi^2}{4}\left(1-\frac{3k^2}{2k_p^2}\right)^2\Theta\left(\frac{2k_p}{\sqrt{3}k}-1\right)
+\left(1+\frac{1}{2}\left(1-\frac{3k^2}{2k_p^2}\right)^2\ln\left|1-\frac{4k_p^2}{3k^2}\right|\right)^2\Bigg\}\,.
\end{align}
One can check that after smearing the resonance using the same methodology we used in \S\ref{subsec:lognormal}, Eq.~\eqref{eq:narrow-averaging}, the adiabatic peak has amplitude
\begin{align}\label{eq:peakad}
\Omega^{\rm peak, ad}_{\rm GW,c}\approx \frac{4}{9}{\cal A}_{\cal R}^2 \left(1 + \frac{\pi^2}{2}+ \left(1 + \frac{1}{2}\ln\left(3\Delta^2\right)\right)^2\right)\,.
\end{align}
One can also check that the infrared tail, i.e. for $k\ll k_p$, goes as
\begin{align}\label{eq:irad}
\Omega_{\rm GW,c}^{\rm ad} (k\ll k_p)\approx 3{\cal A}_{\cal R}^2\frac{k^2}{k_p^2}\ln^2\frac{k}{k_p}\,.
\end{align}

\section{Useful identities and integrals \label{app:usefulformulas}}
In this appendix we present some useful formulas that have been used to derive the results in the main text.
To compute the transfer function of induced GWs we made use of the cosine and sine integrals which are respectively defined as
\begin{align}
{\rm Ci}(z)=-\int_z^\infty\frac{dx}{x}\cos x\quad{\rm and}\quad{\rm Si}(z)=\int_0^\infty\frac{dx}{x}\sin x\,.
\end{align}
Their respective asymptotes at infinity are ${\rm Ci}(z\to\infty)\to 0$ and ${\rm Si}(z\to\pm \infty)\to \pm\pi/2$. For vanishing argument one has ${\rm Ci}(z\to0)\to \gamma_E+\ln z$ and ${\rm Si}(z\to0)\to 0$, where $\gamma_E\approx0.577$ is Euler's constant. We also use that
\begin{align}
\lim_{x\to \infty}\left\{{\rm Si}\left[(1-\beta)x\right]+{\rm Si}\left[(1+\beta)x\right]\right\}=\pi\Theta(1-|\beta|)\,,
\end{align}
where $\Theta(x)$ is the Heaviside step function.

The integrals in the main text can all be written as an integral over a sine or cosine times a negative integer power of the argument. The integrals we need are:
\begin{align}
\int dx \frac{\cos \beta x}{x}={\rm Ci}[|\beta| x]\,,
\end{align}
\begin{align}
\int dx \frac{\sin \beta x}{x}={\rm Si}[\beta x]\,,
\end{align}
\begin{align}
\int dx \frac{\cos \beta x}{x^2}=-\frac{\cos\beta x}{x}-\beta{\rm Si}[\beta x]\,,
\end{align}
\begin{align}
\int dx \frac{\sin \beta x}{x^2}=-\frac{\sin\beta x}{x}+\beta{\rm Ci}[|\beta| x]\,,
\end{align}
\begin{align}
\int dx \frac{\cos \beta x}{x^3}=\frac{\beta}{2}\frac{\sin\beta x}{x}-\frac{\cos\beta x}{2x^2}-\frac{\beta^2}{2}{\rm Ci}[|\beta| x]\,,
\end{align}
\begin{align}
\int dx \frac{\sin \beta x}{x^3}=-\frac{\beta}{2}\frac{\cos\beta x}{x}-\frac{\sin\beta x}{2x^2}-\frac{\beta^2}{2}{\rm Si}[\beta x]\,,
\end{align}
\begin{align}
\int dx \frac{\cos \beta x}{x^4}=\frac{\beta^2}{6}\frac{\cos\beta x}{x}+\frac{\beta}{6}\frac{\sin\beta x}{x^2}-\frac{\cos\beta x}{3x^3}+\frac{\beta^3}{6}{\rm Si}[\beta x]\,,
\end{align}
\begin{align}
\int dx \frac{\sin \beta x}{x^4}=\frac{\beta^2}{6}\frac{\sin\beta x}{x}-\frac{\beta}{6}\frac{\cos\beta x}{x^2}-\frac{\sin\beta x}{3x^3}-\frac{\beta^3}{6}{\rm Ci}[|\beta| x]\,,
\end{align}
\begin{align}
\int dx \frac{\cos \beta x}{x^5}=-\frac{\beta^3}{24}\frac{\sin\beta x}{x}+\frac{\beta^2}{24}\frac{\cos\beta x}{x^2}+\frac{\beta}{12}\frac{\sin\beta x}{x^3}-\frac{\cos\beta x}{4x^4}+\frac{\beta^4}{24}{\rm Ci}[|\beta| x]\,,
\end{align}
\begin{align}
\int dx \frac{\sin \beta x}{x^5}=\frac{\beta^3}{24}\frac{\cos\beta x}{x}+\frac{\beta^2}{24}\frac{\sin\beta x}{x^2}-\frac{\beta}{12}\frac{\cos\beta x}{x^3}-\frac{\sin\beta x}{4x^4}+\frac{\beta^4}{24}{\rm Si}[\beta x]\,.
\end{align}

\section{Other constraints used in this work}
\label{app:other}

In this appendix we explain how we derived the constraints in Figs.~\ref{fig:constraints} and \ref{fig:constraints_bump} for $10^{-3}<k<10^{6}\,{\rm Mpc}^{-1}$. The frequency of the induced GWs for such wavenumbers is too small to be detected by interferometers. However, in this regime, we may directly constrain the amplitude of isocurvature modes using the CMB.

\subsection{CMB anisotropies}

It is known that the main effect of isocurvature modes is to shift the position of the acoustic peaks in CMB anisotropies \cite{Hu:1995em,Langlois:2003fq}. As the position of the peaks is very well measured, isocurvature may only account for less than $1\%$. We show Planck's constraints on isocurvature for an arbitrary spectrum \cite{Akrami:2018odb}, which are roughly given by $\beta_{\rm iso}=\left\{2.5,26,47\right\}\times 10^{-2}$ respectively for $k=\left\{0.002,0.05,0.1\right\}\,{\rm Mpc}^{-1}$ where
\begin{align}
\beta_{\rm iso}=\frac{{\cal P}_S(k)}{{\cal P}_{\cal R}(k)+{\cal P}_S(k)}\,.
\end{align}
These values can be found in Table 14, row 14 of general CDM isocurvature models in \cite{Akrami:2018odb}. Note that we use these values not to constrain our model, which aims to produce GWs from small scale isocurvature fluctuations, but to provide a qualitative comparison of the global constraints on the amplitude of ${\cal P}_S$ in a wide range of scales.

\subsection{CMB \texorpdfstring{$\mu$}{mu}-distortions}

Damping of small-scale fluctuations causes energy release in the early universe and creates CMB spectral distortions. In the case of small-scale CDM isocurvature fluctuations, it is the induced photon density fluctuations which produced the $\mu$-distortions \cite{Hu:1994jd,Dent:2012ne,Chluba:2013dna}. We show current and future constraints on CDM isocurvature due to CMB $\mu$-distortions, following the work of Chluba and Grin \cite{Chluba:2013dna}. The $\mu$-distortions due to CDM isocurvature are given by
\begin{align}
\mu=\int_{k_{\rm min}}^\infty \frac{dk}{k}{\cal P}_S(k)W(k)
\end{align}
where $k_{\rm min}\approx 1\,{\rm Mpc}^{-1}$ and  
\begin{align}
W(k)\approx 5.3\times 10^{-5}\left(\frac{k_{\rm min}}{k}\right)^2\left[\exp\left(-\frac{\left(\frac{k/k_{\rm min}}{1360}\right)^2}{1+\left(\frac{k/k_{\rm min}}{260}\right)^{0.3}+\frac{k/k_{\rm min}}{340}}\right)-\exp\left(-\left(\frac{k/k_{\rm min}}{32}\right)^2\right)\right]\,.
\end{align}
In the main text we consider the COBE experiment ($\mu=10^{-5}$) and a future PIXIE-like experiment ($\mu=10^{-9}$). 

\bibliography{bibliographyIGWs.bib} 

\begin{thebibliography}{132}%
\makeatletter
\providecommand \@ifxundefined [1]{%
 \@ifx{#1\undefined}
}%
\providecommand \@ifnum [1]{%
 \ifnum #1\expandafter \@firstoftwo
 \else \expandafter \@secondoftwo
 \fi
}%
\providecommand \@ifx [1]{%
 \ifx #1\expandafter \@firstoftwo
 \else \expandafter \@secondoftwo
 \fi
}%
\providecommand \natexlab [1]{#1}%
\providecommand \enquote  [1]{``#1''}%
\providecommand \bibnamefont  [1]{#1}%
\providecommand \bibfnamefont [1]{#1}%
\providecommand \citenamefont [1]{#1}%
\providecommand \href@noop [0]{\@secondoftwo}%
\providecommand \href [0]{\begingroup \@sanitize@url \@href}%
\providecommand \@href[1]{\@@startlink{#1}\@@href}%
\providecommand \@@href[1]{\endgroup#1\@@endlink}%
\providecommand \@sanitize@url [0]{\catcode `\\12\catcode `\$12\catcode
  `\&12\catcode `\#12\catcode `\^12\catcode `\_12\catcode `\%12\relax}%
\providecommand \@@startlink[1]{}%
\providecommand \@@endlink[0]{}%
\providecommand \url  [0]{\begingroup\@sanitize@url \@url }%
\providecommand \@url [1]{\endgroup\@href {#1}{\urlprefix }}%
\providecommand \urlprefix  [0]{URL }%
\providecommand \Eprint [0]{\href }%
\providecommand \doibase [0]{https://doi.org/}%
\providecommand \selectlanguage [0]{\@gobble}%
\providecommand \bibinfo  [0]{\@secondoftwo}%
\providecommand \bibfield  [0]{\@secondoftwo}%
\providecommand \translation [1]{[#1]}%
\providecommand \BibitemOpen [0]{}%
\providecommand \bibitemStop [0]{}%
\providecommand \bibitemNoStop [0]{.\EOS\space}%
\providecommand \EOS [0]{\spacefactor3000\relax}%
\providecommand \BibitemShut  [1]{\csname bibitem#1\endcsname}%
\let\auto@bib@innerbib\@empty
%</preamble>
\bibitem [{\citenamefont {Bennett}\ \emph {et~al.}(2003)\citenamefont {Bennett}
  \emph {et~al.}}]{WMAP:2003ivt}%
  \BibitemOpen
  \bibfield  {author} {\bibinfo {author} {\bibfnamefont {C.~L.}\ \bibnamefont
  {Bennett}} \emph {et~al.} (\bibinfo {collaboration} {WMAP}),\ }\bibfield
  {title} {\bibinfo {title} {{First year Wilkinson Microwave Anisotropy Probe
  (WMAP) observations: Preliminary maps and basic results}},\ }\href
  {https://doi.org/10.1086/377253} {\bibfield  {journal} {\bibinfo  {journal}
  {Astrophys. J. Suppl.}\ }\textbf {\bibinfo {volume} {148}},\ \bibinfo {pages}
  {1} (\bibinfo {year} {2003})},\ \Eprint
  {https://arxiv.org/abs/astro-ph/0302207} {arXiv:astro-ph/0302207}
  \BibitemShut {NoStop}%
\bibitem [{\citenamefont {Akrami}\ \emph {et~al.}(2020)\citenamefont {Akrami}
  \emph {et~al.}}]{Akrami:2018odb}%
  \BibitemOpen
  \bibfield  {author} {\bibinfo {author} {\bibfnamefont {Y.}~\bibnamefont
  {Akrami}} \emph {et~al.} (\bibinfo {collaboration} {Planck}),\ }\bibfield
  {title} {\bibinfo {title} {{Planck 2018 results. X. Constraints on
  inflation}},\ }\href {https://doi.org/10.1051/0004-6361/201833887} {\bibfield
   {journal} {\bibinfo  {journal} {Astron. Astrophys.}\ }\textbf {\bibinfo
  {volume} {641}},\ \bibinfo {pages} {A10} (\bibinfo {year} {2020})},\ \Eprint
  {https://arxiv.org/abs/1807.06211} {arXiv:1807.06211 [astro-ph.CO]}
  \BibitemShut {NoStop}%
\bibitem [{\citenamefont {Langlois}(2003)}]{Langlois:2003fq}%
  \BibitemOpen
  \bibfield  {author} {\bibinfo {author} {\bibfnamefont {D.}~\bibnamefont
  {Langlois}},\ }\bibfield  {title} {\bibinfo {title} {{Isocurvature
  cosmological perturbations and the CMB}},\ }\href
  {https://doi.org/10.1016/j.crhy.2003.09.004} {\bibfield  {journal} {\bibinfo
  {journal} {Comptes Rendus Physique}\ }\textbf {\bibinfo {volume} {4}},\
  \bibinfo {pages} {953} (\bibinfo {year} {2003})}\BibitemShut {NoStop}%
\bibitem [{\citenamefont {Kodama}\ and\ \citenamefont
  {Sasaki}(1984)}]{Kodama:1985bj}%
  \BibitemOpen
  \bibfield  {author} {\bibinfo {author} {\bibfnamefont {H.}~\bibnamefont
  {Kodama}}\ and\ \bibinfo {author} {\bibfnamefont {M.}~\bibnamefont
  {Sasaki}},\ }\bibfield  {title} {\bibinfo {title} {{Cosmological Perturbation
  Theory}},\ }\href {https://doi.org/10.1143/PTPS.78.1} {\bibfield  {journal}
  {\bibinfo  {journal} {Prog. Theor. Phys. Suppl.}\ }\textbf {\bibinfo {volume}
  {78}},\ \bibinfo {pages} {1} (\bibinfo {year} {1984})}\BibitemShut {NoStop}%
\bibitem [{\citenamefont {Bucher}\ \emph {et~al.}(2000)\citenamefont {Bucher},
  \citenamefont {Moodley},\ and\ \citenamefont {Turok}}]{Bucher:1999re}%
  \BibitemOpen
  \bibfield  {author} {\bibinfo {author} {\bibfnamefont {M.}~\bibnamefont
  {Bucher}}, \bibinfo {author} {\bibfnamefont {K.}~\bibnamefont {Moodley}},\
  and\ \bibinfo {author} {\bibfnamefont {N.}~\bibnamefont {Turok}},\ }\bibfield
   {title} {\bibinfo {title} {{The General primordial cosmic perturbation}},\
  }\href {https://doi.org/10.1103/PhysRevD.62.083508} {\bibfield  {journal}
  {\bibinfo  {journal} {Phys. Rev. D}\ }\textbf {\bibinfo {volume} {62}},\
  \bibinfo {pages} {083508} (\bibinfo {year} {2000})},\ \Eprint
  {https://arxiv.org/abs/astro-ph/9904231} {arXiv:astro-ph/9904231}
  \BibitemShut {NoStop}%
\bibitem [{\citenamefont {Chluba}\ and\ \citenamefont
  {Grin}(2013)}]{Chluba:2013dna}%
  \BibitemOpen
  \bibfield  {author} {\bibinfo {author} {\bibfnamefont {J.}~\bibnamefont
  {Chluba}}\ and\ \bibinfo {author} {\bibfnamefont {D.}~\bibnamefont {Grin}},\
  }\bibfield  {title} {\bibinfo {title} {{CMB spectral distortions from
  small-scale isocurvature fluctuations}},\ }\href
  {https://doi.org/10.1093/mnras/stt1129} {\bibfield  {journal} {\bibinfo
  {journal} {Mon. Not. Roy. Astron. Soc.}\ }\textbf {\bibinfo {volume} {434}},\
  \bibinfo {pages} {1619} (\bibinfo {year} {2013})},\ \Eprint
  {https://arxiv.org/abs/1304.4596} {arXiv:1304.4596 [astro-ph.CO]}
  \BibitemShut {NoStop}%
\bibitem [{\citenamefont {Chluba}\ \emph {et~al.}(2019)\citenamefont {Chluba}
  \emph {et~al.}}]{Chluba:2019kpb}%
  \BibitemOpen
  \bibfield  {author} {\bibinfo {author} {\bibfnamefont {J.}~\bibnamefont
  {Chluba}} \emph {et~al.},\ }\bibfield  {title} {\bibinfo {title} {{Spectral
  Distortions of the CMB as a Probe of Inflation, Recombination, Structure
  Formation and Particle Physics}: {Astro2020 Science White Paper}},\
  }\href@noop {} {\bibfield  {journal} {\bibinfo  {journal} {Bull. Am. Astron.
  Soc.}\ }\textbf {\bibinfo {volume} {51}},\ \bibinfo {pages} {184} (\bibinfo
  {year} {2019})},\ \Eprint {https://arxiv.org/abs/1903.04218}
  {arXiv:1903.04218 [astro-ph.CO]} \BibitemShut {NoStop}%
\bibitem [{\citenamefont {Inomata}\ \emph {et~al.}(2018)\citenamefont
  {Inomata}, \citenamefont {Kawasaki}, \citenamefont {Kusenko},\ and\
  \citenamefont {Yang}}]{Inomata:2018htm}%
  \BibitemOpen
  \bibfield  {author} {\bibinfo {author} {\bibfnamefont {K.}~\bibnamefont
  {Inomata}}, \bibinfo {author} {\bibfnamefont {M.}~\bibnamefont {Kawasaki}},
  \bibinfo {author} {\bibfnamefont {A.}~\bibnamefont {Kusenko}},\ and\ \bibinfo
  {author} {\bibfnamefont {L.}~\bibnamefont {Yang}},\ }\bibfield  {title}
  {\bibinfo {title} {{Big Bang Nucleosynthesis Constraint on Baryonic
  Isocurvature Perturbations}},\ }\href
  {https://doi.org/10.1088/1475-7516/2018/12/003} {\bibfield  {journal}
  {\bibinfo  {journal} {JCAP}\ }\textbf {\bibinfo {volume} {12}},\ \bibinfo
  {pages} {003}},\ \Eprint {https://arxiv.org/abs/1806.00123} {arXiv:1806.00123
  [astro-ph.CO]} \BibitemShut {NoStop}%
\bibitem [{\citenamefont {Kohri}\ \emph {et~al.}(2014)\citenamefont {Kohri},
  \citenamefont {Nakama},\ and\ \citenamefont {Suyama}}]{Kohri:2014lza}%
  \BibitemOpen
  \bibfield  {author} {\bibinfo {author} {\bibfnamefont {K.}~\bibnamefont
  {Kohri}}, \bibinfo {author} {\bibfnamefont {T.}~\bibnamefont {Nakama}},\ and\
  \bibinfo {author} {\bibfnamefont {T.}~\bibnamefont {Suyama}},\ }\bibfield
  {title} {\bibinfo {title} {{Testing scenarios of primordial black holes being
  the seeds of supermassive black holes by ultracompact minihalos and CMB
  $\mu$-distortions}},\ }\href {https://doi.org/10.1103/PhysRevD.90.083514}
  {\bibfield  {journal} {\bibinfo  {journal} {Phys. Rev. D}\ }\textbf {\bibinfo
  {volume} {90}},\ \bibinfo {pages} {083514} (\bibinfo {year} {2014})},\
  \Eprint {https://arxiv.org/abs/1405.5999} {arXiv:1405.5999 [astro-ph.CO]}
  \BibitemShut {NoStop}%
\bibitem [{\citenamefont {Yang}(2014)}]{Yang:2014lsg}%
  \BibitemOpen
  \bibfield  {author} {\bibinfo {author} {\bibfnamefont {Y.}~\bibnamefont
  {Yang}},\ }\bibfield  {title} {\bibinfo {title} {{Constraints on the
  primordial power spectrum of small scales using the neutrino signals from the
  dark matter decay}},\ }\href {https://doi.org/10.1142/S0217751X14501942}
  {\bibfield  {journal} {\bibinfo  {journal} {Int. J. Mod. Phys. A}\ }\textbf
  {\bibinfo {volume} {29}},\ \bibinfo {pages} {1450194} (\bibinfo {year}
  {2014})},\ \Eprint {https://arxiv.org/abs/1501.00789} {arXiv:1501.00789
  [astro-ph.CO]} \BibitemShut {NoStop}%
\bibitem [{\citenamefont {Nakama}\ \emph {et~al.}(2018)\citenamefont {Nakama},
  \citenamefont {Suyama}, \citenamefont {Kohri},\ and\ \citenamefont
  {Hiroshima}}]{Nakama:2017qac}%
  \BibitemOpen
  \bibfield  {author} {\bibinfo {author} {\bibfnamefont {T.}~\bibnamefont
  {Nakama}}, \bibinfo {author} {\bibfnamefont {T.}~\bibnamefont {Suyama}},
  \bibinfo {author} {\bibfnamefont {K.}~\bibnamefont {Kohri}},\ and\ \bibinfo
  {author} {\bibfnamefont {N.}~\bibnamefont {Hiroshima}},\ }\bibfield  {title}
  {\bibinfo {title} {{Constraints on small-scale primordial power by
  annihilation signals from extragalactic dark matter minihalos}},\ }\href
  {https://doi.org/10.1103/PhysRevD.97.023539} {\bibfield  {journal} {\bibinfo
  {journal} {Phys. Rev. D}\ }\textbf {\bibinfo {volume} {97}},\ \bibinfo
  {pages} {023539} (\bibinfo {year} {2018})},\ \Eprint
  {https://arxiv.org/abs/1712.08820} {arXiv:1712.08820 [astro-ph.CO]}
  \BibitemShut {NoStop}%
\bibitem [{\citenamefont {Kodama}\ and\ \citenamefont
  {Sasaki}(1986)}]{Kodama:1986fg}%
  \BibitemOpen
  \bibfield  {author} {\bibinfo {author} {\bibfnamefont {H.}~\bibnamefont
  {Kodama}}\ and\ \bibinfo {author} {\bibfnamefont {M.}~\bibnamefont
  {Sasaki}},\ }\bibfield  {title} {\bibinfo {title} {{Evolution of Isocurvature
  Perturbations. 1. Photon - Baryon Universe}},\ }\href
  {https://doi.org/10.1142/S0217751X86000137} {\bibfield  {journal} {\bibinfo
  {journal} {Int. J. Mod. Phys. A}\ }\textbf {\bibinfo {volume} {1}},\ \bibinfo
  {pages} {265} (\bibinfo {year} {1986})}\BibitemShut {NoStop}%
\bibitem [{\citenamefont {Passaglia}\ and\ \citenamefont
  {Sasaki}(2021)}]{Passaglia:2021jla}%
  \BibitemOpen
  \bibfield  {author} {\bibinfo {author} {\bibfnamefont {S.}~\bibnamefont
  {Passaglia}}\ and\ \bibinfo {author} {\bibfnamefont {M.}~\bibnamefont
  {Sasaki}},\ }\bibfield  {title} {\bibinfo {title} {{Primordial Black Holes
  from CDM Isocurvature}},\ }\href@noop {} {\  (\bibinfo {year} {2021})},\
  \Eprint {https://arxiv.org/abs/2109.12824} {arXiv:2109.12824 [astro-ph.CO]}
  \BibitemShut {NoStop}%
\bibitem [{\citenamefont {Zel'dovich}(1967)}]{Zeldovich:1967lct}%
  \BibitemOpen
  \bibfield  {author} {\bibinfo {author} {\bibfnamefont {I.~D.}\ \bibnamefont
  {Zel'dovich}, \bibfnamefont {Ya.B.;~Novikov}},\ }\bibfield  {title} {\bibinfo
  {title} {{The Hypothesis of Cores Retarded during Expansion and the Hot
  Cosmological Model}},\ }\href@noop {} {\bibfield  {journal} {\bibinfo
  {journal} {Soviet Astron. AJ (Engl. Transl. ),}\ }\textbf {\bibinfo {volume}
  {10}},\ \bibinfo {pages} {602} (\bibinfo {year} {1967})}\BibitemShut
  {NoStop}%
\bibitem [{\citenamefont {Hawking}(1971)}]{Hawking:1971ei}%
  \BibitemOpen
  \bibfield  {author} {\bibinfo {author} {\bibfnamefont {S.}~\bibnamefont
  {Hawking}},\ }\bibfield  {title} {\bibinfo {title} {{Gravitationally
  collapsed objects of very low mass}},\ }\href@noop {} {\bibfield  {journal}
  {\bibinfo  {journal} {Mon. Not. Roy. Astron. Soc.}\ }\textbf {\bibinfo
  {volume} {152}},\ \bibinfo {pages} {75} (\bibinfo {year} {1971})}\BibitemShut
  {NoStop}%
\bibitem [{\citenamefont {Carr}\ and\ \citenamefont
  {Hawking}(1974)}]{Carr:1974nx}%
  \BibitemOpen
  \bibfield  {author} {\bibinfo {author} {\bibfnamefont {B.~J.}\ \bibnamefont
  {Carr}}\ and\ \bibinfo {author} {\bibfnamefont {S.}~\bibnamefont {Hawking}},\
  }\bibfield  {title} {\bibinfo {title} {{Black holes in the early Universe}},\
  }\href@noop {} {\bibfield  {journal} {\bibinfo  {journal} {Mon. Not. Roy.
  Astron. Soc.}\ }\textbf {\bibinfo {volume} {168}},\ \bibinfo {pages} {399}
  (\bibinfo {year} {1974})}\BibitemShut {NoStop}%
\bibitem [{\citenamefont {Meszaros}(1974)}]{Meszaros:1974tb}%
  \BibitemOpen
  \bibfield  {author} {\bibinfo {author} {\bibfnamefont {P.}~\bibnamefont
  {Meszaros}},\ }\bibfield  {title} {\bibinfo {title} {{The behaviour of point
  masses in an expanding cosmological substratum}},\ }\href@noop {} {\bibfield
  {journal} {\bibinfo  {journal} {Astron. Astrophys.}\ }\textbf {\bibinfo
  {volume} {37}},\ \bibinfo {pages} {225} (\bibinfo {year} {1974})}\BibitemShut
  {NoStop}%
\bibitem [{\citenamefont {Carr}(1975)}]{Carr:1975qj}%
  \BibitemOpen
  \bibfield  {author} {\bibinfo {author} {\bibfnamefont {B.~J.}\ \bibnamefont
  {Carr}},\ }\bibfield  {title} {\bibinfo {title} {{The Primordial black hole
  mass spectrum}},\ }\href {https://doi.org/10.1086/153853} {\bibfield
  {journal} {\bibinfo  {journal} {Astrophys. J.}\ }\textbf {\bibinfo {volume}
  {201}},\ \bibinfo {pages} {1} (\bibinfo {year} {1975})}\BibitemShut {NoStop}%
\bibitem [{\citenamefont {Khlopov}\ \emph {et~al.}(1985)\citenamefont
  {Khlopov}, \citenamefont {Malomed},\ and\ \citenamefont
  {Zeldovich}}]{Khlopov:1985jw}%
  \BibitemOpen
  \bibfield  {author} {\bibinfo {author} {\bibfnamefont {M.}~\bibnamefont
  {Khlopov}}, \bibinfo {author} {\bibfnamefont {B.}~\bibnamefont {Malomed}},\
  and\ \bibinfo {author} {\bibfnamefont {I.}~\bibnamefont {Zeldovich}},\
  }\bibfield  {title} {\bibinfo {title} {{Gravitational instability of scalar
  fields and formation of primordial black holes}},\ }\href@noop {} {\bibfield
  {journal} {\bibinfo  {journal} {Mon. Not. Roy. Astron. Soc.}\ }\textbf
  {\bibinfo {volume} {215}},\ \bibinfo {pages} {575} (\bibinfo {year}
  {1985})}\BibitemShut {NoStop}%
\bibitem [{\citenamefont {Niemeyer}\ and\ \citenamefont
  {Jedamzik}(1999)}]{Niemeyer:1999ak}%
  \BibitemOpen
  \bibfield  {author} {\bibinfo {author} {\bibfnamefont {J.~C.}\ \bibnamefont
  {Niemeyer}}\ and\ \bibinfo {author} {\bibfnamefont {K.}~\bibnamefont
  {Jedamzik}},\ }\bibfield  {title} {\bibinfo {title} {{Dynamics of primordial
  black hole formation}},\ }\href {https://doi.org/10.1103/PhysRevD.59.124013}
  {\bibfield  {journal} {\bibinfo  {journal} {Phys. Rev. D}\ }\textbf {\bibinfo
  {volume} {59}},\ \bibinfo {pages} {124013} (\bibinfo {year} {1999})},\
  \Eprint {https://arxiv.org/abs/astro-ph/9901292} {arXiv:astro-ph/9901292}
  \BibitemShut {NoStop}%
\bibitem [{\citenamefont {Khlopov}(2010)}]{Khlopov:2008qy}%
  \BibitemOpen
  \bibfield  {author} {\bibinfo {author} {\bibfnamefont {M.~Y.}\ \bibnamefont
  {Khlopov}},\ }\bibfield  {title} {\bibinfo {title} {{Primordial Black
  Holes}},\ }\href {https://doi.org/10.1088/1674-4527/10/6/001} {\bibfield
  {journal} {\bibinfo  {journal} {Res. Astron. Astrophys.}\ }\textbf {\bibinfo
  {volume} {10}},\ \bibinfo {pages} {495} (\bibinfo {year} {2010})},\ \Eprint
  {https://arxiv.org/abs/0801.0116} {arXiv:0801.0116 [astro-ph]} \BibitemShut
  {NoStop}%
\bibitem [{\citenamefont {Sasaki}\ \emph {et~al.}(2018)\citenamefont {Sasaki},
  \citenamefont {Suyama}, \citenamefont {Tanaka},\ and\ \citenamefont
  {Yokoyama}}]{Sasaki:2018dmp}%
  \BibitemOpen
  \bibfield  {author} {\bibinfo {author} {\bibfnamefont {M.}~\bibnamefont
  {Sasaki}}, \bibinfo {author} {\bibfnamefont {T.}~\bibnamefont {Suyama}},
  \bibinfo {author} {\bibfnamefont {T.}~\bibnamefont {Tanaka}},\ and\ \bibinfo
  {author} {\bibfnamefont {S.}~\bibnamefont {Yokoyama}},\ }\bibfield  {title}
  {\bibinfo {title} {{Primordial black holes\textemdash{}perspectives in
  gravitational wave astronomy}},\ }\href
  {https://doi.org/10.1088/1361-6382/aaa7b4} {\bibfield  {journal} {\bibinfo
  {journal} {Class. Quant. Grav.}\ }\textbf {\bibinfo {volume} {35}},\ \bibinfo
  {pages} {063001} (\bibinfo {year} {2018})},\ \Eprint
  {https://arxiv.org/abs/1801.05235} {arXiv:1801.05235 [astro-ph.CO]}
  \BibitemShut {NoStop}%
\bibitem [{\citenamefont {Carr}\ \emph {et~al.}(2020)\citenamefont {Carr},
  \citenamefont {Kohri}, \citenamefont {Sendouda},\ and\ \citenamefont
  {Yokoyama}}]{Carr:2020gox}%
  \BibitemOpen
  \bibfield  {author} {\bibinfo {author} {\bibfnamefont {B.}~\bibnamefont
  {Carr}}, \bibinfo {author} {\bibfnamefont {K.}~\bibnamefont {Kohri}},
  \bibinfo {author} {\bibfnamefont {Y.}~\bibnamefont {Sendouda}},\ and\
  \bibinfo {author} {\bibfnamefont {J.}~\bibnamefont {Yokoyama}},\ }\bibfield
  {title} {\bibinfo {title} {{Constraints on Primordial Black Holes}},\
  }\href@noop {} {\  (\bibinfo {year} {2020})},\ \Eprint
  {https://arxiv.org/abs/2002.12778} {arXiv:2002.12778 [astro-ph.CO]}
  \BibitemShut {NoStop}%
\bibitem [{\citenamefont {Carr}\ and\ \citenamefont
  {Kuhnel}(2020)}]{Carr:2020xqk}%
  \BibitemOpen
  \bibfield  {author} {\bibinfo {author} {\bibfnamefont {B.}~\bibnamefont
  {Carr}}\ and\ \bibinfo {author} {\bibfnamefont {F.}~\bibnamefont {Kuhnel}},\
  }\bibfield  {title} {\bibinfo {title} {{Primordial Black Holes as Dark
  Matter: Recent Developments}},\ }\href
  {https://doi.org/10.1146/annurev-nucl-050520-125911} {\bibfield  {journal}
  {\bibinfo  {journal} {Ann. Rev. Nucl. Part. Sci.}\ }\textbf {\bibinfo
  {volume} {70}},\ \bibinfo {pages} {355} (\bibinfo {year} {2020})},\ \Eprint
  {https://arxiv.org/abs/2006.02838} {arXiv:2006.02838 [astro-ph.CO]}
  \BibitemShut {NoStop}%
\bibitem [{\citenamefont {Green}\ and\ \citenamefont
  {Kavanagh}(2021)}]{Green:2020jor}%
  \BibitemOpen
  \bibfield  {author} {\bibinfo {author} {\bibfnamefont {A.~M.}\ \bibnamefont
  {Green}}\ and\ \bibinfo {author} {\bibfnamefont {B.~J.}\ \bibnamefont
  {Kavanagh}},\ }\bibfield  {title} {\bibinfo {title} {{Primordial Black Holes
  as a dark matter candidate}},\ }\href
  {https://doi.org/10.1088/1361-6471/abc534} {\bibfield  {journal} {\bibinfo
  {journal} {J. Phys. G}\ }\textbf {\bibinfo {volume} {48}},\ \bibinfo {pages}
  {4} (\bibinfo {year} {2021})},\ \Eprint {https://arxiv.org/abs/2007.10722}
  {arXiv:2007.10722 [astro-ph.CO]} \BibitemShut {NoStop}%
\bibitem [{\citenamefont {Escriv\`a}(2021)}]{Escriva:2021aeh}%
  \BibitemOpen
  \bibfield  {author} {\bibinfo {author} {\bibfnamefont {A.}~\bibnamefont
  {Escriv\`a}},\ }\bibfield  {title} {\bibinfo {title} {{PBH formation from
  spherically symmetric hydrodynamical perturbations: a review}},\ }\href@noop
  {} {\  (\bibinfo {year} {2021})},\ \Eprint {https://arxiv.org/abs/2111.12693}
  {arXiv:2111.12693 [gr-qc]} \BibitemShut {NoStop}%
\bibitem [{\citenamefont {Dolgov}\ and\ \citenamefont
  {Silk}(1993)}]{Dolgov:1992pu}%
  \BibitemOpen
  \bibfield  {author} {\bibinfo {author} {\bibfnamefont {A.}~\bibnamefont
  {Dolgov}}\ and\ \bibinfo {author} {\bibfnamefont {J.}~\bibnamefont {Silk}},\
  }\bibfield  {title} {\bibinfo {title} {{Baryon isocurvature fluctuations at
  small scales and baryonic dark matter}},\ }\href
  {https://doi.org/10.1103/PhysRevD.47.4244} {\bibfield  {journal} {\bibinfo
  {journal} {Phys. Rev. D}\ }\textbf {\bibinfo {volume} {47}},\ \bibinfo
  {pages} {4244} (\bibinfo {year} {1993})}\BibitemShut {NoStop}%
\bibitem [{\citenamefont {Chung}\ and\ \citenamefont
  {Upadhye}(2018)}]{Chung:2017uzc}%
  \BibitemOpen
  \bibfield  {author} {\bibinfo {author} {\bibfnamefont {D.~J.~H.}\
  \bibnamefont {Chung}}\ and\ \bibinfo {author} {\bibfnamefont
  {A.}~\bibnamefont {Upadhye}},\ }\bibfield  {title} {\bibinfo {title} {{Search
  for strongly blue axion isocurvature}},\ }\href
  {https://doi.org/10.1103/PhysRevD.98.023525} {\bibfield  {journal} {\bibinfo
  {journal} {Phys. Rev. D}\ }\textbf {\bibinfo {volume} {98}},\ \bibinfo
  {pages} {023525} (\bibinfo {year} {2018})},\ \Eprint
  {https://arxiv.org/abs/1711.06736} {arXiv:1711.06736 [astro-ph.CO]}
  \BibitemShut {NoStop}%
\bibitem [{\citenamefont {Chung}\ and\ \citenamefont
  {Tadepalli}(2021)}]{Chung:2021lfg}%
  \BibitemOpen
  \bibfield  {author} {\bibinfo {author} {\bibfnamefont {D.~J.~H.}\
  \bibnamefont {Chung}}\ and\ \bibinfo {author} {\bibfnamefont {S.~C.}\
  \bibnamefont {Tadepalli}},\ }\bibfield  {title} {\bibinfo {title} {{An
  Analytic Treatment of Underdamped Axionic Blue Isocurvature Perturbations}},\
  }\href@noop {} {\  (\bibinfo {year} {2021})},\ \Eprint
  {https://arxiv.org/abs/2110.02272} {arXiv:2110.02272 [astro-ph.CO]}
  \BibitemShut {NoStop}%
\bibitem [{\citenamefont {Cotner}\ \emph {et~al.}(2019)\citenamefont {Cotner},
  \citenamefont {Kusenko}, \citenamefont {Sasaki},\ and\ \citenamefont
  {Takhistov}}]{Cotner:2019ykd}%
  \BibitemOpen
  \bibfield  {author} {\bibinfo {author} {\bibfnamefont {E.}~\bibnamefont
  {Cotner}}, \bibinfo {author} {\bibfnamefont {A.}~\bibnamefont {Kusenko}},
  \bibinfo {author} {\bibfnamefont {M.}~\bibnamefont {Sasaki}},\ and\ \bibinfo
  {author} {\bibfnamefont {V.}~\bibnamefont {Takhistov}},\ }\bibfield  {title}
  {\bibinfo {title} {{Analytic Description of Primordial Black Hole Formation
  from Scalar Field Fragmentation}},\ }\href
  {https://doi.org/10.1088/1475-7516/2019/10/077} {\bibfield  {journal}
  {\bibinfo  {journal} {JCAP}\ }\textbf {\bibinfo {volume} {10}},\ \bibinfo
  {pages} {077}},\ \Eprint {https://arxiv.org/abs/1907.10613} {arXiv:1907.10613
  [astro-ph.CO]} \BibitemShut {NoStop}%
\bibitem [{\citenamefont {Tomita}(1967)}]{Tomita}%
  \BibitemOpen
  \bibfield  {author} {\bibinfo {author} {\bibfnamefont {K.}~\bibnamefont
  {Tomita}},\ }\bibfield  {title} {\bibinfo {title} {{Non-Linear Theory of
  Gravitational Instability in the Expanding Universe}},\ }\href
  {https://doi.org/10.1143/PTP.37.831} {\bibfield  {journal} {\bibinfo
  {journal} {Progress of Theoretical Physics}\ }\textbf {\bibinfo {volume}
  {37}},\ \bibinfo {pages} {831} (\bibinfo {year} {1967})},\ \Eprint
  {https://arxiv.org/abs/https://academic.oup.com/ptp/article-pdf/37/5/831/5234391/37-5-831.pdf}
  {https://academic.oup.com/ptp/article-pdf/37/5/831/5234391/37-5-831.pdf}
  \BibitemShut {NoStop}%
\bibitem [{\citenamefont {Matarrese}\ \emph {et~al.}(1993)\citenamefont
  {Matarrese}, \citenamefont {Pantano},\ and\ \citenamefont
  {Saez}}]{Matarrese:1992rp}%
  \BibitemOpen
  \bibfield  {author} {\bibinfo {author} {\bibfnamefont {S.}~\bibnamefont
  {Matarrese}}, \bibinfo {author} {\bibfnamefont {O.}~\bibnamefont {Pantano}},\
  and\ \bibinfo {author} {\bibfnamefont {D.}~\bibnamefont {Saez}},\ }\bibfield
  {title} {\bibinfo {title} {{A General relativistic approach to the nonlinear
  evolution of collisionless matter}},\ }\href
  {https://doi.org/10.1103/PhysRevD.47.1311} {\bibfield  {journal} {\bibinfo
  {journal} {Phys. Rev. D}\ }\textbf {\bibinfo {volume} {47}},\ \bibinfo
  {pages} {1311} (\bibinfo {year} {1993})}\BibitemShut {NoStop}%
\bibitem [{\citenamefont {Matarrese}\ \emph {et~al.}(1994)\citenamefont
  {Matarrese}, \citenamefont {Pantano},\ and\ \citenamefont
  {Saez}}]{Matarrese:1993zf}%
  \BibitemOpen
  \bibfield  {author} {\bibinfo {author} {\bibfnamefont {S.}~\bibnamefont
  {Matarrese}}, \bibinfo {author} {\bibfnamefont {O.}~\bibnamefont {Pantano}},\
  and\ \bibinfo {author} {\bibfnamefont {D.}~\bibnamefont {Saez}},\ }\bibfield
  {title} {\bibinfo {title} {{General relativistic dynamics of irrotational
  dust: Cosmological implications}},\ }\href
  {https://doi.org/10.1103/PhysRevLett.72.320} {\bibfield  {journal} {\bibinfo
  {journal} {Phys. Rev. Lett.}\ }\textbf {\bibinfo {volume} {72}},\ \bibinfo
  {pages} {320} (\bibinfo {year} {1994})},\ \Eprint
  {https://arxiv.org/abs/astro-ph/9310036} {arXiv:astro-ph/9310036}
  \BibitemShut {NoStop}%
\bibitem [{\citenamefont {Ananda}\ \emph {et~al.}(2007)\citenamefont {Ananda},
  \citenamefont {Clarkson},\ and\ \citenamefont {Wands}}]{Ananda:2006af}%
  \BibitemOpen
  \bibfield  {author} {\bibinfo {author} {\bibfnamefont {K.~N.}\ \bibnamefont
  {Ananda}}, \bibinfo {author} {\bibfnamefont {C.}~\bibnamefont {Clarkson}},\
  and\ \bibinfo {author} {\bibfnamefont {D.}~\bibnamefont {Wands}},\ }\bibfield
   {title} {\bibinfo {title} {{The Cosmological gravitational wave background
  from primordial density perturbations}},\ }\href
  {https://doi.org/10.1103/PhysRevD.75.123518} {\bibfield  {journal} {\bibinfo
  {journal} {Phys. Rev. D}\ }\textbf {\bibinfo {volume} {75}},\ \bibinfo
  {pages} {123518} (\bibinfo {year} {2007})},\ \Eprint
  {https://arxiv.org/abs/gr-qc/0612013} {arXiv:gr-qc/0612013} \BibitemShut
  {NoStop}%
\bibitem [{\citenamefont {Baumann}\ \emph {et~al.}(2007)\citenamefont
  {Baumann}, \citenamefont {Steinhardt}, \citenamefont {Takahashi},\ and\
  \citenamefont {Ichiki}}]{Baumann:2007zm}%
  \BibitemOpen
  \bibfield  {author} {\bibinfo {author} {\bibfnamefont {D.}~\bibnamefont
  {Baumann}}, \bibinfo {author} {\bibfnamefont {P.~J.}\ \bibnamefont
  {Steinhardt}}, \bibinfo {author} {\bibfnamefont {K.}~\bibnamefont
  {Takahashi}},\ and\ \bibinfo {author} {\bibfnamefont {K.}~\bibnamefont
  {Ichiki}},\ }\bibfield  {title} {\bibinfo {title} {{Gravitational Wave
  Spectrum Induced by Primordial Scalar Perturbations}},\ }\href
  {https://doi.org/10.1103/PhysRevD.76.084019} {\bibfield  {journal} {\bibinfo
  {journal} {Phys. Rev. D}\ }\textbf {\bibinfo {volume} {76}},\ \bibinfo
  {pages} {084019} (\bibinfo {year} {2007})},\ \Eprint
  {https://arxiv.org/abs/hep-th/0703290} {arXiv:hep-th/0703290} \BibitemShut
  {NoStop}%
\bibitem [{\citenamefont {Saito}\ and\ \citenamefont
  {Yokoyama}(2009)}]{Saito:2008jc}%
  \BibitemOpen
  \bibfield  {author} {\bibinfo {author} {\bibfnamefont {R.}~\bibnamefont
  {Saito}}\ and\ \bibinfo {author} {\bibfnamefont {J.}~\bibnamefont
  {Yokoyama}},\ }\bibfield  {title} {\bibinfo {title} {{Gravitational wave
  background as a probe of the primordial black hole abundance}},\ }\href
  {https://doi.org/10.1103/PhysRevLett.102.161101} {\bibfield  {journal}
  {\bibinfo  {journal} {Phys. Rev. Lett.}\ }\textbf {\bibinfo {volume} {102}},\
  \bibinfo {pages} {161101} (\bibinfo {year} {2009})},\ \bibinfo {note}
  {[Erratum: Phys.Rev.Lett. 107, 069901 (2011)]},\ \Eprint
  {https://arxiv.org/abs/0812.4339} {arXiv:0812.4339 [astro-ph]} \BibitemShut
  {NoStop}%
\bibitem [{\citenamefont {Saito}\ and\ \citenamefont
  {Yokoyama}(2010)}]{Saito:2009jt}%
  \BibitemOpen
  \bibfield  {author} {\bibinfo {author} {\bibfnamefont {R.}~\bibnamefont
  {Saito}}\ and\ \bibinfo {author} {\bibfnamefont {J.}~\bibnamefont
  {Yokoyama}},\ }\bibfield  {title} {\bibinfo {title} {{Gravitational-Wave
  Constraints on the Abundance of Primordial Black Holes}},\ }\href
  {https://doi.org/10.1143/PTP.126.351} {\bibfield  {journal} {\bibinfo
  {journal} {Prog. Theor. Phys.}\ }\textbf {\bibinfo {volume} {123}},\ \bibinfo
  {pages} {867} (\bibinfo {year} {2010})},\ \bibinfo {note} {[Erratum:
  Prog.Theor.Phys. 126, 351--352 (2011)]},\ \Eprint
  {https://arxiv.org/abs/0912.5317} {arXiv:0912.5317 [astro-ph.CO]}
  \BibitemShut {NoStop}%
\bibitem [{\citenamefont {Yuan}\ and\ \citenamefont
  {Huang}(2021)}]{Yuan:2021qgz}%
  \BibitemOpen
  \bibfield  {author} {\bibinfo {author} {\bibfnamefont {C.}~\bibnamefont
  {Yuan}}\ and\ \bibinfo {author} {\bibfnamefont {Q.-G.}\ \bibnamefont
  {Huang}},\ }\bibfield  {title} {\bibinfo {title} {{A topic review on probing
  primordial black hole dark matter with scalar induced gravitational waves}},\
  }\href@noop {} {\  (\bibinfo {year} {2021})},\ \Eprint
  {https://arxiv.org/abs/2103.04739} {arXiv:2103.04739 [astro-ph.GA]}
  \BibitemShut {NoStop}%
\bibitem [{\citenamefont {Dom\`enech}(2021)}]{Domenech:2021ztg}%
  \BibitemOpen
  \bibfield  {author} {\bibinfo {author} {\bibfnamefont {G.}~\bibnamefont
  {Dom\`enech}},\ }\bibfield  {title} {\bibinfo {title} {{Scalar Induced
  Gravitational Waves Review}},\ }\href
  {https://doi.org/10.3390/universe7110398} {\bibfield  {journal} {\bibinfo
  {journal} {Universe}\ }\textbf {\bibinfo {volume} {7}},\ \bibinfo {pages}
  {398} (\bibinfo {year} {2021})},\ \Eprint {https://arxiv.org/abs/2109.01398}
  {arXiv:2109.01398 [gr-qc]} \BibitemShut {NoStop}%
\bibitem [{\citenamefont {Papanikolaou}\ \emph {et~al.}(2020)\citenamefont
  {Papanikolaou}, \citenamefont {Vennin},\ and\ \citenamefont
  {Langlois}}]{Papanikolaou:2020qtd}%
  \BibitemOpen
  \bibfield  {author} {\bibinfo {author} {\bibfnamefont {T.}~\bibnamefont
  {Papanikolaou}}, \bibinfo {author} {\bibfnamefont {V.}~\bibnamefont
  {Vennin}},\ and\ \bibinfo {author} {\bibfnamefont {D.}~\bibnamefont
  {Langlois}},\ }\bibfield  {title} {\bibinfo {title} {{Gravitational waves
  from a universe filled with primordial black holes}},\ }\href@noop {} {\
  (\bibinfo {year} {2020})},\ \Eprint {https://arxiv.org/abs/2010.11573}
  {arXiv:2010.11573 [astro-ph.CO]} \BibitemShut {NoStop}%
\bibitem [{\citenamefont {Dom\`enech}\ \emph
  {et~al.}(2021{\natexlab{a}})\citenamefont {Dom\`enech}, \citenamefont {Lin},\
  and\ \citenamefont {Sasaki}}]{Domenech:2020ssp}%
  \BibitemOpen
  \bibfield  {author} {\bibinfo {author} {\bibfnamefont {G.}~\bibnamefont
  {Dom\`enech}}, \bibinfo {author} {\bibfnamefont {C.}~\bibnamefont {Lin}},\
  and\ \bibinfo {author} {\bibfnamefont {M.}~\bibnamefont {Sasaki}},\
  }\bibfield  {title} {\bibinfo {title} {{Erratum: Gravitational wave
  constraints on the primordial black hole dominated early universe}},\ }\href
  {https://doi.org/10.1088/1475-7516/2021/11/E01} {\bibfield  {journal}
  {\bibinfo  {journal} {JCAP}\ }\textbf {\bibinfo {volume} {11}},\ \bibinfo
  {pages} {E01}},\ \Eprint {https://arxiv.org/abs/2012.08151} {arXiv:2012.08151
  [gr-qc]} \BibitemShut {NoStop}%
\bibitem [{\citenamefont {Dom\`enech}\ \emph
  {et~al.}(2021{\natexlab{b}})\citenamefont {Dom\`enech}, \citenamefont
  {Takhistov},\ and\ \citenamefont {Sasaki}}]{Domenech:2021wkk}%
  \BibitemOpen
  \bibfield  {author} {\bibinfo {author} {\bibfnamefont {G.}~\bibnamefont
  {Dom\`enech}}, \bibinfo {author} {\bibfnamefont {V.}~\bibnamefont
  {Takhistov}},\ and\ \bibinfo {author} {\bibfnamefont {M.}~\bibnamefont
  {Sasaki}},\ }\bibfield  {title} {\bibinfo {title} {{Exploring evaporating
  primordial black holes with gravitational waves}},\ }\href
  {https://doi.org/10.1016/j.physletb.2021.136722} {\bibfield  {journal}
  {\bibinfo  {journal} {Phys. Lett. B}\ }\textbf {\bibinfo {volume} {823}},\
  \bibinfo {pages} {136722} (\bibinfo {year} {2021}{\natexlab{b}})},\ \Eprint
  {https://arxiv.org/abs/2105.06816} {arXiv:2105.06816 [astro-ph.CO]}
  \BibitemShut {NoStop}%
\bibitem [{\citenamefont {Kozaczuk}\ \emph {et~al.}(2021)\citenamefont
  {Kozaczuk}, \citenamefont {Lin},\ and\ \citenamefont
  {Villarama}}]{Kozaczuk:2021wcl}%
  \BibitemOpen
  \bibfield  {author} {\bibinfo {author} {\bibfnamefont {J.}~\bibnamefont
  {Kozaczuk}}, \bibinfo {author} {\bibfnamefont {T.}~\bibnamefont {Lin}},\ and\
  \bibinfo {author} {\bibfnamefont {E.}~\bibnamefont {Villarama}},\ }\bibfield
  {title} {\bibinfo {title} {{Signals of primordial black holes at
  gravitational wave interferometers}},\ }\href@noop {} {\  (\bibinfo {year}
  {2021})},\ \Eprint {https://arxiv.org/abs/2108.12475} {arXiv:2108.12475
  [astro-ph.CO]} \BibitemShut {NoStop}%
\bibitem [{\citenamefont {Punturo}\ \emph {et~al.}(2010)\citenamefont {Punturo}
  \emph {et~al.}}]{Punturo:2010zz}%
  \BibitemOpen
  \bibfield  {author} {\bibinfo {author} {\bibfnamefont {M.}~\bibnamefont
  {Punturo}} \emph {et~al.},\ }\bibfield  {title} {\bibinfo {title} {{The
  Einstein Telescope: A third-generation gravitational wave observatory}},\
  }\href {https://doi.org/10.1088/0264-9381/27/19/194002} {\bibfield  {journal}
  {\bibinfo  {journal} {Class. Quant. Grav.}\ }\textbf {\bibinfo {volume}
  {27}},\ \bibinfo {pages} {194002} (\bibinfo {year} {2010})}\BibitemShut
  {NoStop}%
\bibitem [{\citenamefont {Hild}\ \emph {et~al.}(2011)\citenamefont {Hild} \emph
  {et~al.}}]{Hild:2010id}%
  \BibitemOpen
  \bibfield  {author} {\bibinfo {author} {\bibfnamefont {S.}~\bibnamefont
  {Hild}} \emph {et~al.},\ }\bibfield  {title} {\bibinfo {title} {{Sensitivity
  Studies for Third-Generation Gravitational Wave Observatories}},\ }\href
  {https://doi.org/10.1088/0264-9381/28/9/094013} {\bibfield  {journal}
  {\bibinfo  {journal} {Class. Quant. Grav.}\ }\textbf {\bibinfo {volume}
  {28}},\ \bibinfo {pages} {094013} (\bibinfo {year} {2011})},\ \Eprint
  {https://arxiv.org/abs/1012.0908} {arXiv:1012.0908 [gr-qc]} \BibitemShut
  {NoStop}%
\bibitem [{\citenamefont {Maggiore}\ \emph {et~al.}(2020)\citenamefont
  {Maggiore} \emph {et~al.}}]{Maggiore:2019uih}%
  \BibitemOpen
  \bibfield  {author} {\bibinfo {author} {\bibfnamefont {M.}~\bibnamefont
  {Maggiore}} \emph {et~al.},\ }\bibfield  {title} {\bibinfo {title} {{Science
  Case for the Einstein Telescope}},\ }\href
  {https://doi.org/10.1088/1475-7516/2020/03/050} {\bibfield  {journal}
  {\bibinfo  {journal} {JCAP}\ }\textbf {\bibinfo {volume} {03}},\ \bibinfo
  {pages} {050}},\ \Eprint {https://arxiv.org/abs/1912.02622} {arXiv:1912.02622
  [astro-ph.CO]} \BibitemShut {NoStop}%
\bibitem [{\citenamefont {Amaro-Seoane}\ \emph {et~al.}(2017)\citenamefont
  {Amaro-Seoane} \emph {et~al.}}]{Audley:2017drz}%
  \BibitemOpen
  \bibfield  {author} {\bibinfo {author} {\bibfnamefont {P.}~\bibnamefont
  {Amaro-Seoane}} \emph {et~al.} (\bibinfo {collaboration} {LISA}),\ }\bibfield
   {title} {\bibinfo {title} {{Laser Interferometer Space Antenna}},\
  }\href@noop {} {\  (\bibinfo {year} {2017})},\ \Eprint
  {https://arxiv.org/abs/1702.00786} {arXiv:1702.00786 [astro-ph.IM]}
  \BibitemShut {NoStop}%
%%CITATION = ARXIV:1702.00786;%%
\bibitem [{\citenamefont {Baker}\ \emph {et~al.}(2019)\citenamefont {Baker}
  \emph {et~al.}}]{Baker:2019nia}%
  \BibitemOpen
  \bibfield  {author} {\bibinfo {author} {\bibfnamefont {J.}~\bibnamefont
  {Baker}} \emph {et~al.},\ }\bibfield  {title} {\bibinfo {title} {{The Laser
  Interferometer Space Antenna: Unveiling the Millihertz Gravitational Wave
  Sky}},\ }\href@noop {} {\  (\bibinfo {year} {2019})},\ \Eprint
  {https://arxiv.org/abs/1907.06482} {arXiv:1907.06482 [astro-ph.IM]}
  \BibitemShut {NoStop}%
\bibitem [{\citenamefont {Carilli}\ and\ \citenamefont
  {Rawlings}(2004)}]{Carilli:2004nx}%
  \BibitemOpen
  \bibfield  {author} {\bibinfo {author} {\bibfnamefont {C.~L.}\ \bibnamefont
  {Carilli}}\ and\ \bibinfo {author} {\bibfnamefont {S.}~\bibnamefont
  {Rawlings}},\ }\bibfield  {title} {\bibinfo {title} {{Science with the Square
  Kilometer Array: Motivation, key science projects, standards and
  assumptions}},\ }\href {https://doi.org/10.1016/j.newar.2004.09.001}
  {\bibfield  {journal} {\bibinfo  {journal} {New Astron. Rev.}\ }\textbf
  {\bibinfo {volume} {48}},\ \bibinfo {pages} {979} (\bibinfo {year} {2004})},\
  \Eprint {https://arxiv.org/abs/astro-ph/0409274} {arXiv:astro-ph/0409274}
  \BibitemShut {NoStop}%
\bibitem [{\citenamefont {Janssen}\ \emph {et~al.}(2015)\citenamefont {Janssen}
  \emph {et~al.}}]{Janssen:2014dka}%
  \BibitemOpen
  \bibfield  {author} {\bibinfo {author} {\bibfnamefont {G.}~\bibnamefont
  {Janssen}} \emph {et~al.},\ }\bibfield  {title} {\bibinfo {title}
  {{Gravitational wave astronomy with the SKA}},\ }\href
  {https://doi.org/10.22323/1.215.0037} {\bibfield  {journal} {\bibinfo
  {journal} {PoS}\ }\textbf {\bibinfo {volume} {AASKA14}},\ \bibinfo {pages}
  {037} (\bibinfo {year} {2015})},\ \Eprint {https://arxiv.org/abs/1501.00127}
  {arXiv:1501.00127 [astro-ph.IM]} \BibitemShut {NoStop}%
\bibitem [{\citenamefont {Weltman}\ \emph {et~al.}(2020)\citenamefont {Weltman}
  \emph {et~al.}}]{Weltman:2018zrl}%
  \BibitemOpen
  \bibfield  {author} {\bibinfo {author} {\bibfnamefont {A.}~\bibnamefont
  {Weltman}} \emph {et~al.},\ }\bibfield  {title} {\bibinfo {title}
  {{Fundamental physics with the Square Kilometre Array}},\ }\href
  {https://doi.org/10.1017/pasa.2019.42} {\bibfield  {journal} {\bibinfo
  {journal} {Publ. Astron. Soc. Austral.}\ }\textbf {\bibinfo {volume} {37}},\
  \bibinfo {pages} {e002} (\bibinfo {year} {2020})},\ \Eprint
  {https://arxiv.org/abs/1810.02680} {arXiv:1810.02680 [astro-ph.CO]}
  \BibitemShut {NoStop}%
\bibitem [{\citenamefont {Garcia-Bellido}\ \emph {et~al.}(2016)\citenamefont
  {Garcia-Bellido}, \citenamefont {Peloso},\ and\ \citenamefont
  {Unal}}]{Garcia-Bellido:2016dkw}%
  \BibitemOpen
  \bibfield  {author} {\bibinfo {author} {\bibfnamefont {J.}~\bibnamefont
  {Garcia-Bellido}}, \bibinfo {author} {\bibfnamefont {M.}~\bibnamefont
  {Peloso}},\ and\ \bibinfo {author} {\bibfnamefont {C.}~\bibnamefont {Unal}},\
  }\bibfield  {title} {\bibinfo {title} {{Gravitational waves at interferometer
  scales and primordial black holes in axion inflation}},\ }\href
  {https://doi.org/10.1088/1475-7516/2016/12/031} {\bibfield  {journal}
  {\bibinfo  {journal} {JCAP}\ }\textbf {\bibinfo {volume} {12}},\ \bibinfo
  {pages} {031}},\ \Eprint {https://arxiv.org/abs/1610.03763} {arXiv:1610.03763
  [astro-ph.CO]} \BibitemShut {NoStop}%
\bibitem [{\citenamefont {Di}\ and\ \citenamefont {Gong}(2018)}]{Gong:2017qlj}%
  \BibitemOpen
  \bibfield  {author} {\bibinfo {author} {\bibfnamefont {H.}~\bibnamefont
  {Di}}\ and\ \bibinfo {author} {\bibfnamefont {Y.}~\bibnamefont {Gong}},\
  }\bibfield  {title} {\bibinfo {title} {{Primordial black holes and second
  order gravitational waves from ultra-slow-roll inflation}},\ }\href
  {https://doi.org/10.1088/1475-7516/2018/07/007} {\bibfield  {journal}
  {\bibinfo  {journal} {JCAP}\ }\textbf {\bibinfo {volume} {07}},\ \bibinfo
  {pages} {007}},\ \Eprint {https://arxiv.org/abs/1707.09578} {arXiv:1707.09578
  [astro-ph.CO]} \BibitemShut {NoStop}%
\bibitem [{\citenamefont {Ando}\ \emph {et~al.}(2018)\citenamefont {Ando},
  \citenamefont {Kawasaki},\ and\ \citenamefont {Nakatsuka}}]{Ando:2018nge}%
  \BibitemOpen
  \bibfield  {author} {\bibinfo {author} {\bibfnamefont {K.}~\bibnamefont
  {Ando}}, \bibinfo {author} {\bibfnamefont {M.}~\bibnamefont {Kawasaki}},\
  and\ \bibinfo {author} {\bibfnamefont {H.}~\bibnamefont {Nakatsuka}},\
  }\bibfield  {title} {\bibinfo {title} {{Formation of primordial black holes
  in an axionlike curvaton model}},\ }\href
  {https://doi.org/10.1103/PhysRevD.98.083508} {\bibfield  {journal} {\bibinfo
  {journal} {Phys. Rev. D}\ }\textbf {\bibinfo {volume} {98}},\ \bibinfo
  {pages} {083508} (\bibinfo {year} {2018})},\ \Eprint
  {https://arxiv.org/abs/1805.07757} {arXiv:1805.07757 [astro-ph.CO]}
  \BibitemShut {NoStop}%
\bibitem [{\citenamefont {Byrnes}\ \emph {et~al.}(2019)\citenamefont {Byrnes},
  \citenamefont {Cole},\ and\ \citenamefont {Patil}}]{Byrnes:2018txb}%
  \BibitemOpen
  \bibfield  {author} {\bibinfo {author} {\bibfnamefont {C.~T.}\ \bibnamefont
  {Byrnes}}, \bibinfo {author} {\bibfnamefont {P.~S.}\ \bibnamefont {Cole}},\
  and\ \bibinfo {author} {\bibfnamefont {S.~P.}\ \bibnamefont {Patil}},\
  }\bibfield  {title} {\bibinfo {title} {{Steepest growth of the power spectrum
  and primordial black holes}},\ }\href
  {https://doi.org/10.1088/1475-7516/2019/06/028} {\bibfield  {journal}
  {\bibinfo  {journal} {JCAP}\ }\textbf {\bibinfo {volume} {06}},\ \bibinfo
  {pages} {028}},\ \Eprint {https://arxiv.org/abs/1811.11158} {arXiv:1811.11158
  [astro-ph.CO]} \BibitemShut {NoStop}%
\bibitem [{\citenamefont {Gao}\ and\ \citenamefont {Yang}(2019)}]{Gao:2019kto}%
  \BibitemOpen
  \bibfield  {author} {\bibinfo {author} {\bibfnamefont {T.-J.}\ \bibnamefont
  {Gao}}\ and\ \bibinfo {author} {\bibfnamefont {X.-Y.}\ \bibnamefont {Yang}},\
  }\bibfield  {title} {\bibinfo {title} {{Gravitational waves induced from
  string axion model of inflation}},\ }\href
  {https://doi.org/10.1142/S0217751X19502130} {\bibfield  {journal} {\bibinfo
  {journal} {Int. J. Mod. Phys. A}\ }\textbf {\bibinfo {volume} {34}},\
  \bibinfo {pages} {1950213} (\bibinfo {year} {2019})}\BibitemShut {NoStop}%
\bibitem [{\citenamefont {Xu}\ \emph {et~al.}(2020)\citenamefont {Xu},
  \citenamefont {Liu}, \citenamefont {Gao},\ and\ \citenamefont
  {Guo}}]{Xu:2019bdp}%
  \BibitemOpen
  \bibfield  {author} {\bibinfo {author} {\bibfnamefont {W.-T.}\ \bibnamefont
  {Xu}}, \bibinfo {author} {\bibfnamefont {J.}~\bibnamefont {Liu}}, \bibinfo
  {author} {\bibfnamefont {T.-J.}\ \bibnamefont {Gao}},\ and\ \bibinfo {author}
  {\bibfnamefont {Z.-K.}\ \bibnamefont {Guo}},\ }\bibfield  {title} {\bibinfo
  {title} {{Gravitational waves from double-inflection-point inflation}},\
  }\href {https://doi.org/10.1103/PhysRevD.101.023505} {\bibfield  {journal}
  {\bibinfo  {journal} {Phys. Rev. D}\ }\textbf {\bibinfo {volume} {101}},\
  \bibinfo {pages} {023505} (\bibinfo {year} {2020})},\ \Eprint
  {https://arxiv.org/abs/1907.05213} {arXiv:1907.05213 [astro-ph.CO]}
  \BibitemShut {NoStop}%
\bibitem [{\citenamefont {Liu}\ \emph {et~al.}(2020)\citenamefont {Liu},
  \citenamefont {Guo},\ and\ \citenamefont {Cai}}]{Liu:2020oqe}%
  \BibitemOpen
  \bibfield  {author} {\bibinfo {author} {\bibfnamefont {J.}~\bibnamefont
  {Liu}}, \bibinfo {author} {\bibfnamefont {Z.-K.}\ \bibnamefont {Guo}},\ and\
  \bibinfo {author} {\bibfnamefont {R.-G.}\ \bibnamefont {Cai}},\ }\bibfield
  {title} {\bibinfo {title} {{Analytical approximation of the scalar spectrum
  in the ultraslow-roll inflationary models}},\ }\href
  {https://doi.org/10.1103/PhysRevD.101.083535} {\bibfield  {journal} {\bibinfo
   {journal} {Phys. Rev. D}\ }\textbf {\bibinfo {volume} {101}},\ \bibinfo
  {pages} {083535} (\bibinfo {year} {2020})},\ \Eprint
  {https://arxiv.org/abs/2003.02075} {arXiv:2003.02075 [astro-ph.CO]}
  \BibitemShut {NoStop}%
\bibitem [{\citenamefont {Cai}\ \emph {et~al.}(2019{\natexlab{a}})\citenamefont
  {Cai}, \citenamefont {Pi}, \citenamefont {Wang},\ and\ \citenamefont
  {Yang}}]{Cai:2019amo}%
  \BibitemOpen
  \bibfield  {author} {\bibinfo {author} {\bibfnamefont {R.-G.}\ \bibnamefont
  {Cai}}, \bibinfo {author} {\bibfnamefont {S.}~\bibnamefont {Pi}}, \bibinfo
  {author} {\bibfnamefont {S.-J.}\ \bibnamefont {Wang}},\ and\ \bibinfo
  {author} {\bibfnamefont {X.-Y.}\ \bibnamefont {Yang}},\ }\bibfield  {title}
  {\bibinfo {title} {{Resonant multiple peaks in the induced gravitational
  waves}},\ }\href {https://doi.org/10.1088/1475-7516/2019/05/013} {\bibfield
  {journal} {\bibinfo  {journal} {JCAP}\ }\textbf {\bibinfo {volume} {05}},\
  \bibinfo {pages} {013}},\ \Eprint {https://arxiv.org/abs/1901.10152}
  {arXiv:1901.10152 [astro-ph.CO]} \BibitemShut {NoStop}%
\bibitem [{\citenamefont {\"Ozsoy}\ and\ \citenamefont
  {Tasinato}(2020)}]{Ozsoy:2019lyy}%
  \BibitemOpen
  \bibfield  {author} {\bibinfo {author} {\bibfnamefont {O.}~\bibnamefont
  {\"Ozsoy}}\ and\ \bibinfo {author} {\bibfnamefont {G.}~\bibnamefont
  {Tasinato}},\ }\bibfield  {title} {\bibinfo {title} {{On the slope of the
  curvature power spectrum in non-attractor inflation}},\ }\href
  {https://doi.org/10.1088/1475-7516/2020/04/048} {\bibfield  {journal}
  {\bibinfo  {journal} {JCAP}\ }\textbf {\bibinfo {volume} {04}},\ \bibinfo
  {pages} {048}},\ \Eprint {https://arxiv.org/abs/1912.01061} {arXiv:1912.01061
  [astro-ph.CO]} \BibitemShut {NoStop}%
\bibitem [{\citenamefont {\"Ozsoy}\ and\ \citenamefont
  {Lalak}(2021)}]{Ozsoy:2020kat}%
  \BibitemOpen
  \bibfield  {author} {\bibinfo {author} {\bibfnamefont {O.}~\bibnamefont
  {\"Ozsoy}}\ and\ \bibinfo {author} {\bibfnamefont {Z.}~\bibnamefont
  {Lalak}},\ }\bibfield  {title} {\bibinfo {title} {{Primordial black holes as
  dark matter and gravitational waves from bumpy axion inflation}},\ }\href
  {https://doi.org/10.1088/1475-7516/2021/01/040} {\bibfield  {journal}
  {\bibinfo  {journal} {JCAP}\ }\textbf {\bibinfo {volume} {01}},\ \bibinfo
  {pages} {040}},\ \Eprint {https://arxiv.org/abs/2008.07549} {arXiv:2008.07549
  [astro-ph.CO]} \BibitemShut {NoStop}%
\bibitem [{\citenamefont {Ragavendra}\ \emph {et~al.}(2020)\citenamefont
  {Ragavendra}, \citenamefont {Saha}, \citenamefont {Sriramkumar},\ and\
  \citenamefont {Silk}}]{Ragavendra:2020sop}%
  \BibitemOpen
  \bibfield  {author} {\bibinfo {author} {\bibfnamefont {H.~V.}\ \bibnamefont
  {Ragavendra}}, \bibinfo {author} {\bibfnamefont {P.}~\bibnamefont {Saha}},
  \bibinfo {author} {\bibfnamefont {L.}~\bibnamefont {Sriramkumar}},\ and\
  \bibinfo {author} {\bibfnamefont {J.}~\bibnamefont {Silk}},\ }\bibfield
  {title} {\bibinfo {title} {{PBHs and secondary GWs from ultra slow roll and
  punctuated inflation}},\ }\href@noop {} {\  (\bibinfo {year} {2020})},\
  \Eprint {https://arxiv.org/abs/2008.12202} {arXiv:2008.12202 [astro-ph.CO]}
  \BibitemShut {NoStop}%
\bibitem [{\citenamefont {Fumagalli}\ \emph {et~al.}(2020)\citenamefont
  {Fumagalli}, \citenamefont {Renaux-Petel},\ and\ \citenamefont
  {Witkowski}}]{Fumagalli:2020nvq}%
  \BibitemOpen
  \bibfield  {author} {\bibinfo {author} {\bibfnamefont {J.}~\bibnamefont
  {Fumagalli}}, \bibinfo {author} {\bibfnamefont {S.}~\bibnamefont
  {Renaux-Petel}},\ and\ \bibinfo {author} {\bibfnamefont {L.~T.}\ \bibnamefont
  {Witkowski}},\ }\bibfield  {title} {\bibinfo {title} {{Oscillations in the
  stochastic gravitational wave background from sharp features and particle
  production during inflation}},\ }\href@noop {} {\  (\bibinfo {year}
  {2020})},\ \Eprint {https://arxiv.org/abs/2012.02761} {arXiv:2012.02761
  [astro-ph.CO]} \BibitemShut {NoStop}%
\bibitem [{\citenamefont {Braglia}\ \emph
  {et~al.}(2020{\natexlab{a}})\citenamefont {Braglia}, \citenamefont {Hazra},
  \citenamefont {Finelli}, \citenamefont {Smoot}, \citenamefont {Sriramkumar},\
  and\ \citenamefont {Starobinsky}}]{Braglia:2020eai}%
  \BibitemOpen
  \bibfield  {author} {\bibinfo {author} {\bibfnamefont {M.}~\bibnamefont
  {Braglia}}, \bibinfo {author} {\bibfnamefont {D.~K.}\ \bibnamefont {Hazra}},
  \bibinfo {author} {\bibfnamefont {F.}~\bibnamefont {Finelli}}, \bibinfo
  {author} {\bibfnamefont {G.~F.}\ \bibnamefont {Smoot}}, \bibinfo {author}
  {\bibfnamefont {L.}~\bibnamefont {Sriramkumar}},\ and\ \bibinfo {author}
  {\bibfnamefont {A.~A.}\ \bibnamefont {Starobinsky}},\ }\bibfield  {title}
  {\bibinfo {title} {{Generating PBHs and small-scale GWs in two-field models
  of inflation}},\ }\href {https://doi.org/10.1088/1475-7516/2020/08/001}
  {\bibfield  {journal} {\bibinfo  {journal} {JCAP}\ }\textbf {\bibinfo
  {volume} {08}},\ \bibinfo {pages} {001}},\ \Eprint
  {https://arxiv.org/abs/2005.02895} {arXiv:2005.02895 [astro-ph.CO]}
  \BibitemShut {NoStop}%
\bibitem [{\citenamefont {Braglia}\ \emph
  {et~al.}(2020{\natexlab{b}})\citenamefont {Braglia}, \citenamefont {Chen},\
  and\ \citenamefont {Hazra}}]{Braglia:2020taf}%
  \BibitemOpen
  \bibfield  {author} {\bibinfo {author} {\bibfnamefont {M.}~\bibnamefont
  {Braglia}}, \bibinfo {author} {\bibfnamefont {X.}~\bibnamefont {Chen}},\ and\
  \bibinfo {author} {\bibfnamefont {D.~K.}\ \bibnamefont {Hazra}},\ }\bibfield
  {title} {\bibinfo {title} {{Probing Primordial Features with the Stochastic
  Gravitational Wave Background}},\ }\href@noop {} {\  (\bibinfo {year}
  {2020}{\natexlab{b}})},\ \Eprint {https://arxiv.org/abs/2012.05821}
  {arXiv:2012.05821 [astro-ph.CO]} \BibitemShut {NoStop}%
\bibitem [{\citenamefont {Fumagalli}\ \emph
  {et~al.}(2021{\natexlab{a}})\citenamefont {Fumagalli}, \citenamefont
  {Renaux-Petel},\ and\ \citenamefont {Witkowski}}]{Fumagalli:2021cel}%
  \BibitemOpen
  \bibfield  {author} {\bibinfo {author} {\bibfnamefont {J.}~\bibnamefont
  {Fumagalli}}, \bibinfo {author} {\bibfnamefont {S.}~\bibnamefont
  {Renaux-Petel}},\ and\ \bibinfo {author} {\bibfnamefont {L.~T.}\ \bibnamefont
  {Witkowski}},\ }\bibfield  {title} {\bibinfo {title} {{Resonant features in
  the stochastic gravitational wave background}},\ }\href@noop {} {\  (\bibinfo
  {year} {2021}{\natexlab{a}})},\ \Eprint {https://arxiv.org/abs/2105.06481}
  {arXiv:2105.06481 [astro-ph.CO]} \BibitemShut {NoStop}%
\bibitem [{\citenamefont {Bastero-Gil}\ and\ \citenamefont
  {D\'\i{}az-Blanco}(2021)}]{Bastero-Gil:2021fac}%
  \BibitemOpen
  \bibfield  {author} {\bibinfo {author} {\bibfnamefont {M.}~\bibnamefont
  {Bastero-Gil}}\ and\ \bibinfo {author} {\bibfnamefont {M.~S.}\ \bibnamefont
  {D\'\i{}az-Blanco}},\ }\bibfield  {title} {\bibinfo {title} {{Gravity Waves
  and Primordial Black Holes in Scalar Warm Little Inflation}},\ }\href@noop {}
  {\  (\bibinfo {year} {2021})},\ \Eprint {https://arxiv.org/abs/2105.08045}
  {arXiv:2105.08045 [hep-ph]} \BibitemShut {NoStop}%
\bibitem [{\citenamefont {Fumagalli}\ \emph
  {et~al.}(2021{\natexlab{b}})\citenamefont {Fumagalli}, \citenamefont {Palma},
  \citenamefont {Renaux-Petel}, \citenamefont {Sypsas}, \citenamefont
  {Witkowski},\ and\ \citenamefont {Zenteno}}]{Fumagalli:2021mpc}%
  \BibitemOpen
  \bibfield  {author} {\bibinfo {author} {\bibfnamefont {J.}~\bibnamefont
  {Fumagalli}}, \bibinfo {author} {\bibfnamefont {G.~A.}\ \bibnamefont
  {Palma}}, \bibinfo {author} {\bibfnamefont {S.}~\bibnamefont {Renaux-Petel}},
  \bibinfo {author} {\bibfnamefont {S.}~\bibnamefont {Sypsas}}, \bibinfo
  {author} {\bibfnamefont {L.~T.}\ \bibnamefont {Witkowski}},\ and\ \bibinfo
  {author} {\bibfnamefont {C.}~\bibnamefont {Zenteno}},\ }\bibfield  {title}
  {\bibinfo {title} {{Primordial gravitational waves from excited states}},\
  }\href@noop {} {\  (\bibinfo {year} {2021}{\natexlab{b}})},\ \Eprint
  {https://arxiv.org/abs/2111.14664} {arXiv:2111.14664 [astro-ph.CO]}
  \BibitemShut {NoStop}%
\bibitem [{\citenamefont {Fumagalli}\ \emph
  {et~al.}(2021{\natexlab{c}})\citenamefont {Fumagalli}, \citenamefont
  {Pieroni}, \citenamefont {Renaux-Petel},\ and\ \citenamefont
  {Witkowski}}]{Fumagalli:2021dtd}%
  \BibitemOpen
  \bibfield  {author} {\bibinfo {author} {\bibfnamefont {J.}~\bibnamefont
  {Fumagalli}}, \bibinfo {author} {\bibfnamefont {M.}~\bibnamefont {Pieroni}},
  \bibinfo {author} {\bibfnamefont {S.}~\bibnamefont {Renaux-Petel}},\ and\
  \bibinfo {author} {\bibfnamefont {L.~T.}\ \bibnamefont {Witkowski}},\
  }\bibfield  {title} {\bibinfo {title} {{Detecting primordial features with
  LISA}},\ }\href@noop {} {\  (\bibinfo {year} {2021}{\natexlab{c}})},\ \Eprint
  {https://arxiv.org/abs/2112.06903} {arXiv:2112.06903 [astro-ph.CO]}
  \BibitemShut {NoStop}%
\bibitem [{\citenamefont {Saikawa}\ and\ \citenamefont
  {Shirai}(2018)}]{Saikawa:2018rcs}%
  \BibitemOpen
  \bibfield  {author} {\bibinfo {author} {\bibfnamefont {K.}~\bibnamefont
  {Saikawa}}\ and\ \bibinfo {author} {\bibfnamefont {S.}~\bibnamefont
  {Shirai}},\ }\bibfield  {title} {\bibinfo {title} {{Primordial gravitational
  waves, precisely: The role of thermodynamics in the Standard Model}},\ }\href
  {https://doi.org/10.1088/1475-7516/2018/05/035} {\bibfield  {journal}
  {\bibinfo  {journal} {JCAP}\ }\textbf {\bibinfo {volume} {05}},\ \bibinfo
  {pages} {035}},\ \Eprint {https://arxiv.org/abs/1803.01038} {arXiv:1803.01038
  [hep-ph]} \BibitemShut {NoStop}%
\bibitem [{\citenamefont {Assadullahi}\ and\ \citenamefont
  {Wands}(2009)}]{Assadullahi:2009nf}%
  \BibitemOpen
  \bibfield  {author} {\bibinfo {author} {\bibfnamefont {H.}~\bibnamefont
  {Assadullahi}}\ and\ \bibinfo {author} {\bibfnamefont {D.}~\bibnamefont
  {Wands}},\ }\bibfield  {title} {\bibinfo {title} {{Gravitational waves from
  an early matter era}},\ }\href {https://doi.org/10.1103/PhysRevD.79.083511}
  {\bibfield  {journal} {\bibinfo  {journal} {Phys. Rev. D}\ }\textbf {\bibinfo
  {volume} {79}},\ \bibinfo {pages} {083511} (\bibinfo {year} {2009})},\
  \Eprint {https://arxiv.org/abs/0901.0989} {arXiv:0901.0989 [astro-ph.CO]}
  \BibitemShut {NoStop}%
\bibitem [{\citenamefont {Inomata}\ \emph
  {et~al.}(2019{\natexlab{a}})\citenamefont {Inomata}, \citenamefont {Kohri},
  \citenamefont {Nakama},\ and\ \citenamefont {Terada}}]{Inomata:2019zqy}%
  \BibitemOpen
  \bibfield  {author} {\bibinfo {author} {\bibfnamefont {K.}~\bibnamefont
  {Inomata}}, \bibinfo {author} {\bibfnamefont {K.}~\bibnamefont {Kohri}},
  \bibinfo {author} {\bibfnamefont {T.}~\bibnamefont {Nakama}},\ and\ \bibinfo
  {author} {\bibfnamefont {T.}~\bibnamefont {Terada}},\ }\bibfield  {title}
  {\bibinfo {title} {{Gravitational Waves Induced by Scalar Perturbations
  during a Gradual Transition from an Early Matter Era to the Radiation Era}},\
  }\href {https://doi.org/10.1088/1475-7516/2019/10/071} {\bibfield  {journal}
  {\bibinfo  {journal} {JCAP}\ }\textbf {\bibinfo {volume} {10}},\ \bibinfo
  {pages} {071}},\ \Eprint {https://arxiv.org/abs/1904.12878} {arXiv:1904.12878
  [astro-ph.CO]} \BibitemShut {NoStop}%
\bibitem [{\citenamefont {Inomata}\ \emph
  {et~al.}(2019{\natexlab{b}})\citenamefont {Inomata}, \citenamefont {Kohri},
  \citenamefont {Nakama},\ and\ \citenamefont {Terada}}]{Inomata:2019ivs}%
  \BibitemOpen
  \bibfield  {author} {\bibinfo {author} {\bibfnamefont {K.}~\bibnamefont
  {Inomata}}, \bibinfo {author} {\bibfnamefont {K.}~\bibnamefont {Kohri}},
  \bibinfo {author} {\bibfnamefont {T.}~\bibnamefont {Nakama}},\ and\ \bibinfo
  {author} {\bibfnamefont {T.}~\bibnamefont {Terada}},\ }\bibfield  {title}
  {\bibinfo {title} {{Enhancement of Gravitational Waves Induced by Scalar
  Perturbations due to a Sudden Transition from an Early Matter Era to the
  Radiation Era}},\ }\href {https://doi.org/10.1103/PhysRevD.100.043532}
  {\bibfield  {journal} {\bibinfo  {journal} {Phys. Rev. D}\ }\textbf {\bibinfo
  {volume} {100}},\ \bibinfo {pages} {043532} (\bibinfo {year}
  {2019}{\natexlab{b}})},\ \Eprint {https://arxiv.org/abs/1904.12879}
  {arXiv:1904.12879 [astro-ph.CO]} \BibitemShut {NoStop}%
\bibitem [{\citenamefont {Inomata}\ \emph {et~al.}(2020)\citenamefont
  {Inomata}, \citenamefont {Kawasaki}, \citenamefont {Mukaida}, \citenamefont
  {Terada},\ and\ \citenamefont {Yanagida}}]{Inomata:2020lmk}%
  \BibitemOpen
  \bibfield  {author} {\bibinfo {author} {\bibfnamefont {K.}~\bibnamefont
  {Inomata}}, \bibinfo {author} {\bibfnamefont {M.}~\bibnamefont {Kawasaki}},
  \bibinfo {author} {\bibfnamefont {K.}~\bibnamefont {Mukaida}}, \bibinfo
  {author} {\bibfnamefont {T.}~\bibnamefont {Terada}},\ and\ \bibinfo {author}
  {\bibfnamefont {T.~T.}\ \bibnamefont {Yanagida}},\ }\bibfield  {title}
  {\bibinfo {title} {{Gravitational Wave Production right after a Primordial
  Black Hole Evaporation}},\ }\href
  {https://doi.org/10.1103/PhysRevD.101.123533} {\bibfield  {journal} {\bibinfo
   {journal} {Phys. Rev. D}\ }\textbf {\bibinfo {volume} {101}},\ \bibinfo
  {pages} {123533} (\bibinfo {year} {2020})},\ \Eprint
  {https://arxiv.org/abs/2003.10455} {arXiv:2003.10455 [astro-ph.CO]}
  \BibitemShut {NoStop}%
\bibitem [{\citenamefont {Dalianis}\ and\ \citenamefont
  {Kouvaris}(2020)}]{Dalianis:2020gup}%
  \BibitemOpen
  \bibfield  {author} {\bibinfo {author} {\bibfnamefont {I.}~\bibnamefont
  {Dalianis}}\ and\ \bibinfo {author} {\bibfnamefont {C.}~\bibnamefont
  {Kouvaris}},\ }\bibfield  {title} {\bibinfo {title} {{Gravitational Waves
  from Density Perturbations in an Early Matter Domination Era}},\ }\href@noop
  {} {\  (\bibinfo {year} {2020})},\ \Eprint {https://arxiv.org/abs/2012.09255}
  {arXiv:2012.09255 [astro-ph.CO]} \BibitemShut {NoStop}%
\bibitem [{\citenamefont {Hajkarim}\ and\ \citenamefont
  {Schaffner-Bielich}(2020)}]{Hajkarim:2019nbx}%
  \BibitemOpen
  \bibfield  {author} {\bibinfo {author} {\bibfnamefont {F.}~\bibnamefont
  {Hajkarim}}\ and\ \bibinfo {author} {\bibfnamefont {J.}~\bibnamefont
  {Schaffner-Bielich}},\ }\bibfield  {title} {\bibinfo {title} {{Thermal
  History of the Early Universe and Primordial Gravitational Waves from Induced
  Scalar Perturbations}},\ }\href {https://doi.org/10.1103/PhysRevD.101.043522}
  {\bibfield  {journal} {\bibinfo  {journal} {Phys. Rev. D}\ }\textbf {\bibinfo
  {volume} {101}},\ \bibinfo {pages} {043522} (\bibinfo {year} {2020})},\
  \Eprint {https://arxiv.org/abs/1910.12357} {arXiv:1910.12357 [hep-ph]}
  \BibitemShut {NoStop}%
\bibitem [{\citenamefont {Bhattacharya}\ \emph {et~al.}(2020)\citenamefont
  {Bhattacharya}, \citenamefont {Mohanty},\ and\ \citenamefont
  {Parashari}}]{Bhattacharya:2019bvk}%
  \BibitemOpen
  \bibfield  {author} {\bibinfo {author} {\bibfnamefont {S.}~\bibnamefont
  {Bhattacharya}}, \bibinfo {author} {\bibfnamefont {S.}~\bibnamefont
  {Mohanty}},\ and\ \bibinfo {author} {\bibfnamefont {P.}~\bibnamefont
  {Parashari}},\ }\bibfield  {title} {\bibinfo {title} {{Primordial black holes
  and gravitational waves in nonstandard cosmologies}},\ }\href
  {https://doi.org/10.1103/PhysRevD.102.043522} {\bibfield  {journal} {\bibinfo
   {journal} {Phys. Rev. D}\ }\textbf {\bibinfo {volume} {102}},\ \bibinfo
  {pages} {043522} (\bibinfo {year} {2020})},\ \Eprint
  {https://arxiv.org/abs/1912.01653} {arXiv:1912.01653 [astro-ph.CO]}
  \BibitemShut {NoStop}%
\bibitem [{\citenamefont {Dom\`enech}(2020)}]{Domenech:2019quo}%
  \BibitemOpen
  \bibfield  {author} {\bibinfo {author} {\bibfnamefont {G.}~\bibnamefont
  {Dom\`enech}},\ }\bibfield  {title} {\bibinfo {title} {{Induced gravitational
  waves in a general cosmological background}},\ }\href
  {https://doi.org/10.1142/S0218271820500285} {\bibfield  {journal} {\bibinfo
  {journal} {Int. J. Mod. Phys. D}\ }\textbf {\bibinfo {volume} {29}},\
  \bibinfo {pages} {2050028} (\bibinfo {year} {2020})},\ \Eprint
  {https://arxiv.org/abs/1912.05583} {arXiv:1912.05583 [gr-qc]} \BibitemShut
  {NoStop}%
\bibitem [{\citenamefont {Dom\`enech}\ \emph {et~al.}(2020)\citenamefont
  {Dom\`enech}, \citenamefont {Pi},\ and\ \citenamefont
  {Sasaki}}]{Domenech:2020kqm}%
  \BibitemOpen
  \bibfield  {author} {\bibinfo {author} {\bibfnamefont {G.}~\bibnamefont
  {Dom\`enech}}, \bibinfo {author} {\bibfnamefont {S.}~\bibnamefont {Pi}},\
  and\ \bibinfo {author} {\bibfnamefont {M.}~\bibnamefont {Sasaki}},\
  }\bibfield  {title} {\bibinfo {title} {{Induced gravitational waves as a
  probe of thermal history of the universe}},\ }\href
  {https://doi.org/10.1088/1475-7516/2020/08/017} {\bibfield  {journal}
  {\bibinfo  {journal} {JCAP}\ }\textbf {\bibinfo {volume} {08}},\ \bibinfo
  {pages} {017}},\ \Eprint {https://arxiv.org/abs/2005.12314} {arXiv:2005.12314
  [gr-qc]} \BibitemShut {NoStop}%
\bibitem [{\citenamefont {Dalianis}\ and\ \citenamefont
  {Kritos}(2021)}]{Dalianis:2020cla}%
  \BibitemOpen
  \bibfield  {author} {\bibinfo {author} {\bibfnamefont {I.}~\bibnamefont
  {Dalianis}}\ and\ \bibinfo {author} {\bibfnamefont {K.}~\bibnamefont
  {Kritos}},\ }\bibfield  {title} {\bibinfo {title} {{Exploring the Spectral
  Shape of Gravitational Waves Induced by Primordial Scalar Perturbations and
  Connection with the Primordial Black Hole Scenarios}},\ }\href
  {https://doi.org/10.1103/PhysRevD.103.023505} {\bibfield  {journal} {\bibinfo
   {journal} {Phys. Rev. D}\ }\textbf {\bibinfo {volume} {103}},\ \bibinfo
  {pages} {023505} (\bibinfo {year} {2021})},\ \Eprint
  {https://arxiv.org/abs/2007.07915} {arXiv:2007.07915 [astro-ph.CO]}
  \BibitemShut {NoStop}%
\bibitem [{\citenamefont {Abe}\ \emph {et~al.}(2020)\citenamefont {Abe},
  \citenamefont {Tada},\ and\ \citenamefont {Ueda}}]{Abe:2020sqb}%
  \BibitemOpen
  \bibfield  {author} {\bibinfo {author} {\bibfnamefont {K.~T.}\ \bibnamefont
  {Abe}}, \bibinfo {author} {\bibfnamefont {Y.}~\bibnamefont {Tada}},\ and\
  \bibinfo {author} {\bibfnamefont {I.}~\bibnamefont {Ueda}},\ }\bibfield
  {title} {\bibinfo {title} {{Induced gravitational waves as a cosmological
  probe of the sound speed during the QCD phase transition}},\ }\href@noop {}
  {\  (\bibinfo {year} {2020})},\ \Eprint {https://arxiv.org/abs/2010.06193}
  {arXiv:2010.06193 [astro-ph.CO]} \BibitemShut {NoStop}%
\bibitem [{\citenamefont {Witkowski}\ \emph {et~al.}(2021)\citenamefont
  {Witkowski}, \citenamefont {Dom\`enech}, \citenamefont {Fumagalli},\ and\
  \citenamefont {Renaux-Petel}}]{Witkowski:2021raz}%
  \BibitemOpen
  \bibfield  {author} {\bibinfo {author} {\bibfnamefont {L.~T.}\ \bibnamefont
  {Witkowski}}, \bibinfo {author} {\bibfnamefont {G.}~\bibnamefont
  {Dom\`enech}}, \bibinfo {author} {\bibfnamefont {J.}~\bibnamefont
  {Fumagalli}},\ and\ \bibinfo {author} {\bibfnamefont {S.}~\bibnamefont
  {Renaux-Petel}},\ }\bibfield  {title} {\bibinfo {title} {{Expansion
  history-dependent oscillations in the scalar-induced gravitational wave
  background}},\ }\href@noop {} {\  (\bibinfo {year} {2021})},\ \Eprint
  {https://arxiv.org/abs/2110.09480} {arXiv:2110.09480 [astro-ph.CO]}
  \BibitemShut {NoStop}%
\bibitem [{\citenamefont {Assadullahi}\ and\ \citenamefont
  {Wands}(2010)}]{Assadullahi:2009jc}%
  \BibitemOpen
  \bibfield  {author} {\bibinfo {author} {\bibfnamefont {H.}~\bibnamefont
  {Assadullahi}}\ and\ \bibinfo {author} {\bibfnamefont {D.}~\bibnamefont
  {Wands}},\ }\bibfield  {title} {\bibinfo {title} {{Constraints on primordial
  density perturbations from induced gravitational waves}},\ }\href
  {https://doi.org/10.1103/PhysRevD.81.023527} {\bibfield  {journal} {\bibinfo
  {journal} {Phys. Rev. D}\ }\textbf {\bibinfo {volume} {81}},\ \bibinfo
  {pages} {023527} (\bibinfo {year} {2010})},\ \Eprint
  {https://arxiv.org/abs/0907.4073} {arXiv:0907.4073 [astro-ph.CO]}
  \BibitemShut {NoStop}%
\bibitem [{\citenamefont {Bugaev}\ and\ \citenamefont
  {Klimai}(2010{\natexlab{a}})}]{Bugaev:2009zh}%
  \BibitemOpen
  \bibfield  {author} {\bibinfo {author} {\bibfnamefont {E.}~\bibnamefont
  {Bugaev}}\ and\ \bibinfo {author} {\bibfnamefont {P.}~\bibnamefont
  {Klimai}},\ }\bibfield  {title} {\bibinfo {title} {{Induced gravitational
  wave background and primordial black holes}},\ }\href
  {https://doi.org/10.1103/PhysRevD.81.023517} {\bibfield  {journal} {\bibinfo
  {journal} {Phys. Rev. D}\ }\textbf {\bibinfo {volume} {81}},\ \bibinfo
  {pages} {023517} (\bibinfo {year} {2010}{\natexlab{a}})},\ \Eprint
  {https://arxiv.org/abs/0908.0664} {arXiv:0908.0664 [astro-ph.CO]}
  \BibitemShut {NoStop}%
\bibitem [{\citenamefont {Bugaev}\ and\ \citenamefont
  {Klimai}(2010{\natexlab{b}})}]{Bugaev:2009kq}%
  \BibitemOpen
  \bibfield  {author} {\bibinfo {author} {\bibfnamefont {E.~V.}\ \bibnamefont
  {Bugaev}}\ and\ \bibinfo {author} {\bibfnamefont {P.~A.}\ \bibnamefont
  {Klimai}},\ }\bibfield  {title} {\bibinfo {title} {{Bound on induced
  gravitational wave background from primordial black holes}},\ }\href
  {https://doi.org/10.1134/S0021364010010017} {\bibfield  {journal} {\bibinfo
  {journal} {JETP Lett.}\ }\textbf {\bibinfo {volume} {91}},\ \bibinfo {pages}
  {1} (\bibinfo {year} {2010}{\natexlab{b}})},\ \Eprint
  {https://arxiv.org/abs/0911.0611} {arXiv:0911.0611 [astro-ph.CO]}
  \BibitemShut {NoStop}%
\bibitem [{\citenamefont {Bugaev}\ and\ \citenamefont
  {Klimai}(2011)}]{Bugaev:2010bb}%
  \BibitemOpen
  \bibfield  {author} {\bibinfo {author} {\bibfnamefont {E.}~\bibnamefont
  {Bugaev}}\ and\ \bibinfo {author} {\bibfnamefont {P.}~\bibnamefont
  {Klimai}},\ }\bibfield  {title} {\bibinfo {title} {{Constraints on the
  induced gravitational wave background from primordial black holes}},\ }\href
  {https://doi.org/10.1103/PhysRevD.83.083521} {\bibfield  {journal} {\bibinfo
  {journal} {Phys. Rev. D}\ }\textbf {\bibinfo {volume} {83}},\ \bibinfo
  {pages} {083521} (\bibinfo {year} {2011})},\ \Eprint
  {https://arxiv.org/abs/1012.4697} {arXiv:1012.4697 [astro-ph.CO]}
  \BibitemShut {NoStop}%
\bibitem [{\citenamefont {Inomata}\ and\ \citenamefont
  {Nakama}(2019)}]{Inomata:2018epa}%
  \BibitemOpen
  \bibfield  {author} {\bibinfo {author} {\bibfnamefont {K.}~\bibnamefont
  {Inomata}}\ and\ \bibinfo {author} {\bibfnamefont {T.}~\bibnamefont
  {Nakama}},\ }\bibfield  {title} {\bibinfo {title} {{Gravitational waves
  induced by scalar perturbations as probes of the small-scale primordial
  spectrum}},\ }\href {https://doi.org/10.1103/PhysRevD.99.043511} {\bibfield
  {journal} {\bibinfo  {journal} {Phys. Rev. D}\ }\textbf {\bibinfo {volume}
  {99}},\ \bibinfo {pages} {043511} (\bibinfo {year} {2019})},\ \Eprint
  {https://arxiv.org/abs/1812.00674} {arXiv:1812.00674 [astro-ph.CO]}
  \BibitemShut {NoStop}%
\bibitem [{\citenamefont {Malik}\ and\ \citenamefont
  {Wands}(2009)}]{Malik:2008im}%
  \BibitemOpen
  \bibfield  {author} {\bibinfo {author} {\bibfnamefont {K.~A.}\ \bibnamefont
  {Malik}}\ and\ \bibinfo {author} {\bibfnamefont {D.}~\bibnamefont {Wands}},\
  }\bibfield  {title} {\bibinfo {title} {{Cosmological perturbations}},\ }\href
  {https://doi.org/10.1016/j.physrep.2009.03.001} {\bibfield  {journal}
  {\bibinfo  {journal} {Phys. Rept.}\ }\textbf {\bibinfo {volume} {475}},\
  \bibinfo {pages} {1} (\bibinfo {year} {2009})},\ \Eprint
  {https://arxiv.org/abs/0809.4944} {arXiv:0809.4944 [astro-ph]} \BibitemShut
  {NoStop}%
\bibitem [{\citenamefont {Mukhanov}(2005)}]{Mukhanov:2005sc}%
  \BibitemOpen
  \bibfield  {author} {\bibinfo {author} {\bibfnamefont {V.}~\bibnamefont
  {Mukhanov}},\ }\href@noop {} {\emph {\bibinfo {title} {{Physical Foundations
  of Cosmology}}}}\ (\bibinfo  {publisher} {Cambridge University Press},\
  \bibinfo {address} {Oxford},\ \bibinfo {year} {2005})\BibitemShut {NoStop}%
\bibitem [{\citenamefont {Kodama}\ and\ \citenamefont
  {Sasaki}(1987)}]{Kodama:1986ud}%
  \BibitemOpen
  \bibfield  {author} {\bibinfo {author} {\bibfnamefont {H.}~\bibnamefont
  {Kodama}}\ and\ \bibinfo {author} {\bibfnamefont {M.}~\bibnamefont
  {Sasaki}},\ }\bibfield  {title} {\bibinfo {title} {{Evolution of Isocurvature
  Perturbations. 2. Radiation Dust Universe}},\ }\href
  {https://doi.org/10.1142/S0217751X8700020X} {\bibfield  {journal} {\bibinfo
  {journal} {Int. J. Mod. Phys. A}\ }\textbf {\bibinfo {volume} {2}},\ \bibinfo
  {pages} {491} (\bibinfo {year} {1987})}\BibitemShut {NoStop}%
\bibitem [{\citenamefont {Espinosa}\ \emph {et~al.}(2018)\citenamefont
  {Espinosa}, \citenamefont {Racco},\ and\ \citenamefont
  {Riotto}}]{Espinosa:2018eve}%
  \BibitemOpen
  \bibfield  {author} {\bibinfo {author} {\bibfnamefont {J.~R.}\ \bibnamefont
  {Espinosa}}, \bibinfo {author} {\bibfnamefont {D.}~\bibnamefont {Racco}},\
  and\ \bibinfo {author} {\bibfnamefont {A.}~\bibnamefont {Riotto}},\
  }\bibfield  {title} {\bibinfo {title} {{A Cosmological Signature of the SM
  Higgs Instability: Gravitational Waves}},\ }\href
  {https://doi.org/10.1088/1475-7516/2018/09/012} {\bibfield  {journal}
  {\bibinfo  {journal} {JCAP}\ }\textbf {\bibinfo {volume} {09}},\ \bibinfo
  {pages} {012}},\ \Eprint {https://arxiv.org/abs/1804.07732} {arXiv:1804.07732
  [hep-ph]} \BibitemShut {NoStop}%
\bibitem [{\citenamefont {Kohri}\ and\ \citenamefont
  {Terada}(2018)}]{Kohri:2018awv}%
  \BibitemOpen
  \bibfield  {author} {\bibinfo {author} {\bibfnamefont {K.}~\bibnamefont
  {Kohri}}\ and\ \bibinfo {author} {\bibfnamefont {T.}~\bibnamefont {Terada}},\
  }\bibfield  {title} {\bibinfo {title} {{Semianalytic calculation of
  gravitational wave spectrum nonlinearly induced from primordial curvature
  perturbations}},\ }\href {https://doi.org/10.1103/PhysRevD.97.123532}
  {\bibfield  {journal} {\bibinfo  {journal} {Phys. Rev. D}\ }\textbf {\bibinfo
  {volume} {97}},\ \bibinfo {pages} {123532} (\bibinfo {year} {2018})},\
  \Eprint {https://arxiv.org/abs/1804.08577} {arXiv:1804.08577 [gr-qc]}
  \BibitemShut {NoStop}%
\bibitem [{\citenamefont {Maggiore}(2007)}]{Maggiore:1900zz}%
  \BibitemOpen
  \bibfield  {author} {\bibinfo {author} {\bibfnamefont {M.}~\bibnamefont
  {Maggiore}},\ }\href@noop {} {\emph {\bibinfo {title} {{Gravitational Waves.
  Vol. 1: Theory and Experiments}}}},\ Oxford Master Series in Physics\
  (\bibinfo  {publisher} {Oxford University Press},\ \bibinfo {year}
  {2007})\BibitemShut {NoStop}%
\bibitem [{\citenamefont {Inomata}\ \emph {et~al.}(2017)\citenamefont
  {Inomata}, \citenamefont {Kawasaki}, \citenamefont {Mukaida}, \citenamefont
  {Tada},\ and\ \citenamefont {Yanagida}}]{Inomata:2016rbd}%
  \BibitemOpen
  \bibfield  {author} {\bibinfo {author} {\bibfnamefont {K.}~\bibnamefont
  {Inomata}}, \bibinfo {author} {\bibfnamefont {M.}~\bibnamefont {Kawasaki}},
  \bibinfo {author} {\bibfnamefont {K.}~\bibnamefont {Mukaida}}, \bibinfo
  {author} {\bibfnamefont {Y.}~\bibnamefont {Tada}},\ and\ \bibinfo {author}
  {\bibfnamefont {T.~T.}\ \bibnamefont {Yanagida}},\ }\bibfield  {title}
  {\bibinfo {title} {{Inflationary primordial black holes for the LIGO
  gravitational wave events and pulsar timing array experiments}},\ }\href
  {https://doi.org/10.1103/PhysRevD.95.123510} {\bibfield  {journal} {\bibinfo
  {journal} {Phys. Rev. D}\ }\textbf {\bibinfo {volume} {95}},\ \bibinfo
  {pages} {123510} (\bibinfo {year} {2017})},\ \Eprint
  {https://arxiv.org/abs/1611.06130} {arXiv:1611.06130 [astro-ph.CO]}
  \BibitemShut {NoStop}%
\bibitem [{\citenamefont {Aghanim}\ \emph {et~al.}(2020)\citenamefont {Aghanim}
  \emph {et~al.}}]{Aghanim:2018eyx}%
  \BibitemOpen
  \bibfield  {author} {\bibinfo {author} {\bibfnamefont {N.}~\bibnamefont
  {Aghanim}} \emph {et~al.} (\bibinfo {collaboration} {Planck}),\ }\bibfield
  {title} {\bibinfo {title} {{Planck 2018 results. VI. Cosmological
  parameters}},\ }\href {https://doi.org/10.1051/0004-6361/201833910}
  {\bibfield  {journal} {\bibinfo  {journal} {Astron. Astrophys.}\ }\textbf
  {\bibinfo {volume} {641}},\ \bibinfo {pages} {A6} (\bibinfo {year} {2020})},\
  \Eprint {https://arxiv.org/abs/1807.06209} {arXiv:1807.06209 [astro-ph.CO]}
  \BibitemShut {NoStop}%
\bibitem [{\citenamefont {Hu}\ and\ \citenamefont
  {Sugiyama}(1995)}]{Hu:1994jd}%
  \BibitemOpen
  \bibfield  {author} {\bibinfo {author} {\bibfnamefont {W.}~\bibnamefont
  {Hu}}\ and\ \bibinfo {author} {\bibfnamefont {N.}~\bibnamefont {Sugiyama}},\
  }\bibfield  {title} {\bibinfo {title} {{Toward understanding CMB anisotropies
  and their implications}},\ }\href {https://doi.org/10.1103/PhysRevD.51.2599}
  {\bibfield  {journal} {\bibinfo  {journal} {Phys. Rev. D}\ }\textbf {\bibinfo
  {volume} {51}},\ \bibinfo {pages} {2599} (\bibinfo {year} {1995})},\ \Eprint
  {https://arxiv.org/abs/astro-ph/9411008} {arXiv:astro-ph/9411008}
  \BibitemShut {NoStop}%
\bibitem [{\citenamefont {Hu}(1995)}]{Hu:1995em}%
  \BibitemOpen
  \bibfield  {author} {\bibinfo {author} {\bibfnamefont {W.~T.}\ \bibnamefont
  {Hu}},\ }\emph {\bibinfo {title} {{Wandering in the Background: A CMB
  Explorer}}},\ \href@noop {} {\bibinfo {type} {Other thesis}} (\bibinfo {year}
  {1995}),\ \Eprint {https://arxiv.org/abs/astro-ph/9508126}
  {arXiv:astro-ph/9508126} \BibitemShut {NoStop}%
\bibitem [{\citenamefont {Pi}\ and\ \citenamefont {Sasaki}(2020)}]{Pi:2020otn}%
  \BibitemOpen
  \bibfield  {author} {\bibinfo {author} {\bibfnamefont {S.}~\bibnamefont
  {Pi}}\ and\ \bibinfo {author} {\bibfnamefont {M.}~\bibnamefont {Sasaki}},\
  }\bibfield  {title} {\bibinfo {title} {{Gravitational Waves Induced by Scalar
  Perturbations with a Lognormal Peak}},\ }\href
  {https://doi.org/10.1088/1475-7516/2020/09/037} {\bibfield  {journal}
  {\bibinfo  {journal} {JCAP}\ }\textbf {\bibinfo {volume} {09}},\ \bibinfo
  {pages} {037}},\ \Eprint {https://arxiv.org/abs/2005.12306} {arXiv:2005.12306
  [gr-qc]} \BibitemShut {NoStop}%
\bibitem [{\citenamefont {Caprini}\ and\ \citenamefont
  {Figueroa}(2018)}]{Caprini:2018mtu}%
  \BibitemOpen
  \bibfield  {author} {\bibinfo {author} {\bibfnamefont {C.}~\bibnamefont
  {Caprini}}\ and\ \bibinfo {author} {\bibfnamefont {D.~G.}\ \bibnamefont
  {Figueroa}},\ }\bibfield  {title} {\bibinfo {title} {{Cosmological
  Backgrounds of Gravitational Waves}},\ }\href
  {https://doi.org/10.1088/1361-6382/aac608} {\bibfield  {journal} {\bibinfo
  {journal} {Class. Quant. Grav.}\ }\textbf {\bibinfo {volume} {35}},\ \bibinfo
  {pages} {163001} (\bibinfo {year} {2018})},\ \Eprint
  {https://arxiv.org/abs/1801.04268} {arXiv:1801.04268 [astro-ph.CO]}
  \BibitemShut {NoStop}%
\bibitem [{\citenamefont {Cai}\ \emph {et~al.}(2019{\natexlab{b}})\citenamefont
  {Cai}, \citenamefont {Pi},\ and\ \citenamefont {Sasaki}}]{Cai:2018dig}%
  \BibitemOpen
  \bibfield  {author} {\bibinfo {author} {\bibfnamefont {R.-g.}\ \bibnamefont
  {Cai}}, \bibinfo {author} {\bibfnamefont {S.}~\bibnamefont {Pi}},\ and\
  \bibinfo {author} {\bibfnamefont {M.}~\bibnamefont {Sasaki}},\ }\bibfield
  {title} {\bibinfo {title} {{Gravitational Waves Induced by non-Gaussian
  Scalar Perturbations}},\ }\href
  {https://doi.org/10.1103/PhysRevLett.122.201101} {\bibfield  {journal}
  {\bibinfo  {journal} {Phys. Rev. Lett.}\ }\textbf {\bibinfo {volume} {122}},\
  \bibinfo {pages} {201101} (\bibinfo {year} {2019}{\natexlab{b}})},\ \Eprint
  {https://arxiv.org/abs/1810.11000} {arXiv:1810.11000 [astro-ph.CO]}
  \BibitemShut {NoStop}%
\bibitem [{\citenamefont {Unal}(2019)}]{Unal:2018yaa}%
  \BibitemOpen
  \bibfield  {author} {\bibinfo {author} {\bibfnamefont {C.}~\bibnamefont
  {Unal}},\ }\bibfield  {title} {\bibinfo {title} {{Imprints of Primordial
  Non-Gaussianity on Gravitational Wave Spectrum}},\ }\href
  {https://doi.org/10.1103/PhysRevD.99.041301} {\bibfield  {journal} {\bibinfo
  {journal} {Phys. Rev. D}\ }\textbf {\bibinfo {volume} {99}},\ \bibinfo
  {pages} {041301} (\bibinfo {year} {2019})},\ \Eprint
  {https://arxiv.org/abs/1811.09151} {arXiv:1811.09151 [astro-ph.CO]}
  \BibitemShut {NoStop}%
\bibitem [{\citenamefont {Atal}\ and\ \citenamefont
  {Dom\`enech}(2021)}]{Atal:2021jyo}%
  \BibitemOpen
  \bibfield  {author} {\bibinfo {author} {\bibfnamefont {V.}~\bibnamefont
  {Atal}}\ and\ \bibinfo {author} {\bibfnamefont {G.}~\bibnamefont
  {Dom\`enech}},\ }\bibfield  {title} {\bibinfo {title} {{Probing
  non-Gaussianities with the high frequency tail of induced gravitational
  waves}},\ }\href@noop {} {\  (\bibinfo {year} {2021})},\ \Eprint
  {https://arxiv.org/abs/2103.01056} {arXiv:2103.01056 [astro-ph.CO]}
  \BibitemShut {NoStop}%
\bibitem [{\citenamefont {Adshead}\ \emph {et~al.}(2021)\citenamefont
  {Adshead}, \citenamefont {Lozanov},\ and\ \citenamefont
  {Weiner}}]{Adshead:2021hnm}%
  \BibitemOpen
  \bibfield  {author} {\bibinfo {author} {\bibfnamefont {P.}~\bibnamefont
  {Adshead}}, \bibinfo {author} {\bibfnamefont {K.~D.}\ \bibnamefont
  {Lozanov}},\ and\ \bibinfo {author} {\bibfnamefont {Z.~J.}\ \bibnamefont
  {Weiner}},\ }\bibfield  {title} {\bibinfo {title} {{Non-Gaussianity and the
  induced gravitational wave background}},\ }\href@noop {} {\  (\bibinfo {year}
  {2021})},\ \Eprint {https://arxiv.org/abs/2105.01659} {arXiv:2105.01659
  [astro-ph.CO]} \BibitemShut {NoStop}%
\bibitem [{\citenamefont {Schmitz}(2021)}]{Schmitz:2020syl}%
  \BibitemOpen
  \bibfield  {author} {\bibinfo {author} {\bibfnamefont {K.}~\bibnamefont
  {Schmitz}},\ }\bibfield  {title} {\bibinfo {title} {{New Sensitivity Curves
  for Gravitational-Wave Signals from Cosmological Phase Transitions}},\ }\href
  {https://doi.org/10.1007/JHEP01(2021)097} {\bibfield  {journal} {\bibinfo
  {journal} {JHEP}\ }\textbf {\bibinfo {volume} {01}},\ \bibinfo {pages}
  {097}},\ \Eprint {https://arxiv.org/abs/2002.04615} {arXiv:2002.04615
  [hep-ph]} \BibitemShut {NoStop}%
\bibitem [{\citenamefont {Robson}\ \emph {et~al.}(2019)\citenamefont {Robson},
  \citenamefont {Cornish},\ and\ \citenamefont {Liu}}]{Robson:2018ifk}%
  \BibitemOpen
  \bibfield  {author} {\bibinfo {author} {\bibfnamefont {T.}~\bibnamefont
  {Robson}}, \bibinfo {author} {\bibfnamefont {N.~J.}\ \bibnamefont
  {Cornish}},\ and\ \bibinfo {author} {\bibfnamefont {C.}~\bibnamefont {Liu}},\
  }\bibfield  {title} {\bibinfo {title} {{The construction and use of LISA
  sensitivity curves}},\ }\href {https://doi.org/10.1088/1361-6382/ab1101}
  {\bibfield  {journal} {\bibinfo  {journal} {Class. Quant. Grav.}\ }\textbf
  {\bibinfo {volume} {36}},\ \bibinfo {pages} {105011} (\bibinfo {year}
  {2019})},\ \Eprint {https://arxiv.org/abs/1803.01944} {arXiv:1803.01944
  [astro-ph.HE]} \BibitemShut {NoStop}%
\bibitem [{\citenamefont {Kuroyanagi}\ \emph {et~al.}(2015)\citenamefont
  {Kuroyanagi}, \citenamefont {Nakayama},\ and\ \citenamefont
  {Yokoyama}}]{Kuroyanagi:2014qza}%
  \BibitemOpen
  \bibfield  {author} {\bibinfo {author} {\bibfnamefont {S.}~\bibnamefont
  {Kuroyanagi}}, \bibinfo {author} {\bibfnamefont {K.}~\bibnamefont
  {Nakayama}},\ and\ \bibinfo {author} {\bibfnamefont {J.}~\bibnamefont
  {Yokoyama}},\ }\bibfield  {title} {\bibinfo {title} {{Prospects of
  determination of reheating temperature after inflation by DECIGO}},\ }\href
  {https://doi.org/10.1093/ptep/ptu176} {\bibfield  {journal} {\bibinfo
  {journal} {PTEP}\ }\textbf {\bibinfo {volume} {2015}},\ \bibinfo {pages}
  {013E02} (\bibinfo {year} {2015})},\ \Eprint
  {https://arxiv.org/abs/1410.6618} {arXiv:1410.6618 [astro-ph.CO]}
  \BibitemShut {NoStop}%
\bibitem [{ET:()}]{ET:sensitivities}%
  \BibitemOpen
  \href@noop {} {}\bibinfo {howpublished} {Einstein Telescope Project, ET
  sensitivities page,
  {\url{http://www.et-gw.eu/index.php/etsensitivities}}.}\BibitemShut {Stop}%
\bibitem [{\citenamefont {Smith}\ \emph {et~al.}(2006)\citenamefont {Smith},
  \citenamefont {Pierpaoli},\ and\ \citenamefont
  {Kamionkowski}}]{Smith:2006nka}%
  \BibitemOpen
  \bibfield  {author} {\bibinfo {author} {\bibfnamefont {T.~L.}\ \bibnamefont
  {Smith}}, \bibinfo {author} {\bibfnamefont {E.}~\bibnamefont {Pierpaoli}},\
  and\ \bibinfo {author} {\bibfnamefont {M.}~\bibnamefont {Kamionkowski}},\
  }\bibfield  {title} {\bibinfo {title} {{A new cosmic microwave background
  constraint to primordial gravitational waves}},\ }\href
  {https://doi.org/10.1103/PhysRevLett.97.021301} {\bibfield  {journal}
  {\bibinfo  {journal} {Phys. Rev. Lett.}\ }\textbf {\bibinfo {volume} {97}},\
  \bibinfo {pages} {021301} (\bibinfo {year} {2006})},\ \Eprint
  {https://arxiv.org/abs/astro-ph/0603144} {arXiv:astro-ph/0603144}
  \BibitemShut {NoStop}%
\bibitem [{\citenamefont {Pagano}\ \emph {et~al.}(2016)\citenamefont {Pagano},
  \citenamefont {Salvati},\ and\ \citenamefont {Melchiorri}}]{Pagano:2015hma}%
  \BibitemOpen
  \bibfield  {author} {\bibinfo {author} {\bibfnamefont {L.}~\bibnamefont
  {Pagano}}, \bibinfo {author} {\bibfnamefont {L.}~\bibnamefont {Salvati}},\
  and\ \bibinfo {author} {\bibfnamefont {A.}~\bibnamefont {Melchiorri}},\
  }\bibfield  {title} {\bibinfo {title} {{New constraints on primordial
  gravitational waves from Planck 2015}},\ }\href
  {https://doi.org/10.1016/j.physletb.2016.07.078} {\bibfield  {journal}
  {\bibinfo  {journal} {Phys. Lett. B}\ }\textbf {\bibinfo {volume} {760}},\
  \bibinfo {pages} {823} (\bibinfo {year} {2016})},\ \Eprint
  {https://arxiv.org/abs/1508.02393} {arXiv:1508.02393 [astro-ph.CO]}
  \BibitemShut {NoStop}%
\bibitem [{\citenamefont {Arzoumanian}\ \emph {et~al.}(2020)\citenamefont
  {Arzoumanian} \emph {et~al.}}]{Arzoumanian:2020vkk}%
  \BibitemOpen
  \bibfield  {author} {\bibinfo {author} {\bibfnamefont {Z.}~\bibnamefont
  {Arzoumanian}} \emph {et~al.} (\bibinfo {collaboration} {NANOGrav}),\
  }\bibfield  {title} {\bibinfo {title} {{The NANOGrav 12.5 yr Data Set: Search
  for an Isotropic Stochastic Gravitational-wave Background}},\ }\href
  {https://doi.org/10.3847/2041-8213/abd401} {\bibfield  {journal} {\bibinfo
  {journal} {Astrophys. J. Lett.}\ }\textbf {\bibinfo {volume} {905}},\
  \bibinfo {pages} {L34} (\bibinfo {year} {2020})},\ \Eprint
  {https://arxiv.org/abs/2009.04496} {arXiv:2009.04496 [astro-ph.HE]}
  \BibitemShut {NoStop}%
\bibitem [{\citenamefont {Abbott}\ \emph {et~al.}(2018)\citenamefont {Abbott}
  \emph {et~al.}}]{KAGRA:2013rdx}%
  \BibitemOpen
  \bibfield  {author} {\bibinfo {author} {\bibfnamefont {B.~P.}\ \bibnamefont
  {Abbott}} \emph {et~al.} (\bibinfo {collaboration} {KAGRA, LIGO Scientific,
  Virgo, VIRGO}),\ }\bibfield  {title} {\bibinfo {title} {{Prospects for
  observing and localizing gravitational-wave transients with Advanced LIGO,
  Advanced Virgo and KAGRA}},\ }\href
  {https://doi.org/10.1007/s41114-020-00026-9} {\bibfield  {journal} {\bibinfo
  {journal} {Living Rev. Rel.}\ }\textbf {\bibinfo {volume} {21}},\ \bibinfo
  {pages} {3} (\bibinfo {year} {2018})},\ \Eprint
  {https://arxiv.org/abs/1304.0670} {arXiv:1304.0670 [gr-qc]} \BibitemShut
  {NoStop}%
\bibitem [{\citenamefont {Blas}\ and\ \citenamefont
  {Jenkins}(2021)}]{Blas:2021mqw}%
  \BibitemOpen
  \bibfield  {author} {\bibinfo {author} {\bibfnamefont {D.}~\bibnamefont
  {Blas}}\ and\ \bibinfo {author} {\bibfnamefont {A.~C.}\ \bibnamefont
  {Jenkins}},\ }\bibfield  {title} {\bibinfo {title} {{Bridging the $\mu$Hz gap
  in the gravitational-wave landscape with binary resonance}},\ }\href@noop {}
  {\  (\bibinfo {year} {2021})},\ \Eprint {https://arxiv.org/abs/2107.04601}
  {arXiv:2107.04601 [astro-ph.CO]} \BibitemShut {NoStop}%
\bibitem [{\citenamefont {Gow}\ \emph {et~al.}(2021)\citenamefont {Gow},
  \citenamefont {Byrnes}, \citenamefont {Cole},\ and\ \citenamefont
  {Young}}]{Gow:2020bzo}%
  \BibitemOpen
  \bibfield  {author} {\bibinfo {author} {\bibfnamefont {A.~D.}\ \bibnamefont
  {Gow}}, \bibinfo {author} {\bibfnamefont {C.~T.}\ \bibnamefont {Byrnes}},
  \bibinfo {author} {\bibfnamefont {P.~S.}\ \bibnamefont {Cole}},\ and\
  \bibinfo {author} {\bibfnamefont {S.}~\bibnamefont {Young}},\ }\bibfield
  {title} {\bibinfo {title} {{The power spectrum on small scales: Robust
  constraints and comparing PBH methodologies}},\ }\href
  {https://doi.org/10.1088/1475-7516/2021/02/002} {\bibfield  {journal}
  {\bibinfo  {journal} {JCAP}\ }\textbf {\bibinfo {volume} {02}},\ \bibinfo
  {pages} {002}},\ \Eprint {https://arxiv.org/abs/2008.03289} {arXiv:2008.03289
  [astro-ph.CO]} \BibitemShut {NoStop}%
\bibitem [{\citenamefont {Kavanagh}\ \emph {et~al.}(2021)\citenamefont
  {Kavanagh}, \citenamefont {Edwards}, \citenamefont {Visinelli},\ and\
  \citenamefont {Weniger}}]{Kavanagh:2020gcy}%
  \BibitemOpen
  \bibfield  {author} {\bibinfo {author} {\bibfnamefont {B.~J.}\ \bibnamefont
  {Kavanagh}}, \bibinfo {author} {\bibfnamefont {T.~D.~P.}\ \bibnamefont
  {Edwards}}, \bibinfo {author} {\bibfnamefont {L.}~\bibnamefont {Visinelli}},\
  and\ \bibinfo {author} {\bibfnamefont {C.}~\bibnamefont {Weniger}},\
  }\bibfield  {title} {\bibinfo {title} {{Stellar disruption of axion
  miniclusters in the Milky~Way}},\ }\href
  {https://doi.org/10.1103/PhysRevD.104.063038} {\bibfield  {journal} {\bibinfo
   {journal} {Phys. Rev. D}\ }\textbf {\bibinfo {volume} {104}},\ \bibinfo
  {pages} {063038} (\bibinfo {year} {2021})},\ \Eprint
  {https://arxiv.org/abs/2011.05377} {arXiv:2011.05377 [astro-ph.GA]}
  \BibitemShut {NoStop}%
\bibitem [{\citenamefont {Hogan}\ and\ \citenamefont
  {Rees}(1988)}]{Hogan:1988mp}%
  \BibitemOpen
  \bibfield  {author} {\bibinfo {author} {\bibfnamefont {C.~J.}\ \bibnamefont
  {Hogan}}\ and\ \bibinfo {author} {\bibfnamefont {M.~J.}\ \bibnamefont
  {Rees}},\ }\bibfield  {title} {\bibinfo {title} {{AXION MINICLUSTERS}},\
  }\href {https://doi.org/10.1016/0370-2693(88)91655-3} {\bibfield  {journal}
  {\bibinfo  {journal} {Phys. Lett. B}\ }\textbf {\bibinfo {volume} {205}},\
  \bibinfo {pages} {228} (\bibinfo {year} {1988})}\BibitemShut {NoStop}%
\bibitem [{\citenamefont {Fairbairn}\ \emph {et~al.}(2018)\citenamefont
  {Fairbairn}, \citenamefont {Marsh}, \citenamefont {Quevillon},\ and\
  \citenamefont {Rozier}}]{Fairbairn:2017sil}%
  \BibitemOpen
  \bibfield  {author} {\bibinfo {author} {\bibfnamefont {M.}~\bibnamefont
  {Fairbairn}}, \bibinfo {author} {\bibfnamefont {D.~J.~E.}\ \bibnamefont
  {Marsh}}, \bibinfo {author} {\bibfnamefont {J.}~\bibnamefont {Quevillon}},\
  and\ \bibinfo {author} {\bibfnamefont {S.}~\bibnamefont {Rozier}},\
  }\bibfield  {title} {\bibinfo {title} {{Structure formation and microlensing
  with axion miniclusters}},\ }\href
  {https://doi.org/10.1103/PhysRevD.97.083502} {\bibfield  {journal} {\bibinfo
  {journal} {Phys. Rev. D}\ }\textbf {\bibinfo {volume} {97}},\ \bibinfo
  {pages} {083502} (\bibinfo {year} {2018})},\ \Eprint
  {https://arxiv.org/abs/1707.03310} {arXiv:1707.03310 [astro-ph.CO]}
  \BibitemShut {NoStop}%
\bibitem [{\citenamefont {Eggemeier}\ \emph {et~al.}(2020)\citenamefont
  {Eggemeier}, \citenamefont {Redondo}, \citenamefont {Dolag}, \citenamefont
  {Niemeyer},\ and\ \citenamefont {Vaquero}}]{Eggemeier:2019khm}%
  \BibitemOpen
  \bibfield  {author} {\bibinfo {author} {\bibfnamefont {B.}~\bibnamefont
  {Eggemeier}}, \bibinfo {author} {\bibfnamefont {J.}~\bibnamefont {Redondo}},
  \bibinfo {author} {\bibfnamefont {K.}~\bibnamefont {Dolag}}, \bibinfo
  {author} {\bibfnamefont {J.~C.}\ \bibnamefont {Niemeyer}},\ and\ \bibinfo
  {author} {\bibfnamefont {A.}~\bibnamefont {Vaquero}},\ }\bibfield  {title}
  {\bibinfo {title} {{First Simulations of Axion Minicluster Halos}},\ }\href
  {https://doi.org/10.1103/PhysRevLett.125.041301} {\bibfield  {journal}
  {\bibinfo  {journal} {Phys. Rev. Lett.}\ }\textbf {\bibinfo {volume} {125}},\
  \bibinfo {pages} {041301} (\bibinfo {year} {2020})},\ \Eprint
  {https://arxiv.org/abs/1911.09417} {arXiv:1911.09417 [astro-ph.CO]}
  \BibitemShut {NoStop}%
\bibitem [{\citenamefont {Savastano}\ \emph {et~al.}(2019)\citenamefont
  {Savastano}, \citenamefont {Amendola}, \citenamefont {Rubio},\ and\
  \citenamefont {Wetterich}}]{Savastano:2019zpr}%
  \BibitemOpen
  \bibfield  {author} {\bibinfo {author} {\bibfnamefont {S.}~\bibnamefont
  {Savastano}}, \bibinfo {author} {\bibfnamefont {L.}~\bibnamefont {Amendola}},
  \bibinfo {author} {\bibfnamefont {J.}~\bibnamefont {Rubio}},\ and\ \bibinfo
  {author} {\bibfnamefont {C.}~\bibnamefont {Wetterich}},\ }\bibfield  {title}
  {\bibinfo {title} {{Primordial dark matter halos from fifth forces}},\ }\href
  {https://doi.org/10.1103/PhysRevD.100.083518} {\bibfield  {journal} {\bibinfo
   {journal} {Phys. Rev. D}\ }\textbf {\bibinfo {volume} {100}},\ \bibinfo
  {pages} {083518} (\bibinfo {year} {2019})},\ \Eprint
  {https://arxiv.org/abs/1906.05300} {arXiv:1906.05300 [astro-ph.CO]}
  \BibitemShut {NoStop}%
\bibitem [{\citenamefont {Flores}\ and\ \citenamefont
  {Kusenko}(2021)}]{Flores:2020drq}%
  \BibitemOpen
  \bibfield  {author} {\bibinfo {author} {\bibfnamefont {M.~M.}\ \bibnamefont
  {Flores}}\ and\ \bibinfo {author} {\bibfnamefont {A.}~\bibnamefont
  {Kusenko}},\ }\bibfield  {title} {\bibinfo {title} {{Primordial Black Holes
  from Long-Range Scalar Forces and Scalar Radiative Cooling}},\ }\href
  {https://doi.org/10.1103/PhysRevLett.126.041101} {\bibfield  {journal}
  {\bibinfo  {journal} {Phys. Rev. Lett.}\ }\textbf {\bibinfo {volume} {126}},\
  \bibinfo {pages} {041101} (\bibinfo {year} {2021})},\ \Eprint
  {https://arxiv.org/abs/2008.12456} {arXiv:2008.12456 [astro-ph.CO]}
  \BibitemShut {NoStop}%
\bibitem [{\citenamefont {Erickcek}\ \emph {et~al.}(2021)\citenamefont
  {Erickcek}, \citenamefont {Ralegankar},\ and\ \citenamefont
  {Shelton}}]{Erickcek:2020wzd}%
  \BibitemOpen
  \bibfield  {author} {\bibinfo {author} {\bibfnamefont {A.~L.}\ \bibnamefont
  {Erickcek}}, \bibinfo {author} {\bibfnamefont {P.}~\bibnamefont
  {Ralegankar}},\ and\ \bibinfo {author} {\bibfnamefont {J.}~\bibnamefont
  {Shelton}},\ }\bibfield  {title} {\bibinfo {title} {{Cannibal domination and
  the matter power spectrum}},\ }\href
  {https://doi.org/10.1103/PhysRevD.103.103508} {\bibfield  {journal} {\bibinfo
   {journal} {Phys. Rev. D}\ }\textbf {\bibinfo {volume} {103}},\ \bibinfo
  {pages} {103508} (\bibinfo {year} {2021})},\ \Eprint
  {https://arxiv.org/abs/2008.04311} {arXiv:2008.04311 [astro-ph.CO]}
  \BibitemShut {NoStop}%
\bibitem [{\citenamefont {Blinov}\ \emph {et~al.}(2021)\citenamefont {Blinov},
  \citenamefont {Dolan}, \citenamefont {Draper},\ and\ \citenamefont
  {Shelton}}]{Blinov:2021axd}%
  \BibitemOpen
  \bibfield  {author} {\bibinfo {author} {\bibfnamefont {N.}~\bibnamefont
  {Blinov}}, \bibinfo {author} {\bibfnamefont {M.~J.}\ \bibnamefont {Dolan}},
  \bibinfo {author} {\bibfnamefont {P.}~\bibnamefont {Draper}},\ and\ \bibinfo
  {author} {\bibfnamefont {J.}~\bibnamefont {Shelton}},\ }\bibfield  {title}
  {\bibinfo {title} {{Dark Matter Microhalos From Simplified Models}},\ }\href
  {https://doi.org/10.1103/PhysRevD.103.103514} {\bibfield  {journal} {\bibinfo
   {journal} {Phys. Rev. D}\ }\textbf {\bibinfo {volume} {103}},\ \bibinfo
  {pages} {103514} (\bibinfo {year} {2021})},\ \Eprint
  {https://arxiv.org/abs/2102.05070} {arXiv:2102.05070 [astro-ph.CO]}
  \BibitemShut {NoStop}%
\bibitem [{\citenamefont {Dai}\ and\ \citenamefont
  {Miralda-Escud\'e}(2020)}]{Dai:2019lud}%
  \BibitemOpen
  \bibfield  {author} {\bibinfo {author} {\bibfnamefont {L.}~\bibnamefont
  {Dai}}\ and\ \bibinfo {author} {\bibfnamefont {J.}~\bibnamefont
  {Miralda-Escud\'e}},\ }\bibfield  {title} {\bibinfo {title} {{Gravitational
  Lensing Signatures of Axion Dark Matter Minihalos in Highly Magnified
  Stars}},\ }\href {https://doi.org/10.3847/1538-3881/ab5e83} {\bibfield
  {journal} {\bibinfo  {journal} {Astron. J.}\ }\textbf {\bibinfo {volume}
  {159}},\ \bibinfo {pages} {49} (\bibinfo {year} {2020})},\ \Eprint
  {https://arxiv.org/abs/1908.01773} {arXiv:1908.01773 [astro-ph.CO]}
  \BibitemShut {NoStop}%
\bibitem [{\citenamefont {Dror}\ \emph {et~al.}(2019)\citenamefont {Dror},
  \citenamefont {Ramani}, \citenamefont {Trickle},\ and\ \citenamefont
  {Zurek}}]{Dror:2019twh}%
  \BibitemOpen
  \bibfield  {author} {\bibinfo {author} {\bibfnamefont {J.~A.}\ \bibnamefont
  {Dror}}, \bibinfo {author} {\bibfnamefont {H.}~\bibnamefont {Ramani}},
  \bibinfo {author} {\bibfnamefont {T.}~\bibnamefont {Trickle}},\ and\ \bibinfo
  {author} {\bibfnamefont {K.~M.}\ \bibnamefont {Zurek}},\ }\bibfield  {title}
  {\bibinfo {title} {{Pulsar Timing Probes of Primordial Black Holes and
  Subhalos}},\ }\href {https://doi.org/10.1103/PhysRevD.100.023003} {\bibfield
  {journal} {\bibinfo  {journal} {Phys. Rev. D}\ }\textbf {\bibinfo {volume}
  {100}},\ \bibinfo {pages} {023003} (\bibinfo {year} {2019})},\ \Eprint
  {https://arxiv.org/abs/1901.04490} {arXiv:1901.04490 [astro-ph.CO]}
  \BibitemShut {NoStop}%
\bibitem [{\citenamefont {Yoo}\ \emph {et~al.}(2021)\citenamefont {Yoo},
  \citenamefont {Harada}, \citenamefont {Hirano}, \citenamefont {Okawa},\ and\
  \citenamefont {Sasaki}}]{Yoo:2021fxs}%
  \BibitemOpen
  \bibfield  {author} {\bibinfo {author} {\bibfnamefont {C.-M.}\ \bibnamefont
  {Yoo}}, \bibinfo {author} {\bibfnamefont {T.}~\bibnamefont {Harada}},
  \bibinfo {author} {\bibfnamefont {S.}~\bibnamefont {Hirano}}, \bibinfo
  {author} {\bibfnamefont {H.}~\bibnamefont {Okawa}},\ and\ \bibinfo {author}
  {\bibfnamefont {M.}~\bibnamefont {Sasaki}},\ }\bibfield  {title} {\bibinfo
  {title} {{Primordial black hole formation from massless scalar
  isocurvature}},\ }\href@noop {} {\  (\bibinfo {year} {2021})},\ \Eprint
  {https://arxiv.org/abs/2112.12335} {arXiv:2112.12335 [gr-qc]} \BibitemShut
  {NoStop}%
\bibitem [{\citenamefont {Gurian}\ \emph {et~al.}(2021)\citenamefont {Gurian},
  \citenamefont {Jeong}, \citenamefont {Hwang},\ and\ \citenamefont
  {Noh}}]{Gurian:2021rfv}%
  \BibitemOpen
  \bibfield  {author} {\bibinfo {author} {\bibfnamefont {J.}~\bibnamefont
  {Gurian}}, \bibinfo {author} {\bibfnamefont {D.}~\bibnamefont {Jeong}},
  \bibinfo {author} {\bibfnamefont {J.-C.}\ \bibnamefont {Hwang}},\ and\
  \bibinfo {author} {\bibfnamefont {H.}~\bibnamefont {Noh}},\ }\bibfield
  {title} {\bibinfo {title} {{Gauge-Invariant Tensor Perturbations Induced from
  Baryon-CDM Relative Velocity and the B-mode Polarization of the CMB}},\
  }\href@noop {} {\  (\bibinfo {year} {2021})},\ \Eprint
  {https://arxiv.org/abs/2104.03330} {arXiv:2104.03330 [astro-ph.CO]}
  \BibitemShut {NoStop}%
\bibitem [{\citenamefont {Hu}\ \emph {et~al.}(1995)\citenamefont {Hu},
  \citenamefont {Bunn},\ and\ \citenamefont {Sugiyama}}]{Hu:1995xs}%
  \BibitemOpen
  \bibfield  {author} {\bibinfo {author} {\bibfnamefont {W.}~\bibnamefont
  {Hu}}, \bibinfo {author} {\bibfnamefont {E.~F.}\ \bibnamefont {Bunn}},\ and\
  \bibinfo {author} {\bibfnamefont {N.}~\bibnamefont {Sugiyama}},\ }\bibfield
  {title} {\bibinfo {title} {{COBE constraints on baryon isocurvature
  models}},\ }\href {https://doi.org/10.1086/309562} {\bibfield  {journal}
  {\bibinfo  {journal} {Astrophys. J. Lett.}\ }\textbf {\bibinfo {volume}
  {447}},\ \bibinfo {pages} {L59} (\bibinfo {year} {1995})},\ \Eprint
  {https://arxiv.org/abs/astro-ph/9501034} {arXiv:astro-ph/9501034}
  \BibitemShut {NoStop}%
\bibitem [{\citenamefont {Bucher}\ and\ \citenamefont
  {Zhu}(1997)}]{Bucher:1996gg}%
  \BibitemOpen
  \bibfield  {author} {\bibinfo {author} {\bibfnamefont {M.}~\bibnamefont
  {Bucher}}\ and\ \bibinfo {author} {\bibfnamefont {Y.}~\bibnamefont {Zhu}},\
  }\bibfield  {title} {\bibinfo {title} {{NonGaussian isocurvature
  perturbations from Goldstone modes generated during inflation}},\ }\href
  {https://doi.org/10.1103/PhysRevD.55.7415} {\bibfield  {journal} {\bibinfo
  {journal} {Phys. Rev. D}\ }\textbf {\bibinfo {volume} {55}},\ \bibinfo
  {pages} {7415} (\bibinfo {year} {1997})},\ \Eprint
  {https://arxiv.org/abs/astro-ph/9610223} {arXiv:astro-ph/9610223}
  \BibitemShut {NoStop}%
\bibitem [{\citenamefont {Linde}\ and\ \citenamefont
  {Mukhanov}(1997)}]{Linde:1996gt}%
  \BibitemOpen
  \bibfield  {author} {\bibinfo {author} {\bibfnamefont {A.~D.}\ \bibnamefont
  {Linde}}\ and\ \bibinfo {author} {\bibfnamefont {V.~F.}\ \bibnamefont
  {Mukhanov}},\ }\bibfield  {title} {\bibinfo {title} {{Nongaussian
  isocurvature perturbations from inflation}},\ }\href
  {https://doi.org/10.1103/PhysRevD.56.R535} {\bibfield  {journal} {\bibinfo
  {journal} {Phys. Rev. D}\ }\textbf {\bibinfo {volume} {56}},\ \bibinfo
  {pages} {R535} (\bibinfo {year} {1997})},\ \Eprint
  {https://arxiv.org/abs/astro-ph/9610219} {arXiv:astro-ph/9610219}
  \BibitemShut {NoStop}%
\bibitem [{\citenamefont {Peebles}(1997)}]{Peebles:1997av}%
  \BibitemOpen
  \bibfield  {author} {\bibinfo {author} {\bibfnamefont {P.~J.~E.}\
  \bibnamefont {Peebles}},\ }\bibfield  {title} {\bibinfo {title} {{An
  isocurvature model for early galaxy assembly}},\ }\href
  {https://doi.org/10.1086/310738} {\bibfield  {journal} {\bibinfo  {journal}
  {Astrophys. J. Lett.}\ }\textbf {\bibinfo {volume} {483}},\ \bibinfo {pages}
  {L1} (\bibinfo {year} {1997})}\BibitemShut {NoStop}%
\bibitem [{\citenamefont {Komatsu}(2001)}]{Komatsu:2001ysk}%
  \BibitemOpen
  \bibfield  {author} {\bibinfo {author} {\bibfnamefont {E.}~\bibnamefont
  {Komatsu}},\ }\emph {\bibinfo {title} {{The pursuit of non-gaussian
  fluctuations in the cosmic microwave background}}},\ \href@noop {} {Ph.D.
  thesis},\ \bibinfo  {school} {Tohoku U.} (\bibinfo {year} {2001}),\ \Eprint
  {https://arxiv.org/abs/astro-ph/0206039} {arXiv:astro-ph/0206039}
  \BibitemShut {NoStop}%
\bibitem [{\citenamefont {Garcia-Bellido}\ \emph {et~al.}(2017)\citenamefont
  {Garcia-Bellido}, \citenamefont {Peloso},\ and\ \citenamefont
  {Unal}}]{Garcia-Bellido:2017aan}%
  \BibitemOpen
  \bibfield  {author} {\bibinfo {author} {\bibfnamefont {J.}~\bibnamefont
  {Garcia-Bellido}}, \bibinfo {author} {\bibfnamefont {M.}~\bibnamefont
  {Peloso}},\ and\ \bibinfo {author} {\bibfnamefont {C.}~\bibnamefont {Unal}},\
  }\bibfield  {title} {\bibinfo {title} {{Gravitational Wave signatures of
  inflationary models from Primordial Black Hole Dark Matter}},\ }\href
  {https://doi.org/10.1088/1475-7516/2017/09/013} {\bibfield  {journal}
  {\bibinfo  {journal} {JCAP}\ }\textbf {\bibinfo {volume} {09}},\ \bibinfo
  {pages} {013}},\ \Eprint {https://arxiv.org/abs/1707.02441} {arXiv:1707.02441
  [astro-ph.CO]} \BibitemShut {NoStop}%
\bibitem [{\citenamefont {Dent}\ \emph {et~al.}(2012)\citenamefont {Dent},
  \citenamefont {Easson},\ and\ \citenamefont {Tashiro}}]{Dent:2012ne}%
  \BibitemOpen
  \bibfield  {author} {\bibinfo {author} {\bibfnamefont {J.~B.}\ \bibnamefont
  {Dent}}, \bibinfo {author} {\bibfnamefont {D.~A.}\ \bibnamefont {Easson}},\
  and\ \bibinfo {author} {\bibfnamefont {H.}~\bibnamefont {Tashiro}},\
  }\bibfield  {title} {\bibinfo {title} {{Cosmological constraints from CMB
  distortion}},\ }\href {https://doi.org/10.1103/PhysRevD.86.023514} {\bibfield
   {journal} {\bibinfo  {journal} {Phys. Rev. D}\ }\textbf {\bibinfo {volume}
  {86}},\ \bibinfo {pages} {023514} (\bibinfo {year} {2012})},\ \Eprint
  {https://arxiv.org/abs/1202.6066} {arXiv:1202.6066 [astro-ph.CO]}
  \BibitemShut {NoStop}%
\end{thebibliography}%

\end{document}